\documentclass{aa}

\usepackage{txfonts}
\usepackage{natbib}
\usepackage{graphicx}

\bibpunct{(}{)}{;}{a}{}{,} 

\begin{document}

\title{The dust distribution in edge-on galaxies}
\subtitle{Radiative transfer fits of V and K'-band images\thanks{
Based on observations made with the Italian Telescopio Nazionale 
Galileo (TNG) operated on the island of La Palma by the Fundación 
Galileo Galilei of the INAF (Istituto Nazionale di Astrofisica) at 
the Spanish Observatorio del Roque de los Muchachos of the Instituto 
de Astrofisica de Canarias}\fnmsep\thanks{Figures \ref{n4217_f} to 
\ref{u4277_p} are only available in electronic form via 
http://www.edpsciences.org}}

\author{Simone Bianchi}
\institute{INAF-Istituto di Radioastronomia, Sezione di Firenze,  
           Largo Enrico Fermi 5, 50125 Firenze, Italy \\
           \email{sbianchi@arcetri.astro.it}
          }

\date{Received 15 April 2007 / Accepted 3 May 2007}

\abstract
{}
{I have analyzed a sample of seven nearby edge-on galaxies observed in the V and
K'-band, in order to infer the properties of the dust distribution.}
{A radiative transfer model, including scattering, have been used to decompose
each image into a stellar disk, a bulge, and a dust disk. The parameters describing
the distributions have been obtained through standard $\chi^2$ minimization 
techniques.
}
{The dust disks fitted to the V-band images are consistent with previous work
in literature: the radial scalelength of dust is larger than that
for stars ($h_\mathrm{d}/h_\mathrm{s} \sim 1.5$); the dust disk has a smaller 
vertical scalelength than the stellar ($z_\mathrm{d}/z_\mathrm{s} \sim 1/3$); 
the dust disk is almost transparent when seen face-on (central, face-on, optical
depth $\tau_0 =0.5-1.5$). Faster radiative transfer models which neglect scattering
can produce equivalent fits, with changes in the derived parameters within the 
accuracy of full fits including scattering. In the K'-band, no trace is found of a 
second, massive, dust disk which has been invoked to explain observations of dust 
emission in the submillimeter. I discuss the model degeneracies and the effect of 
complex structures on the fitted distributions. In particular, most bulges in the
sample show a box/peanuts morphology with large residuals; two lower-inclination 
galaxies show a dust ring distribution, which could be the cause for the large 
fitted dust scalelengths.
}
{}

\keywords{dust, extinction -- galaxies: ISM - galaxies: stellar content - galaxies: spiral}

\maketitle

\section{Introduction}

Edge-on spiral galaxies offer a unique opportunity to study the 
three dimensional structure of disks: the vertical and radial
behavior of the stellar distribution can be analyzed in a rather 
direct manner, provided the heavily extinguished dust lane is 
avoided and/or observations are taken in a band not strongly
affected by dust \citep[see, e.g.][]{VanDerKruitA&A1982,
DeGrijsA&AS1996,PohlenA&A2000b,FloridoA&A2001}; the properties of 
the dust distribution can be inferred from the extinction lane 
itself. In this second case, however, the analysis requires 
three-dimensional models for the radiative transfer of starlight 
through dust, which are computationally demanding, especially 
when proper geometries and scattering are taken into account.

To date, only \citet{KylafisApJ1987} and 
\citet{XilourisA&A1997,XilourisA&A1998,XilourisSub1998} have 
fitted the surface brightness distribution in edge-on spiral galaxies 
by using realistic radiative transfer models. In the works of
\citeauthor{XilourisSub1998}, two independent disks were used to 
describe the stellar and dust distribution (both having a radial and
vertical exponential fall-off), to which a spheroidal distribution was 
added to model the bulge. After analyzing a sample of seven galaxies,
\citet{XilourisSub1998} concluded that the dust disk is thinner (vertically)
but larger (radially) than the stellar; its optical depth perpendicular
to the disk is smaller than one in optical wavebands, making the disk 
almost transparent when seen face-on.

The dust disk emerging from the work of \citeauthor{XilourisSub1998} poses
a problem, since such a structure cannot absorb more than 10\% of the
stellar radiation. Observations in the infrared, instead, show that 
$\sim$30\% of the total bolometric emission of a spiral galaxy is
radiated by dust \citep{PopescuMNRAS2002} and thus have been absorbed 
from starlight. The conservation of this {\em energy balance} demands
a larger amount of dust in the disk than what derived from optical 
images \citep{BianchiA&A2000b}. The deficit in absorption is mitigated 
if a sizable fraction of starlight is assumed to suffer extinction
within localized areas (i.e.\ star forming region in molecular clouds) 
and not only diffuse extinction within the disk. Even taking this 
into account, models fail to predict the emission from the colder dust 
component in the FIR/sub-mm and a supplemental dust distribution is 
needed \citep{PopescuA&A2000,MisiriotisA&A2001}.

In the models for the edge-ons NGC~891 and NGC~5907, \citet{PopescuA&A2000} 
and \citet{MisiriotisA&A2001} include a second dust disk, associated with
the younger stellar population. Being thinner than the disk resulting
from the analysis of \citeauthor{XilourisSub1998}, the second disk is
not detectable in optical images, as its effects occurs in a region 
already heavily affected by the extinction from the first, thicker disk.
Nevertheless, the second dust disk is required to be a factor 2-3 more
massive than the first. An alternative explanation for the model failures
in the FIR/sub-mm is that the adopted dust emissivity, derived from
the diffuse emission in the Milky Way, is underestimated by
about a factor 3 \citep{DasyraA&A2005}.

To test the reliability of radiative transfer fitting techniques 
and ascertain if an extra dust disk component is indeed present, 
I analyze in this paper a sample of seven edge-on galaxies, observed
in the V and in the K' bands. The second dust disk should be 
discernible in NIR observations, due to its higher optical depth
with respect to the thicker disk: \citet{DasyraA&A2005} analyzed
a K$_n$-band image of NGC~891 and found an extinction lane consistent
with a single disk model. In this work I will be able to carry the 
same analysis on a higher resolution and deeper set of NIR images.
The galaxies were selected because of the presence of an evident dust
lane, their symmetrical appearance and their closeness, so as to 
provide optimal ground for the application of radiative transfer 
fitting techniques. They all are of morphological type Sb to Sc.
The sample is listed in Tab.~\ref{galdata}. Two objects (NGC4013 and 
NGC5529) have been extensively analyzed in \citet{XilourisSub1998}: 
this will allow to check on the uniqueness of the results of fitting 
procedures.

The paper is structured as follows: the observations are described
in Sect.~\ref{obs}; the radiative transfer model and the procedure 
adopted for fitting are presented in Sect.~\ref{model}; the results 
of fits to observations are shown in Sect.~\ref{results}; finally, 
the main points are summarized and discussed in Sect.~\ref{summary}.

\begin{table}
\begin{minipage}[t]{\columnwidth}
\caption{The sample}
\centering
\label{galdata}
\renewcommand{\footnoterule}{}  
\begin{tabular}{lccrc}
name  
&$T$\footnote{morphology parameter (RC3).}
&$D$\footnote{distance in Mpc, assuming $H_0 = 73$ km s$^{-1}$ Mpc$^{-1}$
and velocities w. r. t. the CMB dipole (from NED).} 
&$PA$\footnote{Position angle in degrees, from the V-band fit.}
&$RA$ \& $Dec$ (J2000) \footnote{Galactic centre, from the V-band fit.}
\\ \hline\\
\object{NGC 4013}&3&14.5&244.8& 11 58 31.3  +43 56 50.8\\
\object{NGC 4217}&3&16.9& 49.3& 12 15 51.1  +47 05 29.8\\
\object{NGC 4302}&5&20.2& 50.8& 12 21 42.3  +14 35 53.9\\
\object{NGC 5529}&5&41.9&294.1& 14 15 34.2  +36 13 37.5\\
\object{NGC 5746}&3&26.6&350.2& 14 44 56.1  +01 57 18.8\\
\object{NGC 5965}&3&47.4& 52.6& 15 34 02.4  +56 41 07.5\\
\object{UGC 4277}&6&76.5&109.5& 08 13 57.1  +52 38 53.0\\
\end{tabular}
\end{minipage}
\end{table}

\section{Observations \& data reduction}
\label{obs}

Observations were carried out in March and April 2006, at the 3.5~m 
TNG (Telescopio Nazionale Galileo) telescope, located at the Roque de 
Los Muchachos Observatory in La Palma, Canary Islands.

V-band images were obtained in dark time using the DOLORES instrument
in imaging mode \citep{MolinariMmSAI1997}. The pixel scale is 0\farcs275 
and the field of view 9\farcm4 x 9\farcm4, a size well suited to the 
galaxies in the sample. For each object, three dithered exposures were 
taken, for a total of 750 seconds. Standard data reduction was carried out 
using the STARLINK package CCDPACK \citep{DraperMan2002}. As conditions 
during the observing nights were not photometric, calibration was achieved 
using the total V-band magnitude of the galaxies from RC3 \citep{RC3}. 
The sky noise in the V-band images is typically 25.8 mag arcsec$^{-2}$ 
(1-$\sigma$) and the seeing $1\farcs8$ (FWHM).

K'-band images were obtained with the NICS instrument in wide-field imaging
\citep{BaffaA&A2001}. Because of the smaller field of view (4\farcm2 x 
4\farcm2, with pixel scale 0\farcs25), the camera was aligned with the 
galactic plane of each galaxy using the RC3 position angle; two 
overlapping fields were observed, each with an offset of 100'' from the 
galactic center. Each field was observed with a dithered pattern of
12 on and off positions, each consisting of 6 short exposure of 20 sec 
on the source (and on the sky) for a total exposure of 1440 second on 
source. Due to its smaller extent on the sky, only UGC~4277 required
a single telescope pointing, with the galactic plane aligned with the 
camera diagonal. Data reduction was carried out with the dedicated 
software SNAP (Speedy  Near-infrared data Automatic Pipeline; Mannucci 
et al., in preparation) which takes care of flat-fielding, sky subtraction, 
corrections for geometrical distortion and electronic effects, and final 
image mosaicing. Calibration was obtained from objects in the 2MASS point 
source catalog \citep{SkrutskieAJ2006}. The sky noise in the K'-band 
images is typically 21.2 mag arcsec$^{-2}$ (1-$\sigma$), about a
magnitude deeper than in 2MASS Large Galaxy Atlas images
\citep{JarrettAJ2003}. The seeing during observations was $1\farcs0$ (FWHM).

\section{Modelling \& fitting}
\label{model}

The analysis presented in this work consists in producing a mock image 
of a galaxy and comparing it with the observed image by means of $\chi^2$ 
minimization techniques, with the aim of deriving the parameters that 
better describe the object.

\subsection{The galactic model}
\label{geometry}

A standard description is adopted for the disk 
\citep[see, e.g.\ ][]{XilourisSub1998}, an exponential (both along the radial 
coordinate $r$ and the vertical coordinate $z$) with luminosity density (per 
unit solid angle)
\begin{equation}
\rho^\mathrm{disk}(r,z)=\frac{I_0^\mathrm{disk}}{2 \; z_\mathrm{s}} 
  \exp\left[-\frac{r}{h_\mathrm{s}} -\frac{|z|}{z_\mathrm{s}}\right],
\label{stardisk}
\end{equation}
where $I_0^\mathrm{disk}$ is the disk {\em face-on} surface brightness through the
galactic centre and $h_\mathrm{s}$ and $z_\mathrm{s}$ are the radial and vertical 
scalelength.

For the bulge I have used a de-projected Sersic profile of index $n$, 
with luminosity density given by
\begin{equation}
\rho^\mathrm{bulge}(r,z)=\frac{I_0^\mathrm{bulge}}{A_n\; R_\mathrm{e}\; b/a} 
\frac{\exp(-b_n\; B^{1/n})}{B^\alpha},
\label{sersic}
\end{equation}
where $I_0^\mathrm{bulge}$ is the bulge {\em face-on} surface brightness through the
galactic centre, $R_\mathrm{e}$ is the effective radius, $b/a$ is the minor/major axis 
ratio of the bulge,
\[
B=\frac{\sqrt{r^2+z^2/(b/a)}}{R_\mathrm{e}},
\]
\[
b_n=2n-1/3-0.009876/n, \qquad \alpha=(2n-1)/2n
\]
and
$A_n=2.73, 3.70, 4.47, 5.12$ for $n=1, 2, 3, 4$, respectively 
\citep{PrugnielA&A1997}. When projected on the sky plane, Eq.~\ref{sersic}
corresponds to a \citet{SersicBook1968} profile of index $n$. Traditionally, 
bulges in spiral galaxies have been described with the $R^{1/4}$ 
\citep{DeVaucouleursBook1959} profile typical of elliptical galaxies ($n=4$).
Recent work, however \citep[see, e.g.\ ][]{HuntA&A2004}, suggests that bulges 
in late type galaxies follow preferentially Sersic profiles with smaller 
$n$. Here I use $n=2$ and 4.

The model includes a single dust disk with extinction coefficient given by
\begin{equation}
\kappa(r,z)=\frac{\tau_0}{2 \; z_\mathrm{d}} 
  \exp\left[-\frac{r}{h_\mathrm{d}} -\frac{|z|}{z_\mathrm{d}}\right],
\label{dustdisk}
\end{equation}
where $\tau_0$ is the central, face-on, optical depth of the dust disk
and $h_\mathrm{d}$ and $z_\mathrm{d}$ are the radial and vertical scalelength. 

For computational reasons, the stellar disk is generally truncated along 
the radius at $4 h_\mathrm{s}$ and the dust 
disk at $4 h_\mathrm{d}$. Observations suggest that stellar disks follow the 
simple exponential decline of Eq.~\ref{stardisk} up to about $4 h_\mathrm{s}$, 
although the truncation is not sharp and there is a large scatter in 
the measures \citep[see the review in][]{PohlenProc2004}. No indications
are available for the dust disk. The stellar and dust disks are truncated 
vertically at $6 z_\mathrm{s}$ and at $6 z_\mathrm{d}$, while the bulge 
extends to $10 R_\mathrm{e}$. 

The simulated image is created with the same pixel resolution and extent
as the observed image with which it is compared. The geometrical parameters 
describing the appearance of the model galaxy on the image are $\theta$, the 
inclination of the galactic z-axis with respect to the line of sight (l.o.s; 
$\theta =90^\circ$ for the pure edge-on case), the position of the projection
of the galactic center on the sky, and the position angle ($PA$). $PA$ is 
defined so that, by rotating the galaxy counterclockwise by (90-$PA$) 
degrees, the projection on the sky of the positive $z$-axis of the galaxy 
lies along the positive $y$-axis on the image.

I have used two radiative transfer models to produce simulated images.
In the first, faster, model scattering is not taken into account (the 
extinction coefficient in Eq.~\ref{dustdisk} is taken to be entirely
due to absorption). The image surface brightness is derived by analytic 
integration of the dust attenuated stellar luminosity $\rho e^{-\tau}$
along the l.o.s.  passing through the center of each image pixel
(with the optical depth $\tau$ the integral of Eq.~\ref{dustdisk} 
from the location of stellar emission to the observer).

The second model is a Monte Carlo (MC) radiative transfer code including 
scattering. The basics of the method are presented in \citet{BianchiApJ1996}.
The code has been rewritten and optimized along the lines described in
\citet{BaesMNRAS2003}. In particular, the {\em peeling-off} technique
\citep{YusefZadehApJ1984} has been implemented, allowing to produce images 
at a specific inclination $\theta$ rather than for a broad inclination band 
as in our original paper.  As for the dust scattering properties, the albedo
$\omega$ and asymmetry parameter $g$ have been taken from the Milky Way 
dust grain model of \citet{WeingartnerApJ2001a}: for the V-band it is
$\omega=0.67$ and $g=0.54$, for the K'-band $\omega=0.45$ and $g=0.14$.
The extinction law gives $A_\mathrm{K'}/A_\mathrm{V}=0.12$.
The \citet{HenyeyApJ1941} phase function for scattering is used.

Finally, the model image is convolved with a gaussian of appropriate FWHM
to simulate the effect of seeing. This is important to reduce the central 
spike of the bulge, especially for the case of the analytic calculation 
where, contrary to MC images, the surface brightness is not smoothed
over the pixel area.

\subsection{The fitting procedure}
\label{proc}

Before proceeding to the fitting, stars and image defects have been masked
out from observations, together with all pixels laying out of the 3-$\sigma$ 
isophote. The sky level, derived from regions free of sources, has been
subtracted. 
To speed up the fitting process, images are further smoothed and rebinned
over 2x2 or 3x3 pixel, typically leaving 50000 and 25000 pixels available 
for the fit, in the V- and K'-band, respectively. Tests show that this further
rebinning does not affect the fit.

For fits with the analytic no-scattering model, I have tested
both the Levenberg-Marquardt and the downhill simplex (the {\em amoeba} 
algorithm) minimization techniques \citep{NumRec}. The first is quicker, 
but only when the initial guesses for the parameters are close to the final 
values, otherwise the technique tends to be trapped into local minima of 
the complex $\chi^2$ surface. The {\em amoeba} algorithm, instead, is able, 
after successive restarts of the procedure, to converge over minima quite 
distant, in parameter space, from the initial guess. When using the MC model,
the {\em amoeba} method is mandatory, because the statistic nature of MC 
images does not allow to quickly compute the derivatives required by the 
Levenberg-Marquardt method. In all cases, the fitting codes have been
implemented on a parallel machine: for the analytic method, calculations 
on groups of pixels are distributed among different {\em cpu}s and the
results collected in the final image; for the MC method, each {\em cpu} 
produces a whole image starting from an independent random seed, and all 
images are finally summed up to produce a higher signal-to-noise result.

To resume, the parameters obtained through the fit are twelve: 
$I_0^\mathrm{disk}$, $h_\mathrm{s}$, $z_\mathrm{s}$, $I_0^\mathrm{bulge}$, 
$R_\mathrm{e}$, $b/a$, $h_\mathrm{d}$, $z_\mathrm{d}$, $\tau_0$, $\theta$, 
$PA$ and the centre position. First, their values have been estimated from 
cuts parallel and perpendicular to the disk plane, as described in 
\citet{XilourisA&A1997}. Then, a no scattering fit was done, with successive 
calls of the {\em amoeba} and Levenberg-Marquardt methods, until a stable 
minimum was found. The final fit was achieved with the MC method, again 
with various calls of the {\em amoeba} method.

\subsection{Testing the procedure}
\label{tests}

The fitting techniques were extensively tested on a large set of
simulated images, covering a wide range of parameters. Simulated images
were produced using the full MC model including dust scattering. The
surface brightness in the models was scaled to the typical values
in observations. The pixel size was also chosen to match the 
observations, and appropriate noise and seeing were added.
Finally, the simulated images were prepared for fitting as described
in Sect.~\ref{proc}. 
For illustrative purposes, I discuss here the 
results of tests on a model based on the NGC~891 fit of 
\citet{XilourisSub1998}. Simulations were made for the the V- 
and K'-band (the dust disk fitted to the V-band image was used in both 
cases, with $\tau_0^{K'}$ scaled according to the Milky Way extinction law) 
and for inclinations of 89.8$^\circ$ (the fitted galaxy inclination), 
88$^\circ$ and 86$^\circ$. 

\begin{figure*}
\centering
\includegraphics[height=4.5cm]{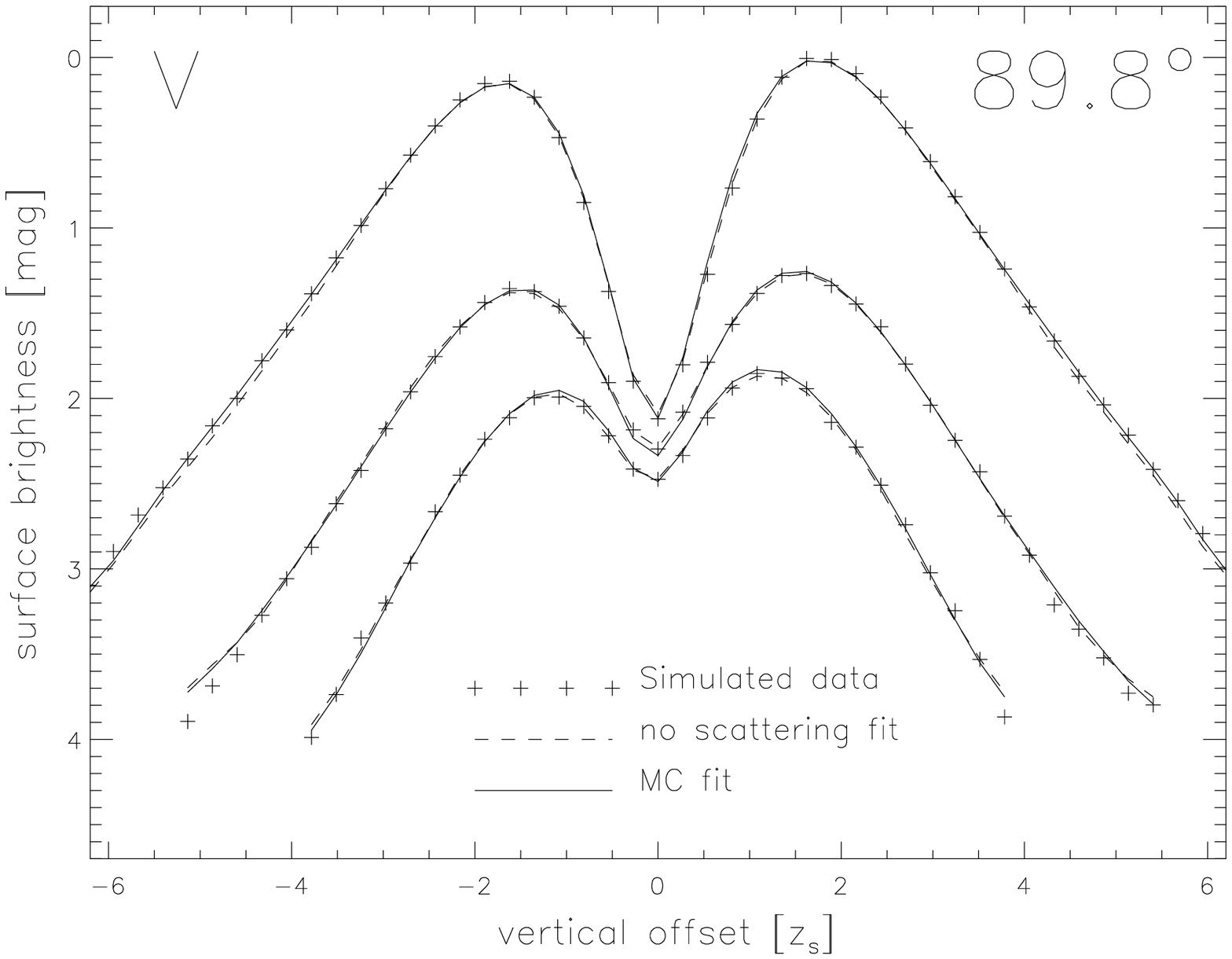}
\includegraphics[height=4.5cm]{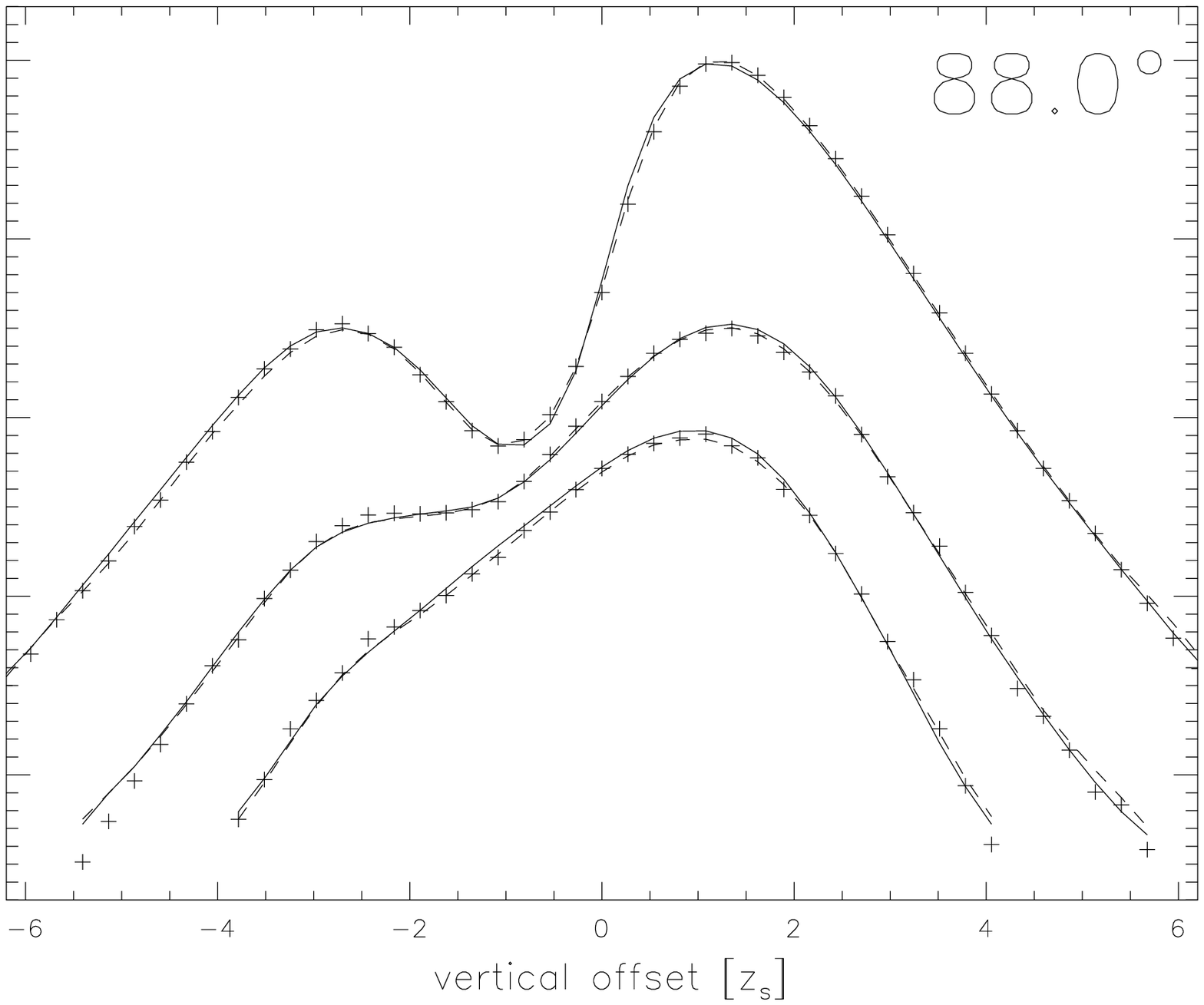}
\includegraphics[height=4.5cm]{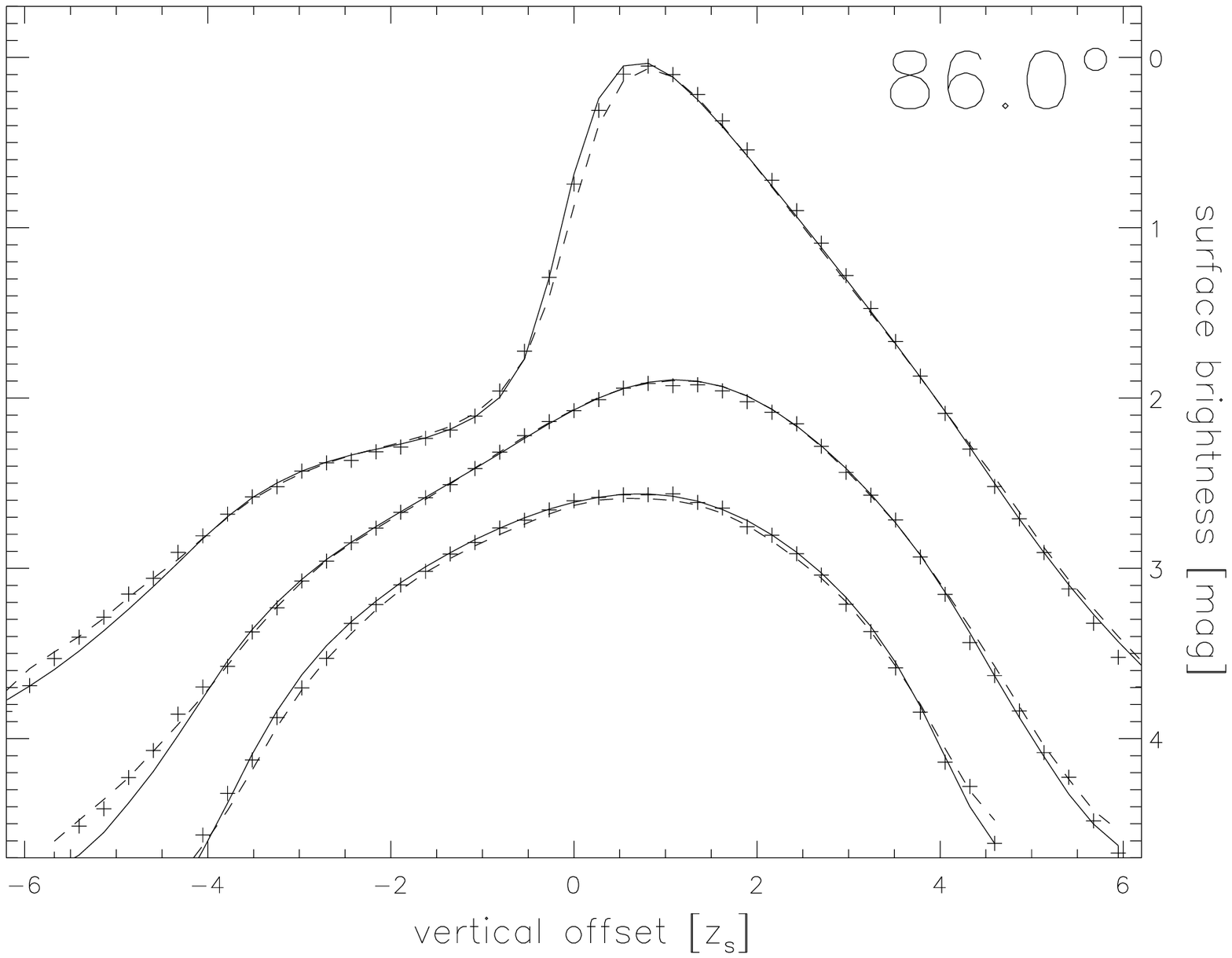}
\caption{V-band surface brightness profiles perpendicular to the galactic 
disk for the model and the fits described in Sect.~\ref{tests}. Three cuts 
are shown, at distances 0 (brightest profile), 1 and 2 $h_\mathrm{s}$ 
(dimmest profile) from the center along the galaxy's major axis.}
\label{testV}
\end{figure*}

In Fig.~\ref{testV} I show the V-band surface brightness profiles 
of models (crosses) and fits (lines) for the three inclination cases.
In each case, the profiles are obtained along three vertical cuts
(perpendicular to the galactic plane) passing through the model centre 
and at distances from the model of $h_\mathrm{s}$ and $2 h_\mathrm{s}$.
As outlined in Sect.~\ref{proc}, fits were first produced with the
no-scattering analytical model (dashed line), then the results used 
as starting point for the MC fit (solid line). The MC fits reproduce 
very well the data and most of the input parameters can be retrieved 
with accuracies within 10\%, gradually worsening for images of lower
$\theta$ (whose value, even in the worst cases, is retrieved with 
an error of 2\%). The values of $R_\mathrm{e}$ and $I_0^\mathrm{bulge}$ 
suffer the largest error (up to 20\% and 40\%, respectively, in the 
image with $\theta=86^\circ$). 

Fig.~\ref{testV} also shows that the data can be reproduced equally 
well by the faster analytical fits which neglect scattering. The
no scattering fit can reproduce the surface brightness of a model
including scattering with slight changes in the parameters, which are
still derived with the same accuracy discussed before (apart from
$z_\mathrm{d}$, $I_0^\mathrm{disk}$ and $I_0^\mathrm{bulge}$, which 
can have errors as high as 20\%, 20\% and 50\%, respectively). 
The optical depth $\tau_0$ is underestimated \citep[for a given 
optical depth, neglecting scattering results in an overestimate 
of the attenuation;][]{BaesMNRAS2001b}, but no more than 20\%. 
In fact, the effects of scattering are minimal in edge-on galaxies
\citep[see, e.g. ][]{BianchiApJ1996}. 

\begin{figure*}
\centering
\includegraphics[height=4.5cm]{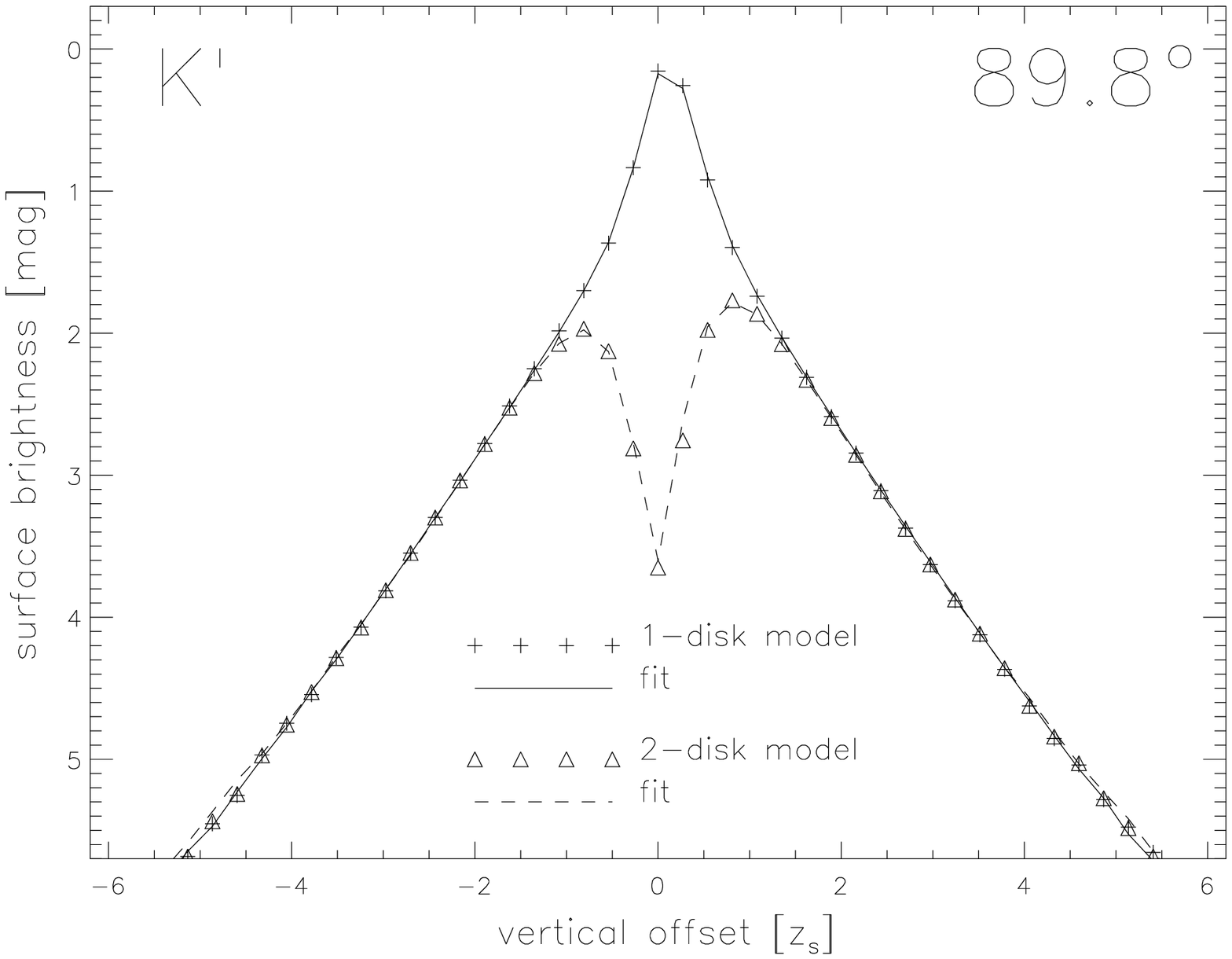}
\includegraphics[height=4.5cm]{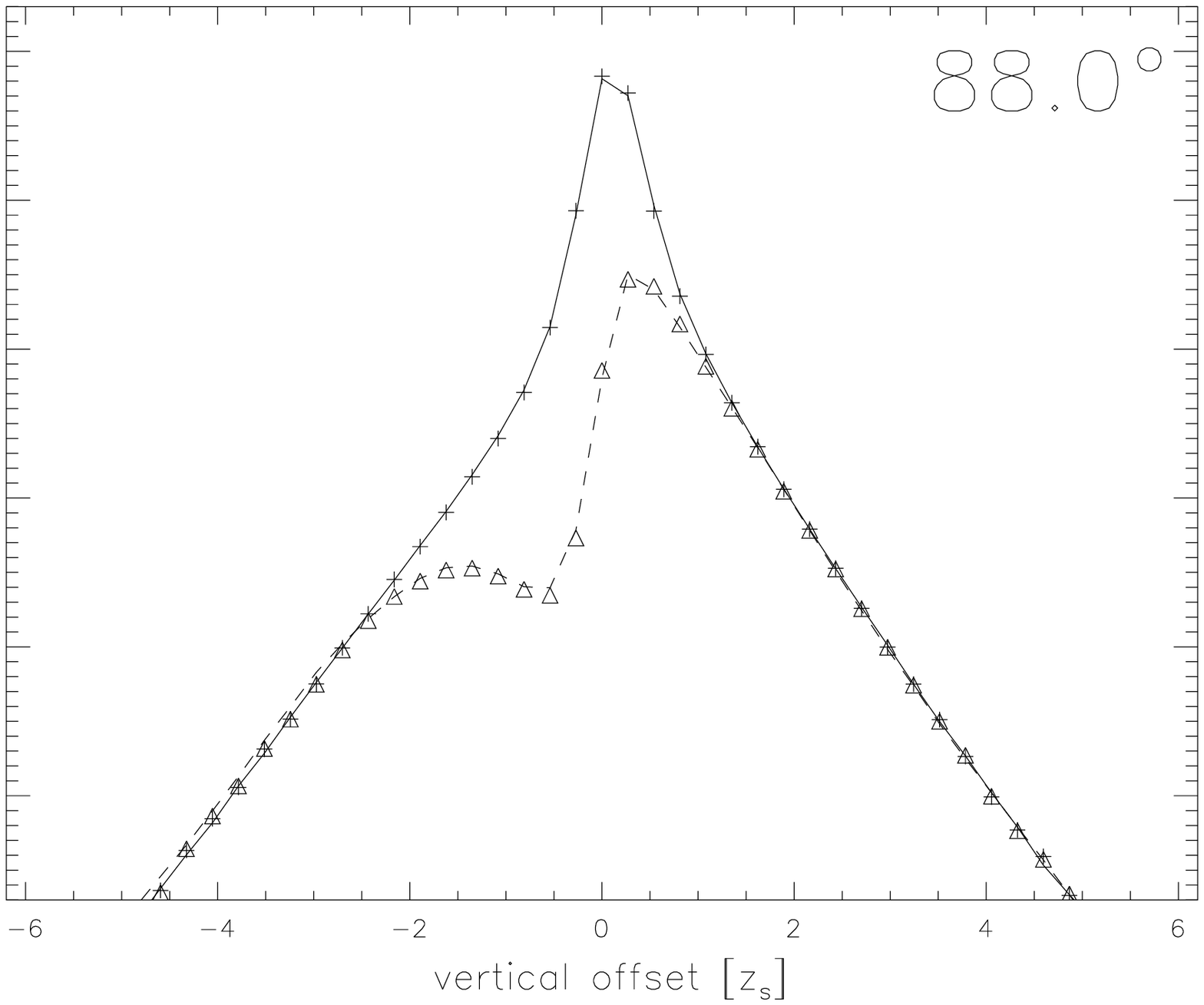}
\includegraphics[height=4.5cm]{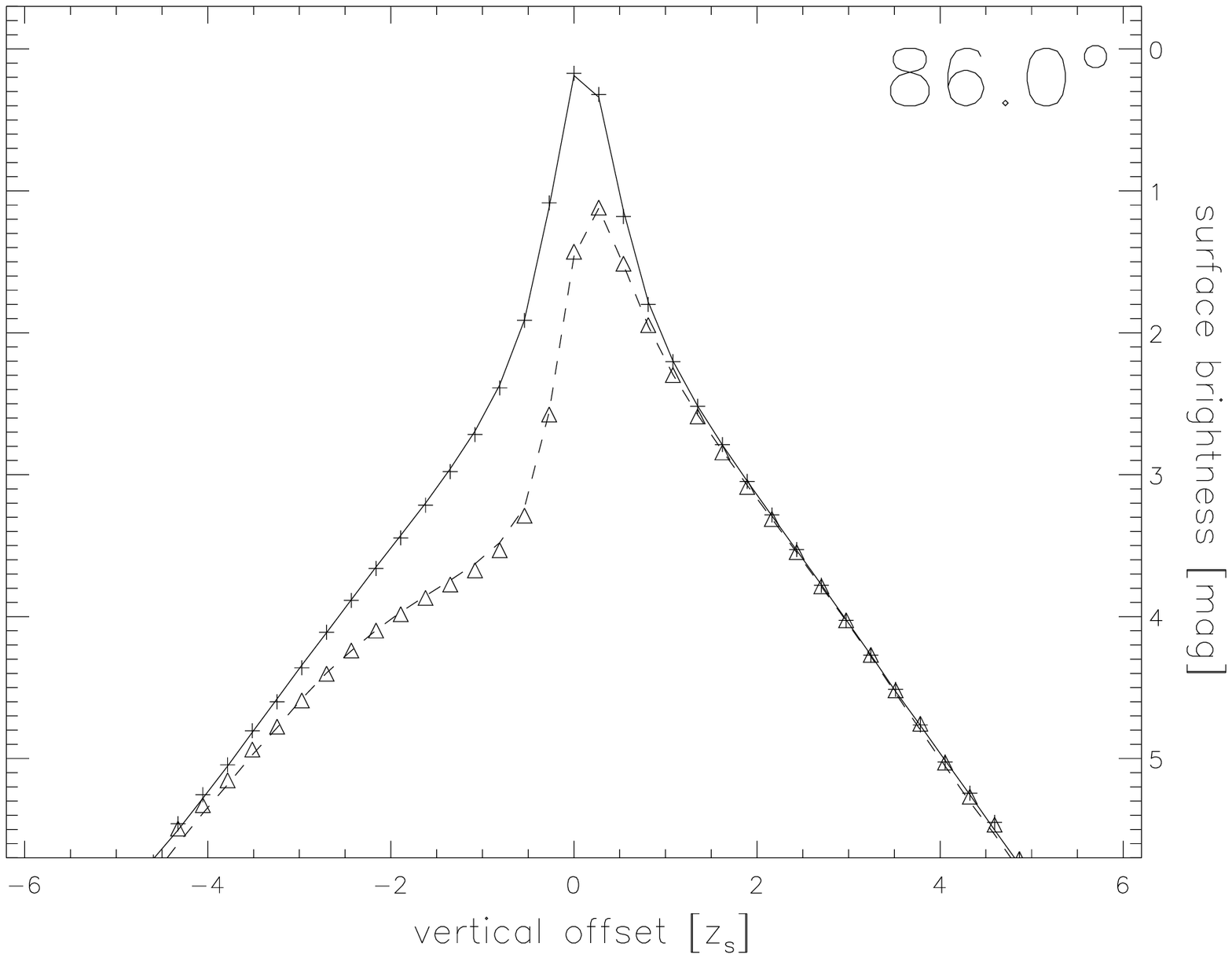}
\caption{K'-band surface brightness profiles along the minor axis
for the models with a single dust disk and the model with two dust disks,
together with their fits.}
\label{testK}
\end{figure*}

Models in K'-band (see Fig.~\ref{testK}) have reduced extinction and the
presence of dust can only be inferred from small asymmetries along
the minor axis direction. Therefore, it is more difficult to obtain 
the parameters of the dust disk: the fitted $h_\mathrm{d}$, $z_\mathrm{d}$ 
and $\tau_0$ are within 20\% of the input values. For the same reason, 
the parameters defining the stellar distributions are retrieved 
with slightly better accuracy than for the V-band case.

Other systematic effects were also explored using these simulations. 
Since only pixels with high S/N were chosen, a constant sky was 
subtracted from the images without attempting to fit its level. 
However, even assuming that the sky has been miscalculated by 
$\pm 3\sigma$, the parameters resulting from the fits are still 
very close to the input values, with the uncertainties quoted above.
The same is true if the radial truncation of the stellar disk
is varied within the estimates currently available from 
observations. I confirm the correlation observed by 
\citet{PohlenA&A2000a}: an underestimate in the truncation 
of the stellar disk causes an overestimate of $h_\mathrm{s}$. 
Asymmetries and dishomogeneities in real images may hamper the
correct determination of the inclination angle, especially for
objects very close to edge-on: an overestimate of $\theta$ (so 
that the fit is more edge-on than the galaxy) would cause an 
overestimate of $z_\mathrm{s}$ and $z_\mathrm{d}$, leading also 
to larger $h_\mathrm{d}$ and smaller $\tau_0$. It is worth noting 
that in the cases discussed here, where the parameters are 
deliberately kept fixed on values different from the input ones, 
the residuals in the fit are generally smaller than 20\% for most 
(90\%) of the data points. When dealing with real images, the 
deviations from the smooth model adopted here cause residuals 
of the same order or larger (Sect.~\ref{results}). A large model 
degeneracy is thus possible.

Finally, I have added to the simulated images of NGC~891 a second
dust disk, following the recipes of \citet{TuffsA&A2004}. As a 
derivation of the parameters describing the second disk is not 
feasible, I have studied its effects on the single dust disk fitting 
model. The second, thin, dust disk has a central face-on, optical depth 
which is about 2.5 times that of the thick disk ($\tau_0=$0.22 in 
the thin disk vs $\tau_0=$0.09 in the thick disk). As pointed out by 
\citep{PopescuA&A2000}, its effects would not be easily discerned 
from those of the thick disk in the edge-on case. However, the second 
disk would be easily seen in all cases where the first produces a 
lower extinction, i.e. at lower $\theta$ and larger wavelength. 
Indeed, a deeper absorption trough is visible in V-band simulated 
images, at all inclinations. It is also visible in K'-band simulations, 
as a deeper trough within 2$^\circ$ from the edge-on case, and as a 
stronger minor axis asymmetry at lower inclination (see Fig.~\ref{testK}). 
As Fig.~\ref{testK} shows, low-residual fits can be achieved in 
the K'-band, even if the simulated image include a second dust disk:
the parameters retrieved for the dust disk are now close to those
of the second, more opaque, disk.

\section{Results}
\label{results}

\begin{figure*}
\centering
\includegraphics[width=17cm]{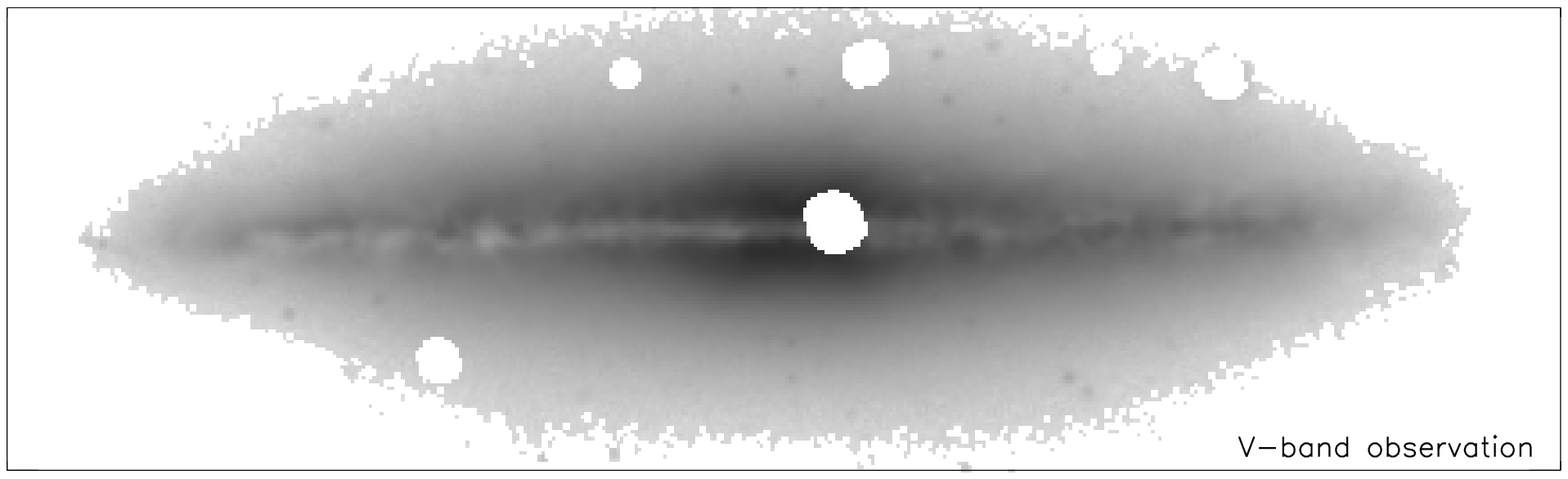}

\vspace{0.05cm}
\includegraphics[width=17cm]{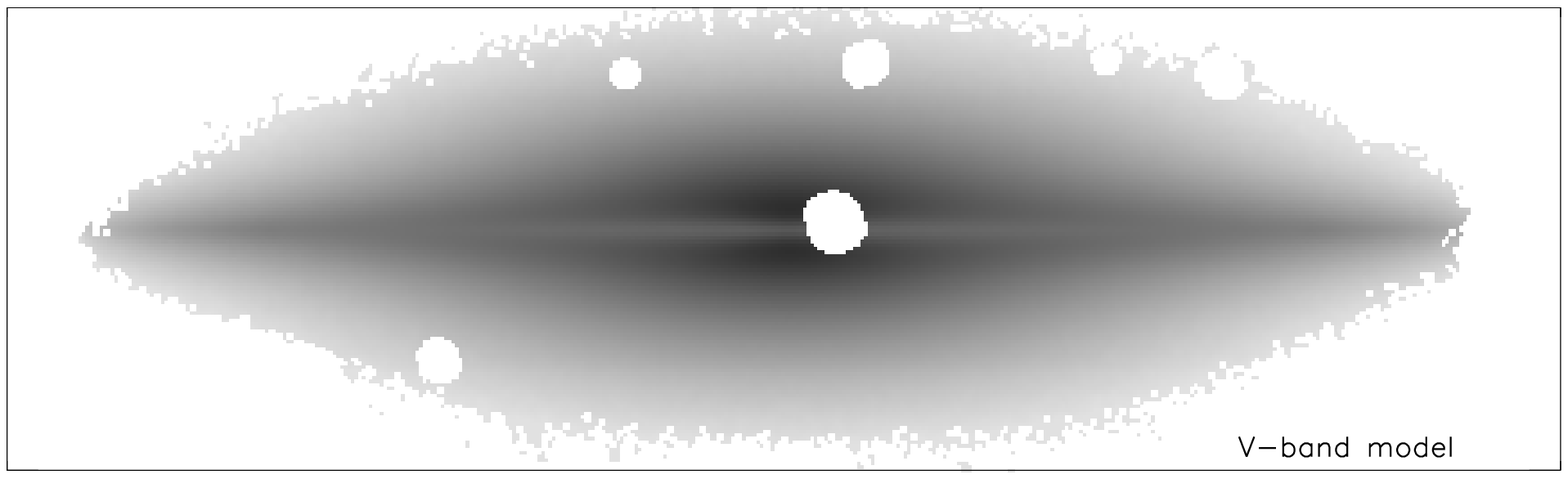}

\vspace{0.05cm}
\includegraphics[width=17cm]{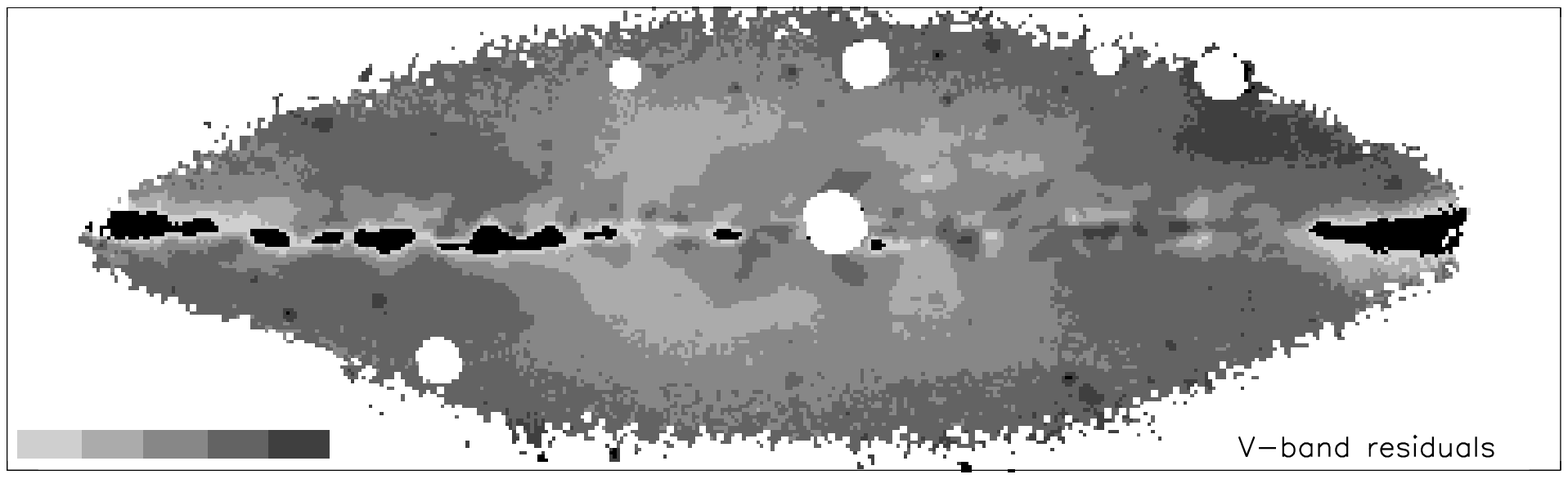}

\vspace{0.3cm}
\includegraphics[width=17cm]{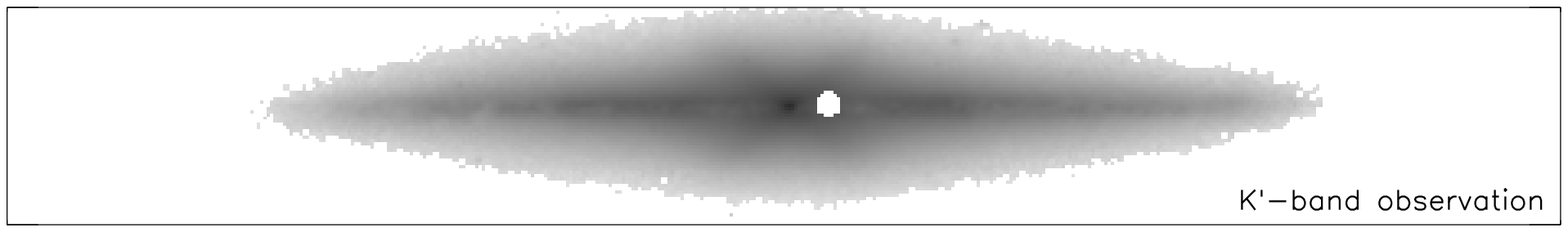}

\vspace{0.05cm}
\includegraphics[width=17cm]{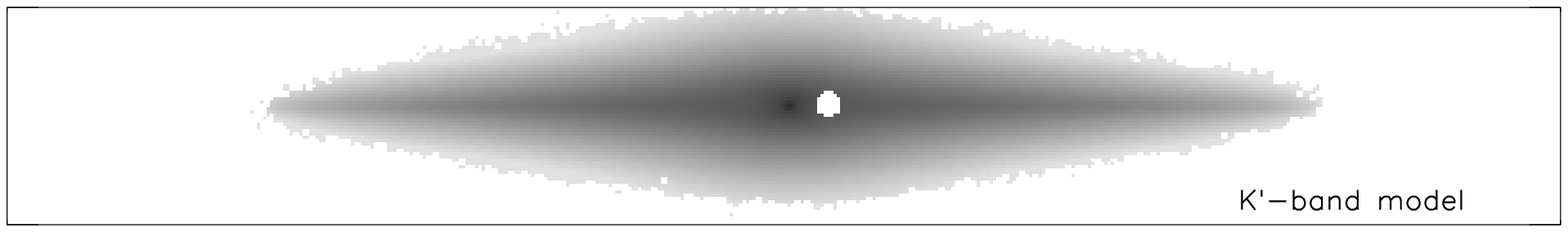}

\vspace{0.05cm}
\includegraphics[width=17cm]{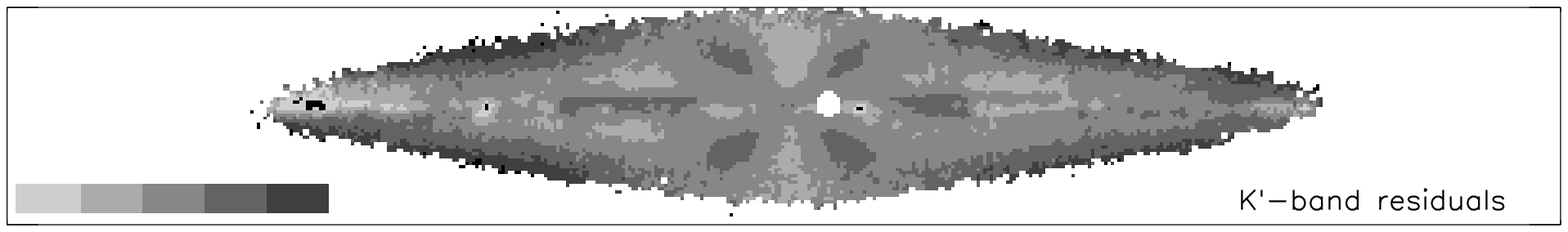}
\caption{NGC 4013. From top to bottom: the V-band image, the fit 
to the V-band image and the residuals; the K'-band image, the fit 
to the K'-band image and the residuals.  Images and models are 
displayed with the modified logarithmic visualization method of
\citet{JarrettAJ2003}. The scale of residuals is linear:
the bar with gray shades helps to distinguish regions with values 
in the ranges [-50\%,-30\%], [-30\%,-10\%], [-10\%,10\%], [10\%,30\%]
and [30\%,50\%] from lighter to darker, respectively. All regions 
with values outside these ranges are coded black. The horizontal 
size of the box is 6\arcmin.}
\label{n4013_f}
\end{figure*}

\begin{figure*}
\centering
\includegraphics[width=8.5cm]{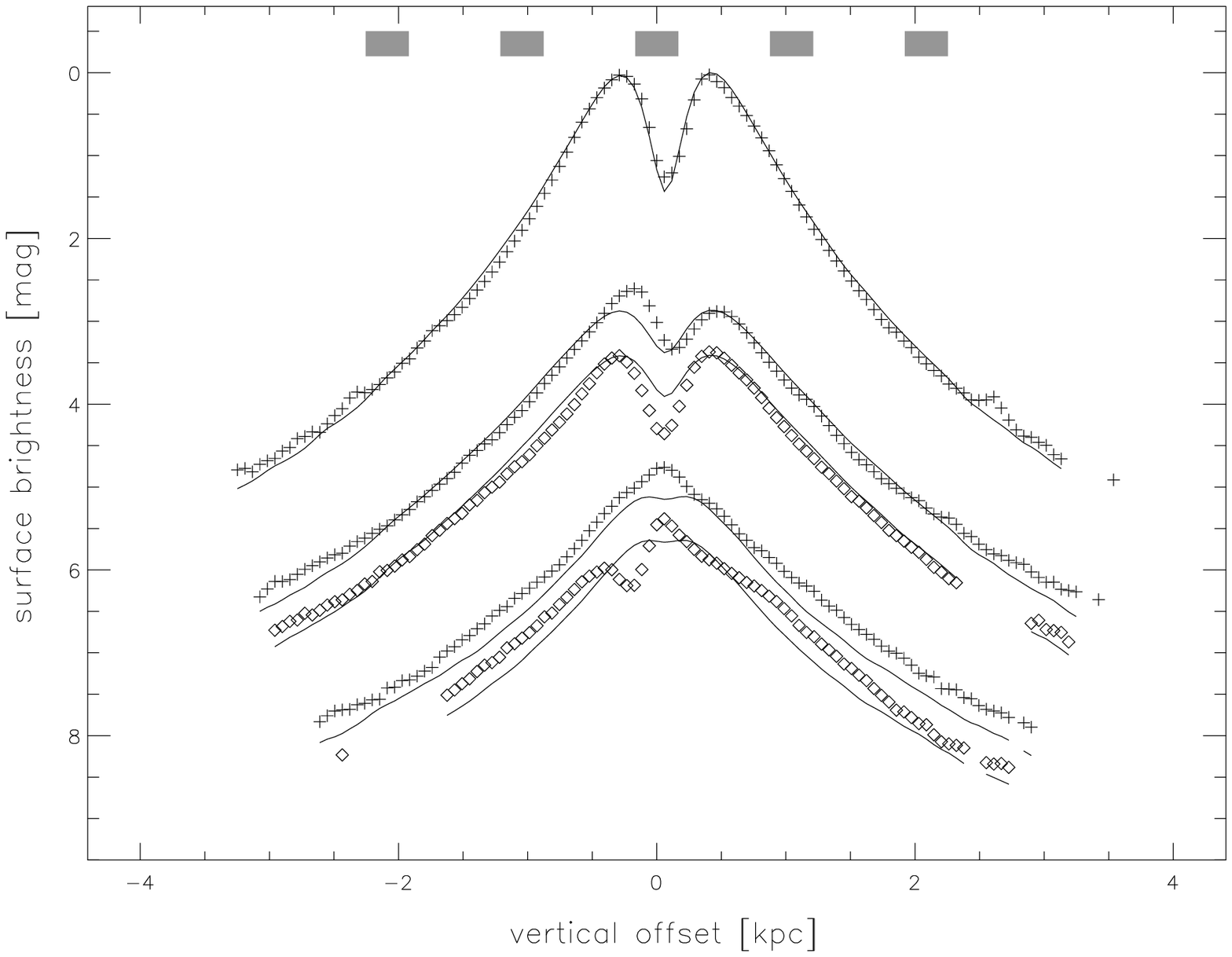}
\includegraphics[width=8.5cm]{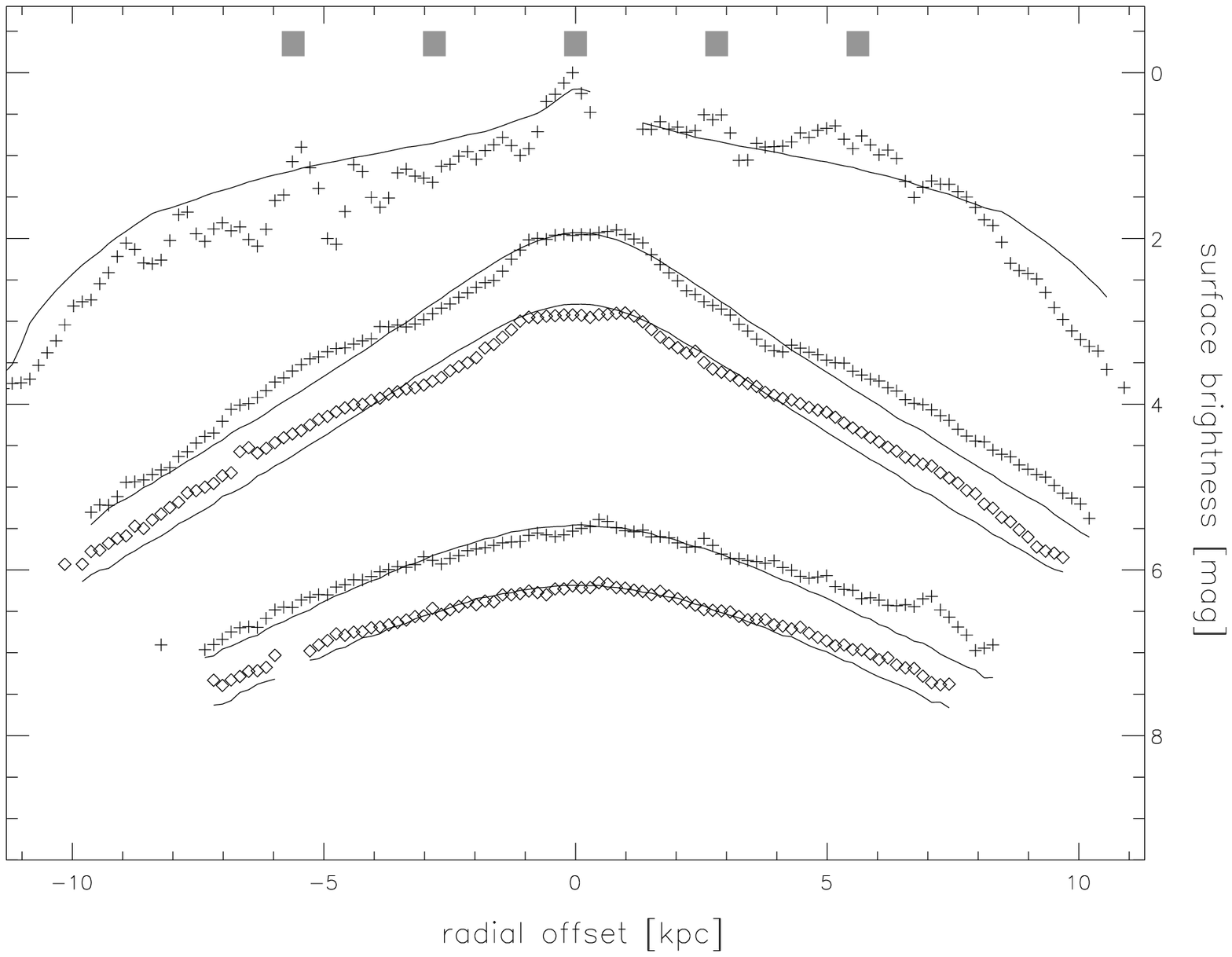}
\vspace{0.05cm}
\includegraphics[width=8.5cm]{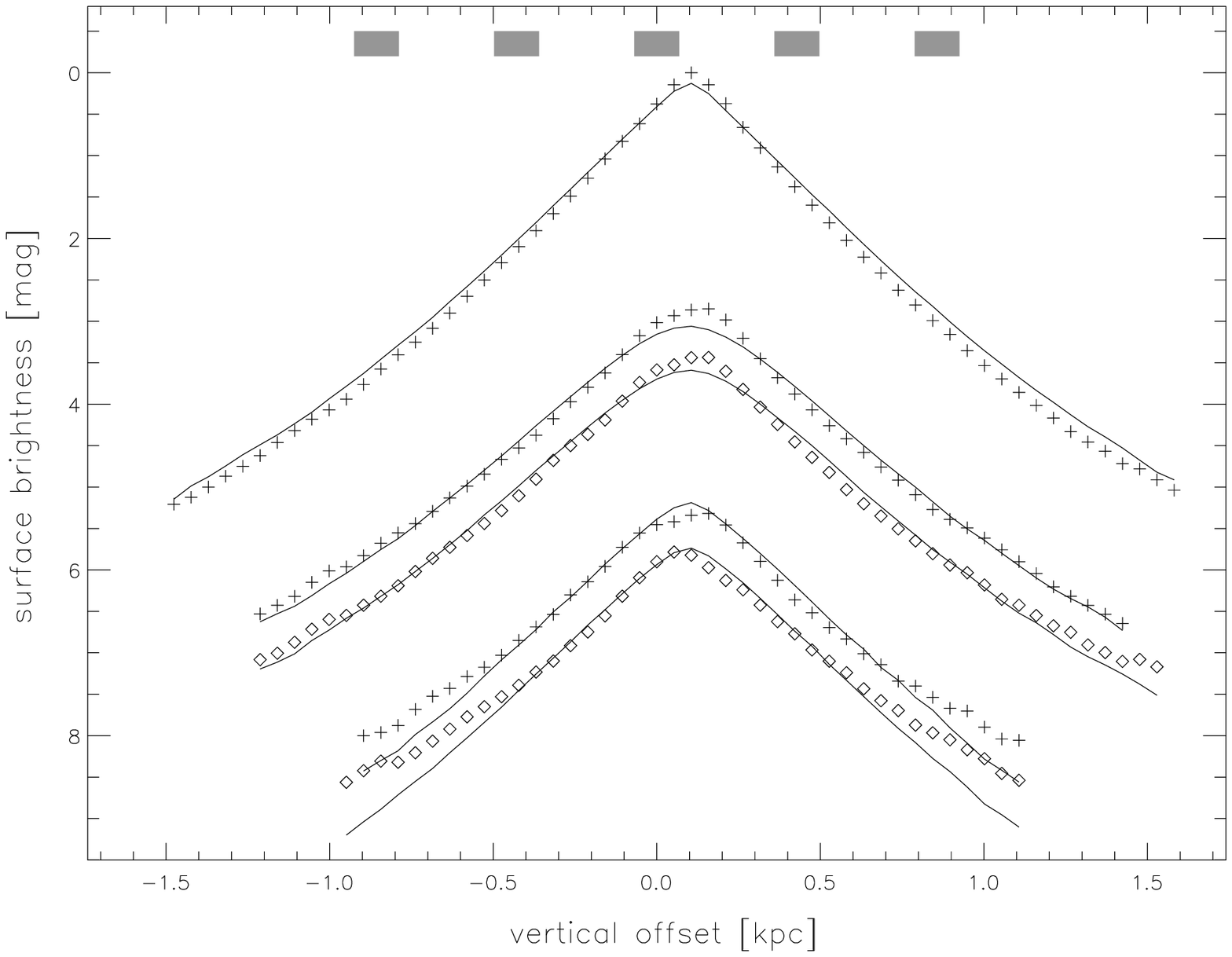}
\includegraphics[width=8.5cm]{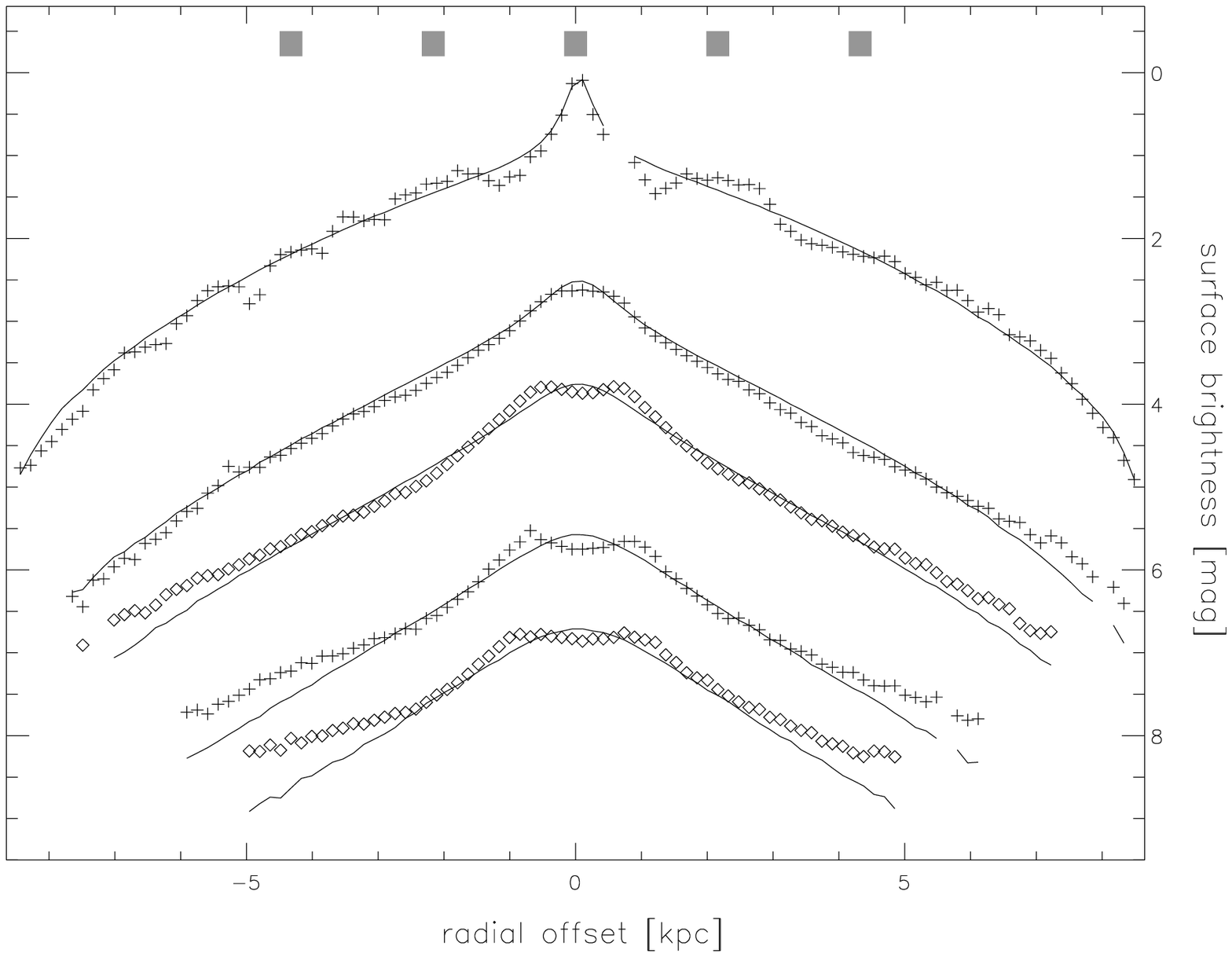}
\caption{NGC 4013. Surface brightness profiles for cuts perpendicular 
(left panels) and parallel (right panels) to the galactic disk, for 
the V-band (top panels) and the K'-band (bottom panels) images. In the 
left panels, profiles are shown for distances from the fitted center 
along the galactic plane corresponding to 0 (minor axis), $\pm$1/8 and
$\pm$1/4 of the horizontal extent of the fitted area. Median values 
have been taken on a cut width of 2\% that extent. The topmost data 
(crosses) refer to the minor axis, whose maximum has been
assigned magnitude zero. Crosses refer to profiles on the right side of 
the images (with respect to Fig.~\ref{n4013_f}) and are progressively 
displaced downward of 1.5 magnitudes when moving from the center outward. 
Diamonds refer to the left side profiles, and are further displaced 
by 0.5 magnitude from their specular profile on the right side.
The same scheme is adopted for the surface brightness profiles 
parallel to the major axis, shown in the right panels: profiles
are shown for cuts at 0 (major axis; crosses), 1/8 and 1/4 of the 
vertical extent of the fitted area, above (crosses) and below (diamonds) 
the galactic plane. Median values have been taken over a cut width of 
4\% that extent. To reduce the cluttering of symbols only one point 
every three is plotted in horizontal profiles. Profiles are displaced in 
the same manner as in the left panels. In all panels, the curves refer to 
analogous cuts on the fitted images. Gaps in both model and data correspond 
to masked areas (i.e.\ stars). The shaded boxes in the upper part of the
left (right) panels show the position and extent of the cuts displayed 
in the right (left) panels.
}
\label{n4013_p}
\end{figure*}

For each galaxy in the sample, I show in Fig.~\ref{n4013_f} and
\ref{n4217_f} to \ref{u4277_f} the image, the fit and its residual 
for the V and  K'-band, respectively. 
Images have been processed as
described in Sect.~\ref{proc} and rotated counterclockwise by 90-$PA$.
Cuts perpendicular and 
parallel to the galactic plane, through the center and at different 
positions on the images, are shown in Fig.~\ref{n4013_p} and 
\ref{n4217_p} to \ref{u4277_p}, for both bands. The best-fit parameters 
describing the stellar disk, the bulge and the dust disk are listed in 
Table~\ref{fit_table}, together with the galaxy's inclination $\theta$.
The galactic centre and the $PA$ fitted to V and K' images generally
agree within 1'' and 0\fdg5, respectively: those derived for V 
images are shown in Table~\ref{galdata}.

\begin{table*}
\caption{Fit results. $I_0^\mathrm{disk}$ and $I_0^\mathrm{bulge}$ are in 
mag arcsec$^{-2}$, the V-band $h_\mathrm{s}$ is in kpc, the K'-band 
$h_\mathrm{s}$ and all values for $z_\mathrm{s}$, $R_\mathrm{e}$, 
$h_\mathrm{d}$ and $z_\mathrm{d}$ are in units of the V-band $h_\mathrm{s}$. 
The Sersic index $n$ is not fitted. The bulge-to-total light ratio $B/T$ 
is computed from the stellar disk and bulge fit. The values for 
$h_\mathrm{d}$, $z_\mathrm{d}$ and $\theta$ are derived from the V-band 
fit and kept fixed in the K'-band fit.}
\label{fit_table}
\centering
\begin{tabular}{r|ccc|ccccc|ccc|c}
& \multicolumn{3}{c|}{stellar disk} 
& \multicolumn{5}{c|}{bulge} 
& \multicolumn{3}{c|}{dust disk} & \\
& $I_0^\mathrm{disk}$ & $h_\mathrm{s}$ & $z_\mathrm{s}$ &
  $I_0^\mathrm{bulge}$ & $R_\mathrm{e}$ & $b/a$ & $B/T$ & $n$ &
  $h_\mathrm{d}$ & $z_\mathrm{d}$ & $\tau_0$ & $\theta$ 
\\ \hline &&&&&&&&&&&& \\
NGC 4013 V &21.7& 2.89&0.13&12.1&0.66&0.37&0.68&4&0.72&0.05&1.46&89.89
\\
         K' &16.9& 0.75&0.09& 9.1&0.49&0.55&0.27&4&&&0.15&
\\ &&&&&&&&&&&& \\
NGC 4217 V &20.7& 3.31&0.06&13.0&1.31&0.29&0.59&4&1.75&0.10&1.26&88.01
\\
         K' &17.5& 0.89&0.06&11.0&1.91&0.47&0.55&4&&&0.17&
\\ &&&&&&&&&&&& \\
NGC 4302 V &22.3& 6.02&0.17&17.8&0.17&0.68&0.07&2&1.52&0.03&0.11&89.63
\\
         K' &17.4& 0.56&0.08&10.1&0.01&0.54&0.02&2&&&0.01&
\\ &&&&&&&&&&&& \\
NGC 5529 V &21.2& 7.26&0.08&17.4&0.47&0.49&0.21&2&1.36&0.03&0.68&86.94
\\
         K' &18.2& 0.89&0.08&13.0&0.15&0.71&0.11&2&&&0.02&
\\ &&&&&&&&&&&& \\
NGC 5746 V &20.4& 7.21&0.12&14.4&0.23&0.71&0.34&2&2.54&0.07&0.52&86.80
\\
         K' &17.7& 0.99&0.13&11.7&0.18&0.65&0.24&2&&&0.05&
\\ &&&&&&&&&&&& \\
NGC 5965 V &21.6& 9.69&0.10&15.2&0.21&0.72&0.39&2&1.82&0.02&0.62&84.43
\\
         K' &17.3& 0.56&0.14&11.6&0.11&0.52&0.21&2&&&0.02&
\\ &&&&&&&&&&&& \\
UGC 4277 V &23.4&12.27&0.09&18.2&0.38&0.41&0.41&2&1.02&0.02&0.49&88.89
\\
         K' &18.4& 0.61&0.07&12.8&0.11&0.71&0.19&2&&&0.02&
\\
\end{tabular}
\end{table*}

\subsection{Global trends}

The asymmetries and dishomogeneities in the stellar and dust distributions 
result in large residuals, which also highlight the presence of structures 
different from those adopted in Sect.~\ref{geometry}. From a qualitative
analysis of the fitted images and of the profiles, it appears that in four 
objects (NGC~4013, NGC~4217, NGC~5529 and UGC~4277) the model can describe 
reasonably well the V images, both for what concern the stellar components
and the extinction lanes produced by the dust disk. In two objects
(NGC~5746 and NGC~5965) the fit is marginal, with a good description
of the general appearance of the stellar distributions, but with extinction
features that are not reproduced by the model. In one object, NGC~4302,
the fit is not able to describe the stellar distribution out of the
galactic plane. In the K'-band, the general quality of the fits is similar 
to that in the V-band.

In a fit, the parameter that most quickly converge to the final value
is the radial scalelength $h_\mathrm{s}$. In Table~\ref{fit_table} the
V-band $h_\mathrm{s}$ is given in kpc. To allow a ready comparison,
all other scalelengths in Table~\ref{fit_table} have been normalized
to the V-band $h_\mathrm{s}$. The V-band $h_\mathrm{s}$ goes
from about 3 kpc in the two bulge-dominated galaxies NGC~4013 and 
NGC~4217 to about 12 kpc in UGC~4277. Indeed, such a wide range for
$h_\mathrm{s}$ has been found when fitting the surface brightness
distribution in large samples of less inclined galaxies 
\citep{DeJongA&A1996a,MacArthurApJ2003}. 

The K'-band $h_\mathrm{s}$ is generally smaller than in the V-band, 
with a median ratio 0.75. This is in agreement with several works 
in literature, which suggest that the reduction in $h_\mathrm{s}$ 
observed in low inclination galaxies when going from optical to NIR 
images is due to an intrinsic color gradient in the stellar population 
and not to the effects of dust extinction \citep{DeJongA&A1996b,
MacArthurApJ2003, MoellenhoffA&A2004,CunowMNRAS2004}. A notable 
exception is that of \citet{MoellenhoffA&A2006}, which ascribe the 
increase of radial scalelengths from B to I to the optically thick 
second dust disk of \citet{PopescuA&A2000}. There is no evidence,
instead, of a systematic difference between the vertical scalelength 
$z_\mathrm{s}$ in the V and K'-bands. In the V-band, it is 
$z_\mathrm{s}/h_\mathrm{s}\approx 0.1$, a result similar
to what obtained by \citet{XilourisSub1998}. 

More delicate is the derivation of bulge parameters. For most object,
I have used a spheroidal bulge with Sersic index $n=2$, a choice 
appropriate for the morphological type of the galaxies considered here
\citep{HuntA&A2004}. However, for two objects (NGC~4013 and NGC~4217)
the $n=2$ bulge was too steep to fit both the central peak and the
ellipsoidal appearance of more external isophotes: $n=4$ was used
in these cases. A spheroidal bulge appears to be a poor approximation
to the real stellar distribution: in all but one object, positive
residuals (i.e.\ model smaller than data) show 'X'-like features
in the fit to the K' or both bands. These features reveal families of 
orbits in which disk material is trapped by the perturbations caused 
by a bar \citep{PatsisMNRAS2006}, whose edge-on view causes a 
{\em boxy/peanut} appearance \citep{BureauMNRAS2006}. The deviations
from the assumed bulge model are likely to affect also the derivation
of $z_\mathrm{s}$ (because bulge and vertical disk structure contributes 
together to the surface brightness profile out of the disk plane near 
the centre) and of the parameters describing the dust disk (because the 
effects of dust extinction are more evident in the central part of the 
disk where the bulge contribution is sizable, if not dominant). 

Excluding the bad, almost no-bulge, NGC~4302 fit, the $n=2$ 
bulges have B/T ratios in the range 0.2-0.4 in the V-band, while the 
bulge contribution is larger in the two $n=4$ fits. In the K'-band fit,
the B/T ratio is always smaller than in the V-band, thus implying
that the color of the bulges is {\em bluer} than the color of the disks.
This is puzzling, being in contrast to what generally found in the 
bulge/disk decomposition of less inclined objects 
\citep[see, e.g. ][]{MoellenhoffA&A2004}. However, a similar variation
of the B/T ratios can be inferred from the fit of edge-on galaxies
in \citet{XilourisSub1998}.

The dust disks fitted to the V-band images appear to be consistent with 
those derived by \citet{XilourisSub1998}, although the spread in values 
is larger. Generally, the radial scalelength $h_\mathrm{d}$ is larger than 
the stellar ($h_\mathrm{d}/h_\mathrm{s} \sim 1.5$), while the vertical 
scalelength $z_\mathrm{d}$ is smaller than $z_\mathrm{s}$ 
($z_\mathrm{s}/z_\mathrm{d} \sim 3$). This last condition is necessary to 
produce the well defined extinction lanes along the galactic disk. The disk 
opacity is moderate, with a  V-band $\tau_0$ in the range 0.5-1.5, making
the galaxies almost transparent when seen face-on \citep{XilourisSub1998}. 

The extinction features in K' images are not prominent, and result mostly
in weak asymmetries along the minor axis and in clumpy structures. The 
asymmetries due to dust cannot be easily distinguished by those due to 
deviations from the adopted models. In most cases, a free fit would converge 
to a dust free model. Thus, the fits have been achieved by allowing $\tau_0$ 
only to vary, while fixing both dust disk scalelengths, 
$h_\mathrm{d}$ and 
$z_\mathrm{d}$, and the inclination 
$\theta$ to the result of the V-band fit. Small values for $\tau_0$, 
in the range 0.02-0.15 have been found. Such opacities only produces
minimal extinction effects on the model (see, e.g. the solid lines in
Fig.~\ref{testK}). 

Instead, no extinction lanes associated to a thin, massive, dust disk are 
found. If the extra dust component supposed by \citep{PopescuA&A2000}
exists, it does not appear to be in the form of a second, smooth, dust 
disk, which would produce more easily discernible features (see, e.g., 
the dashed lines in Fig.~\ref{testK}). 
The result is in agreement with 
the analysis of a K$_n$-band image of NGC~891 by \citet{DasyraA&A2005}.

\subsection{Comments on individual galaxies\protect\footnote{All literature 
data have been scaled to the distances listed in Table~\ref{galdata}.}}

\noindent {\em NGC 4013:} the fit is generally good for both bands, with 
residuals smaller than 30\% in 85\% of the fitted data points. The dust lane 
can be easily seen in both bands. It shows a clumpy structure which lead to 
larger residuals along the major axis, especially in the V-band. A bulge 
with Sersic index $n=4$ was used for this object. The V-band fit has parameters 
similar to those obtained by \citet{XilourisSub1998}, with the exception of 
$h_\mathrm{d}$, which I find to be smaller than $h_\mathrm{s}$, rather then 
larger. This also lead to a larger $\tau_0^V$ obtained here.

\noindent {\em NGC 4217:} the same considerations as for NGC~4013 apply to 
this object. The geometrical parameters of the stellar and dust disk,
together with $\theta$, are given for the B-band by \citet{AltonA&AS2000}: 
the fit here is more edge-on and $h_\mathrm{d}$ is larger by 30\%, while 
$h_\mathrm{s}$ is consistent with their fit. The difference in $\theta$ 
is also the cause of the smaller $z_\mathrm{s}$ I obtain. For this object 
I also have $z_\mathrm{d} > z_\mathrm{s}$: since the bulge dominates the 
surface brightness apart from regions close to the galactic plane, an 
extinction lane can appear even if the dust disk is thicker that the 
stellar. The major axis profiles clearly show a steepening of the profile 
beyond 10 kpc: this is the signature of a Type II truncation
\citep[in the notation of][]{PohlenA&A2006}, which cannot be reproduced 
by the sharp cut used in the model. A truncation at the same distance is 
also found in the K$_s$-band by \citet{FloridoA&A2001}, although the value
of $h_\mathrm{s}$ they obtain is larger by 40\% than in the present fit.
Dust extinction can be seen in the K'-band image, although not in the
form of a dust lane but rather as a y-axis asymmetry in images, because 
of the inclination smaller than edge-on.
 
\noindent {\em NGC 4302:} this is the worst object to fit, because of
the major axis asymmetry around the center, the presence of a surface
brightness contribution from the nearby face on galaxy NGC~4298 (whose
centre is at about 2\farcm3 from the galactic plane in the direction of 
the negative y axis in Fig.~\ref{n4302_f}) and because of the complex 
vertical structure. Indeed the model does not represent well the
observed surface brightness, especially in the V-band. The contribution 
of the fitted bulge is negligible. It appears that there is a second, 
thicker stellar disk (of V-band vertical scalelength 1.5 kpc, a quarter 
of the fitted $h_\mathrm{s}$), with radial scalelength increasing with 
the distance from the galactic plane.  Extinction in K' appears clumpy 
and does not alter significantly the symmetry above the plane.

\noindent {\em NGC 5529:} apart from a slight warp in the external part
of the disk, a fit can be achieved, with residuals smaller than 30\% in 
70\% of the data points. The fit of the disk is consistent with the 
analysis of a V-band image in \citet{XilourisSub1998}. Only tenuous, 
clumpy extinction can be seen in the K' image.

\noindent {\em NGC 5746:} the fit quality is the same as for NGC 5529,
in both bands. The fit to the extinction lane in the V-band, though, is 
poorer.  The parameters for the stellar and dust disks are not consistent 
with those reported for the B-band by \citet{AltonA&AS2000}, with 
$h_\mathrm{s}$ larger by 40\% in the present case. Extinction is
located in a clumpy ring structure of radius $\approx$1\farcm3, which
is evident in the K' image. From the ring in K', an inclination 
$\theta=86^\circ\pm 1^\circ$ can be estimated, which is consistent with
the fitted value.

\noindent {\em NGC 5965:} the galaxy is warped, with the outer disk and
a ring-like dust lane on a different plane with respect to the inner
peanut-shaped bulge. The stellar ring is evident in the K' image,
while no extinction features can be detected. As with NGC 5746, the fit
to the extinction lane in the V-band is poor, despite the global
fit quality is not much different from the other objects. An estimate 
for $h_\mathrm{s}$ is available for the K$_s$-band which is about 20\% smaller 
than the present value \citep{BizyaevA&A2002}. The fitted $\theta$ is 
consistent with the aspect of the K' ring, which suggests 
$\theta=83.5^\circ\pm 1.5^\circ$.

\noindent {\em UGC 4277:} the fit is good, with 90\% of the data points
fitted by the model with residuals smaller than 30\%. While the dust
lane is evident for this nearly edge-on object in the V-band, in the
K'-band only a weak asymmetry can be seen for the bulge above and 
below the galactic plane. The K-$_s$band value for $h_\mathrm{s}$ derived 
by \citet{BizyaevA&A2002} is consistent with the result of the present 
fit.


\onlfig{5}{
\begin{figure*}
\centering
\includegraphics[width=17cm]{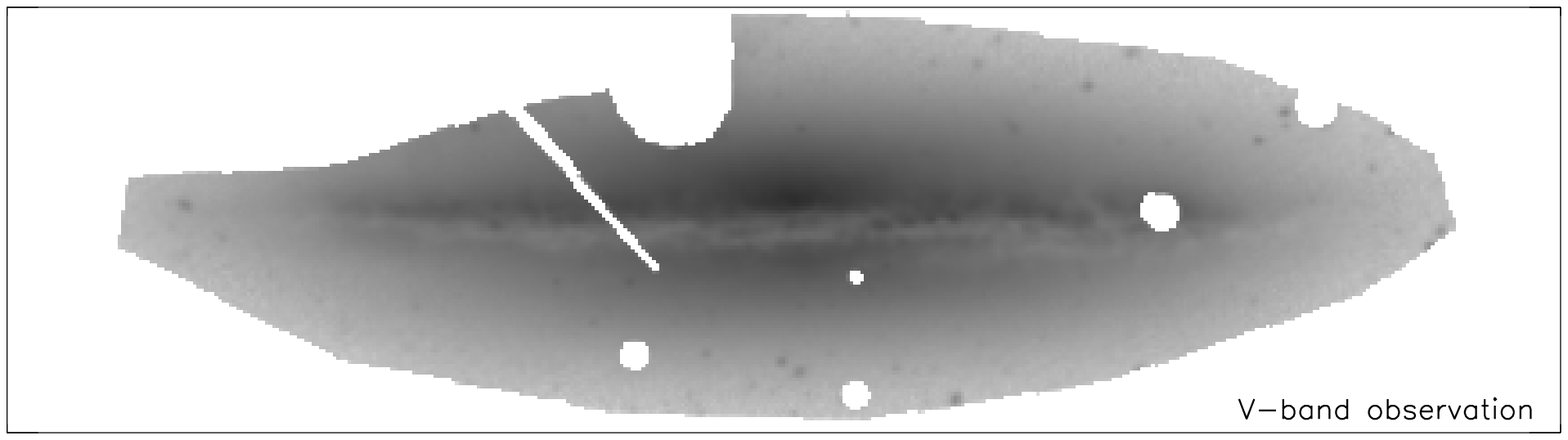}

\vspace{0.05cm}
\includegraphics[width=17cm]{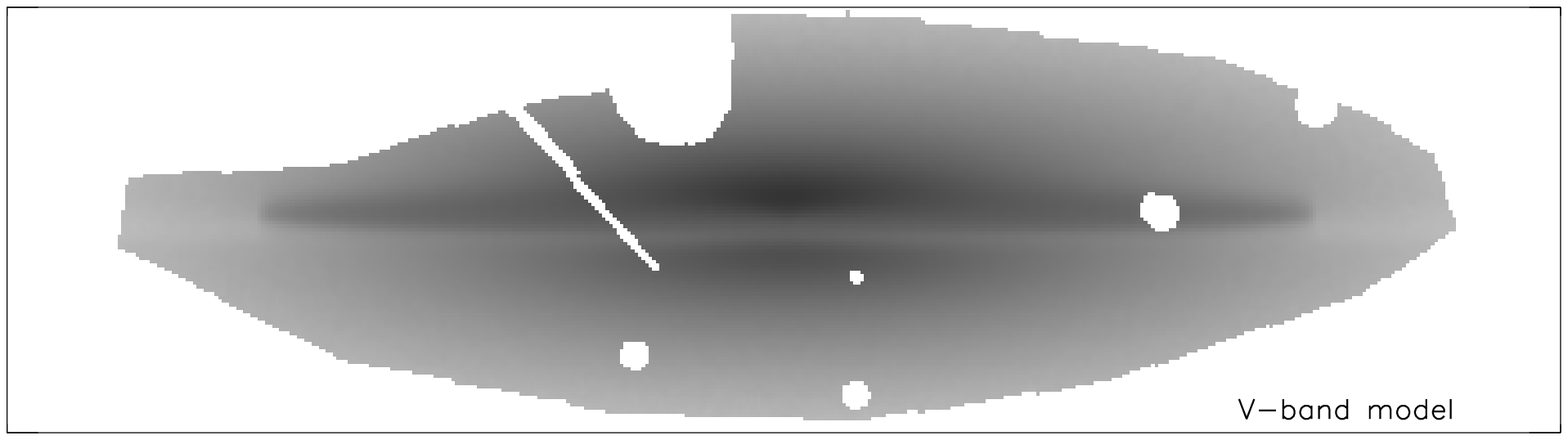}

\vspace{0.05cm}
\includegraphics[width=17cm]{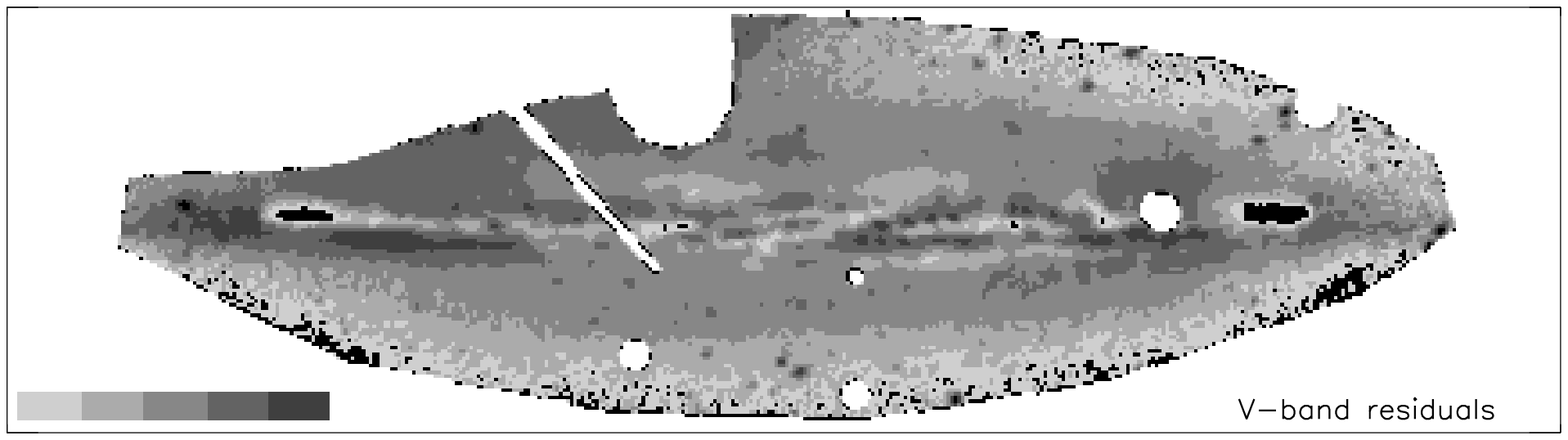}

\vspace{0.3cm}
\includegraphics[width=17cm]{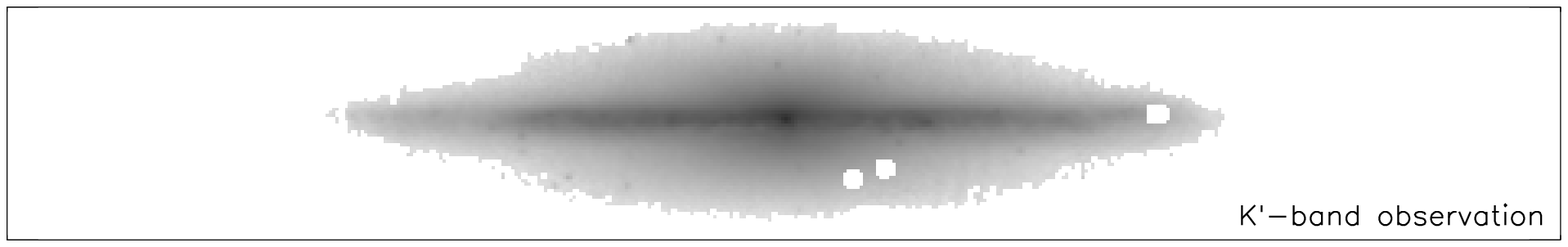}

\vspace{0.05cm}
\includegraphics[width=17cm]{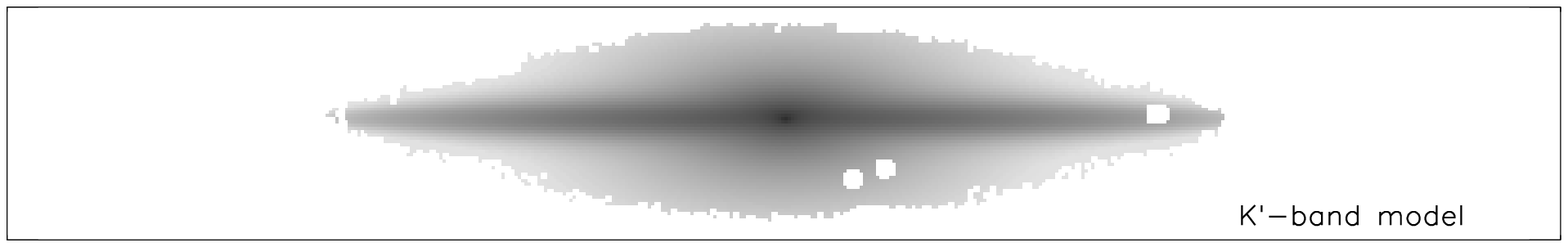}

\vspace{0.05cm}
\includegraphics[width=17cm]{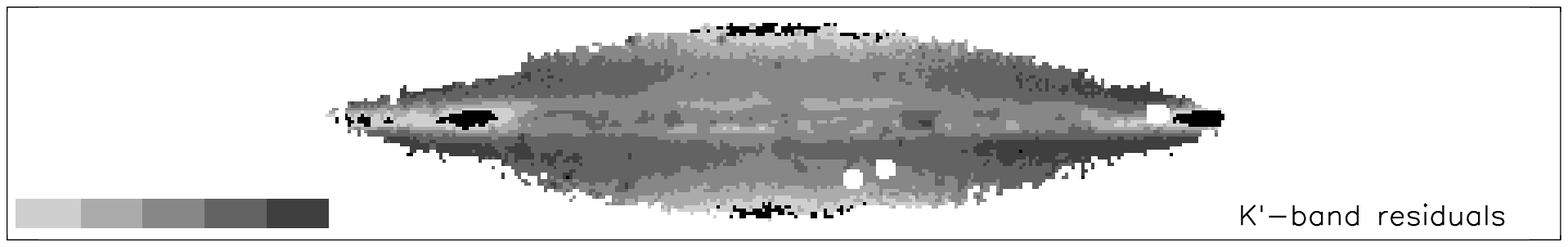}
\caption{NGC 4217. Image order is the same as in Fig.~\ref{n4013_f}.
The horizontal size of the box is 8\arcmin.}
\label{n4217_f}
\end{figure*}
}

\onlfig{6}{
\begin{figure*}
\centering
\includegraphics[width=17cm]{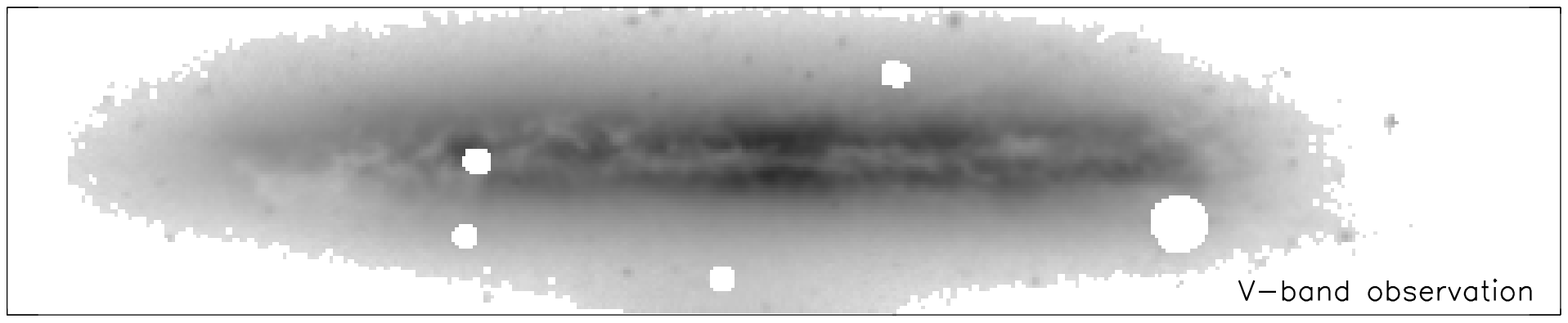}

\vspace{0.05cm}
\includegraphics[width=17cm]{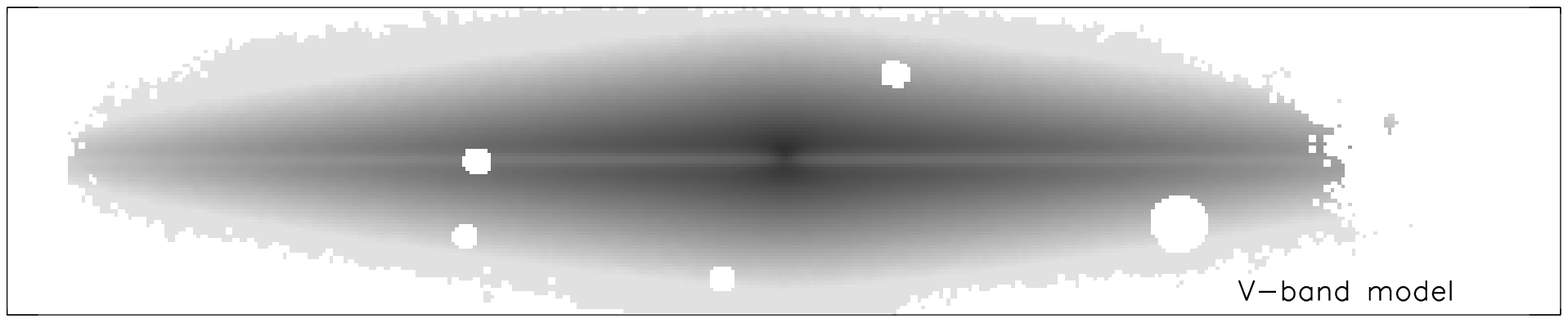}

\vspace{0.05cm}
\includegraphics[width=17cm]{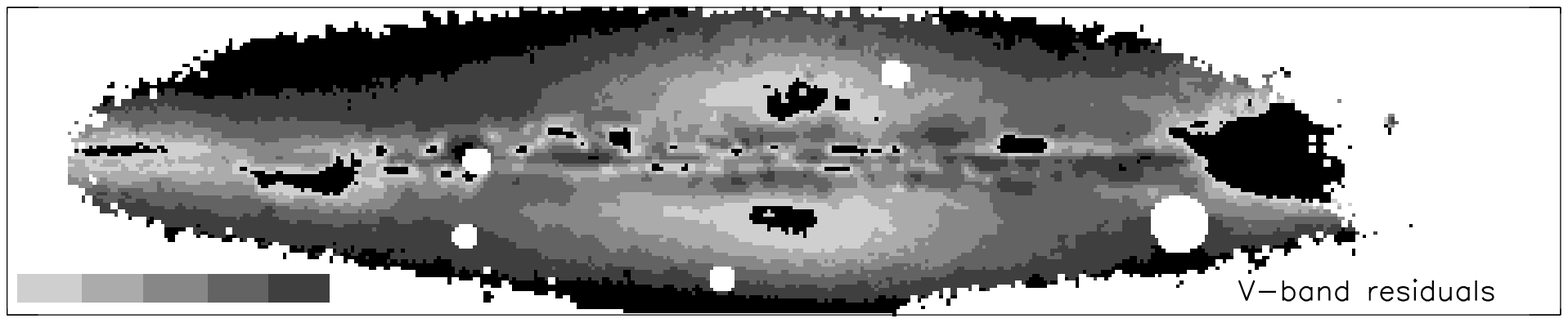}

\vspace{0.3cm}
\includegraphics[width=17cm]{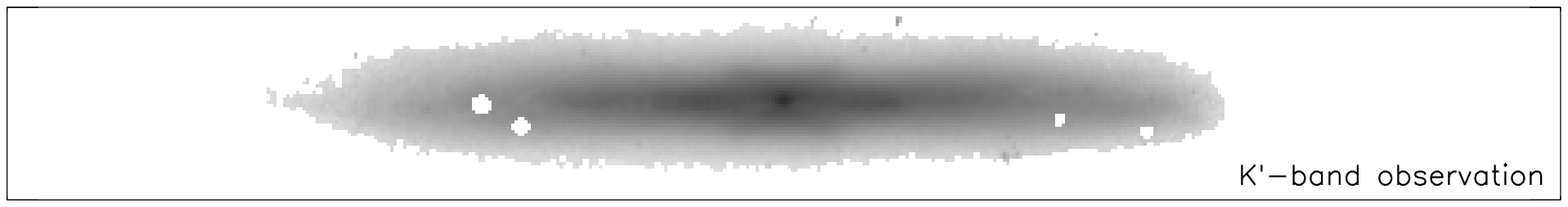}

\vspace{0.05cm}
\includegraphics[width=17cm]{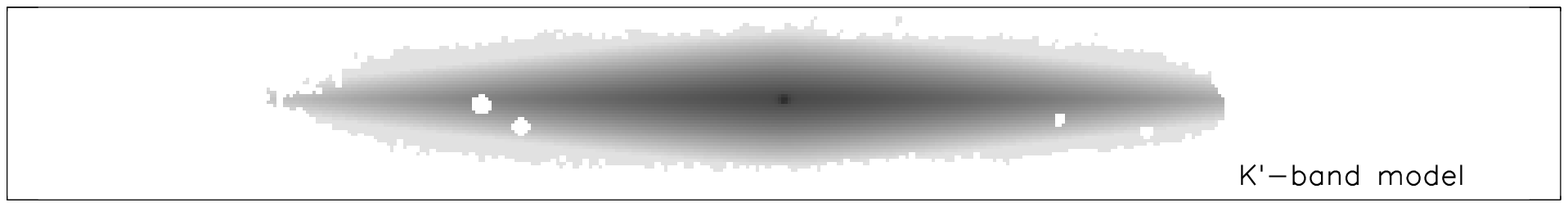}

\vspace{0.05cm}
\includegraphics[width=17cm]{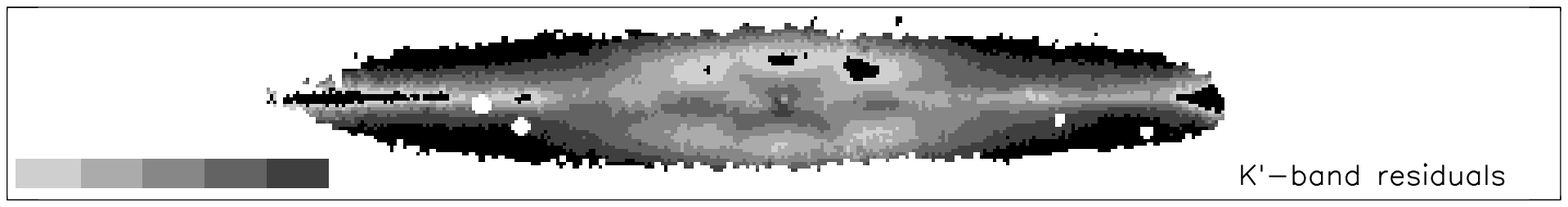}
\caption{NGC 4302. Image order is the same as in Fig.~\ref{n4013_f}.
The horizontal size of the box is 8\arcmin.}
\label{n4302_f}
\end{figure*}
}

\onlfig{7}{
\begin{figure*}
\centering
\includegraphics[width=17cm]{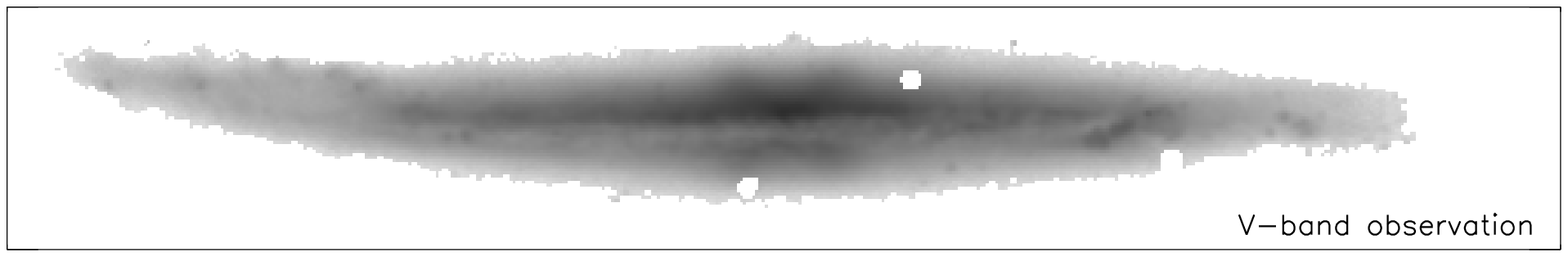}

\vspace{0.05cm}
\includegraphics[width=17cm]{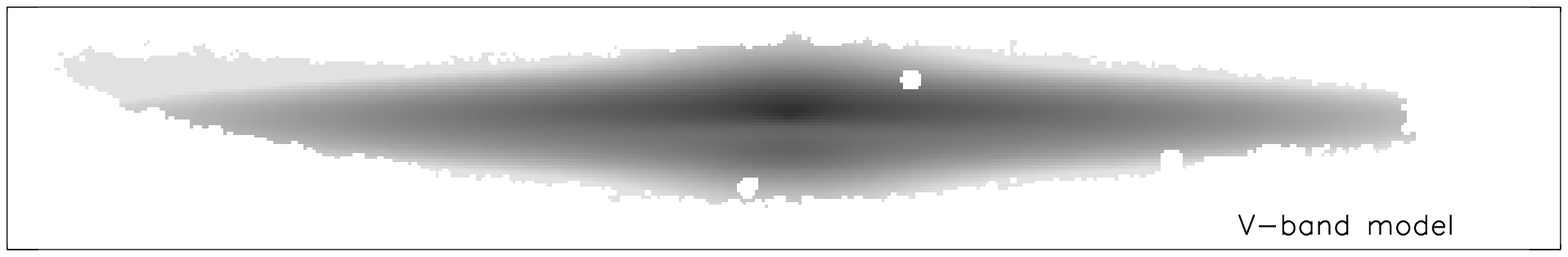}

\vspace{0.05cm}
\includegraphics[width=17cm]{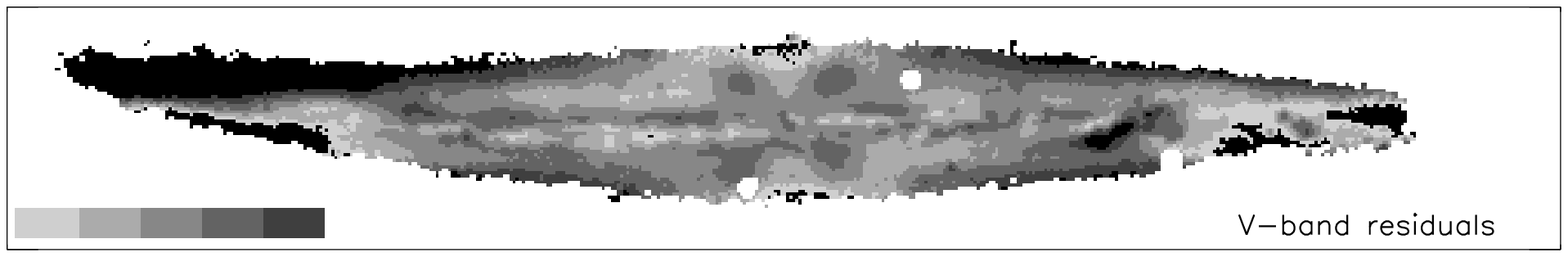}

\vspace{0.3cm}
\includegraphics[width=17cm]{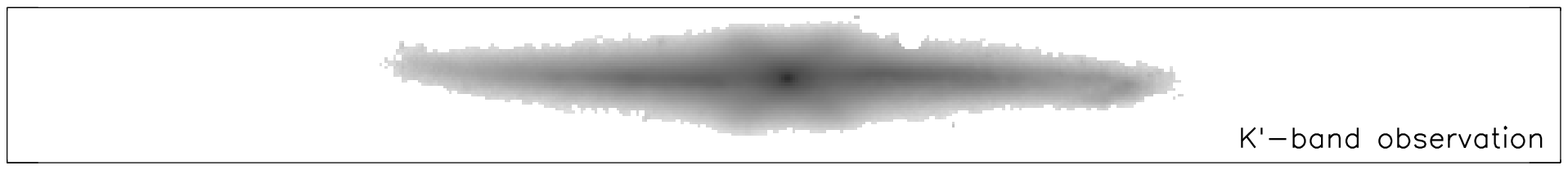}

\vspace{0.05cm}
\includegraphics[width=17cm]{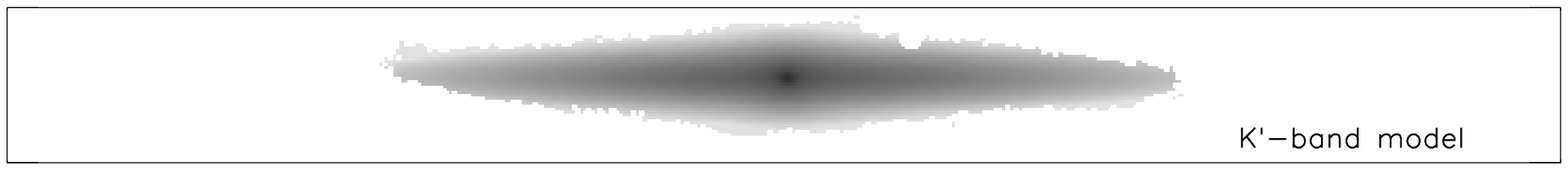}

\vspace{0.05cm}
\includegraphics[width=17cm]{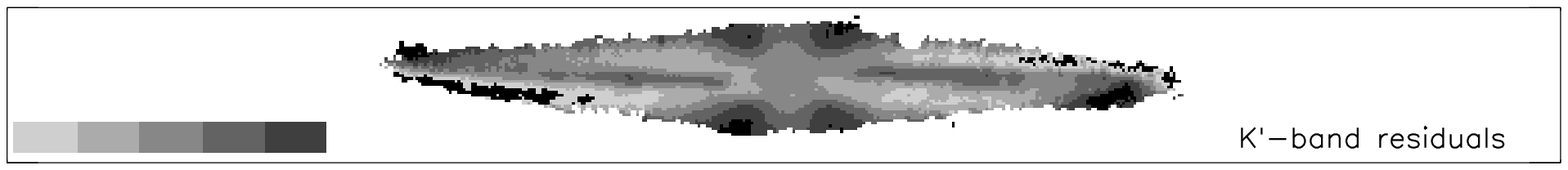}
\caption{NGC 5529. Image order is the same as in Fig.~\ref{n4013_f}.
The horizontal size of the box is 7\arcmin.}
\end{figure*}
}

\onlfig{8}{
\begin{figure*}
\centering
\includegraphics[width=17cm]{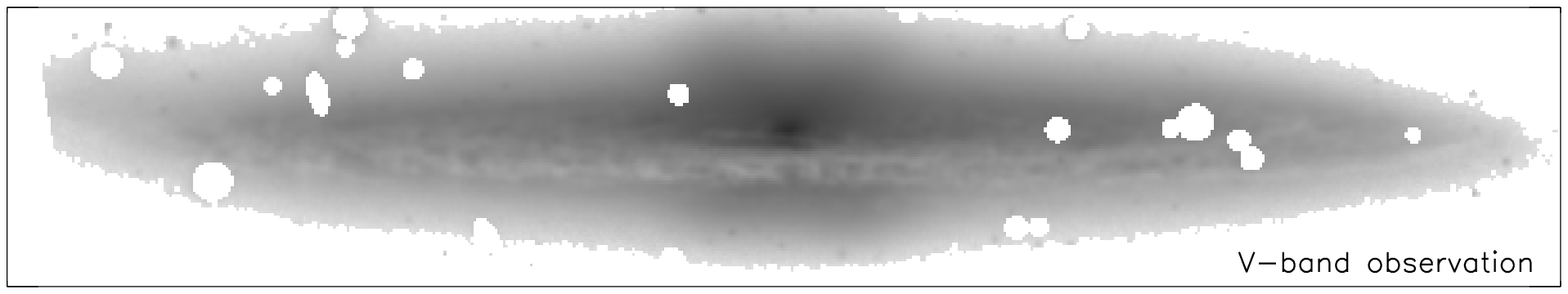}

\vspace{0.05cm}
\includegraphics[width=17cm]{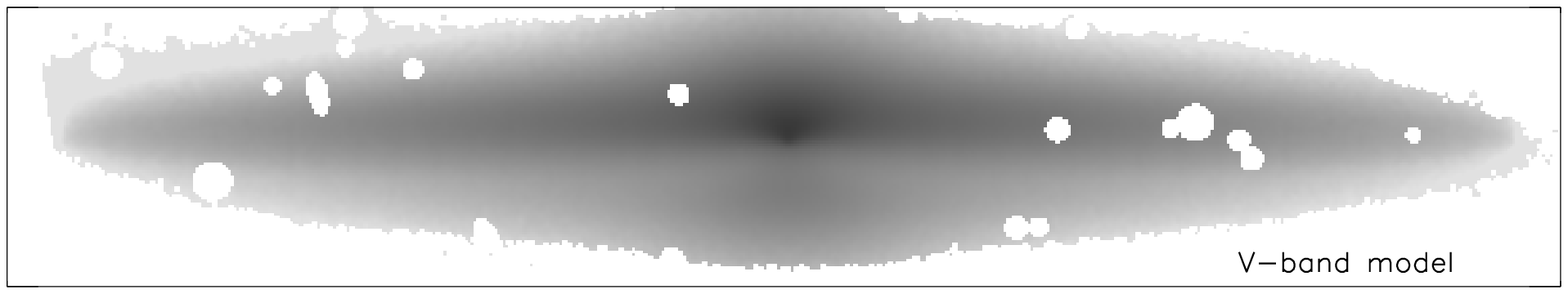}

\vspace{0.05cm}
\includegraphics[width=17cm]{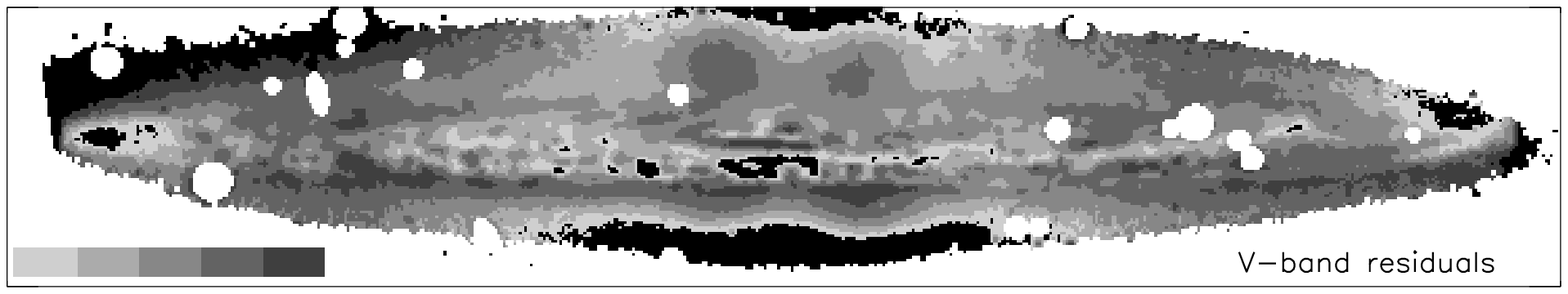}

\vspace{0.3cm}
\includegraphics[width=17cm]{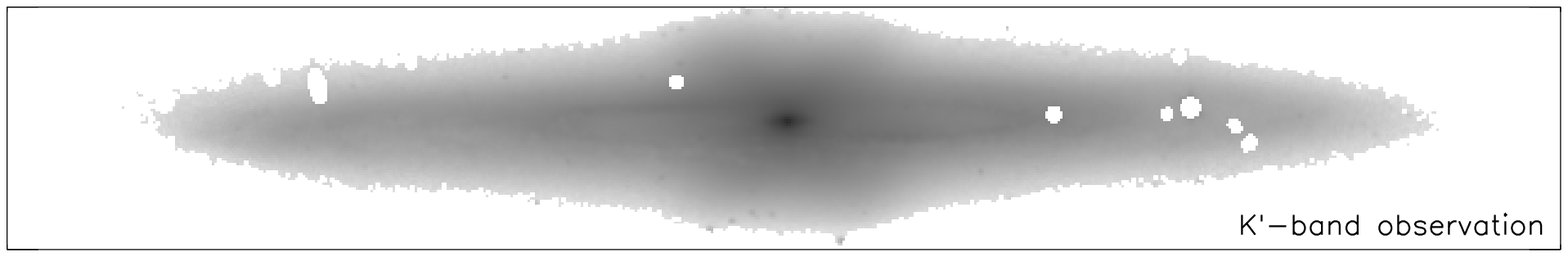}

\vspace{0.05cm}
\includegraphics[width=17cm]{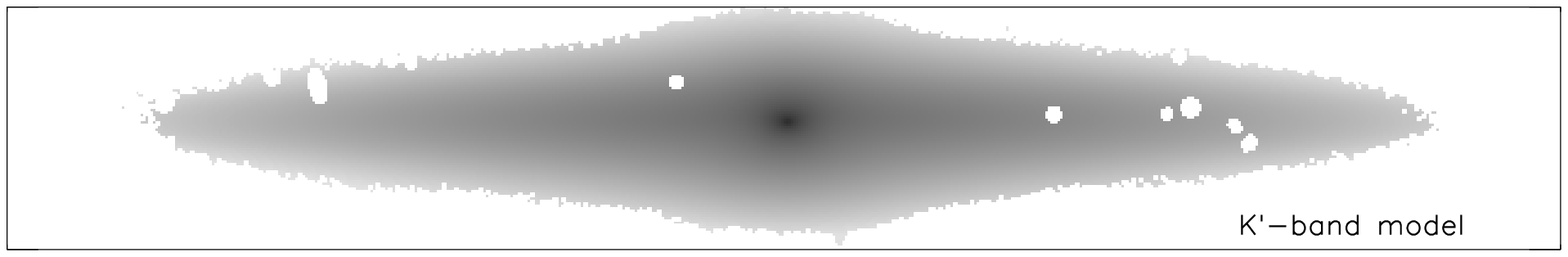}

\vspace{0.05cm}
\includegraphics[width=17cm]{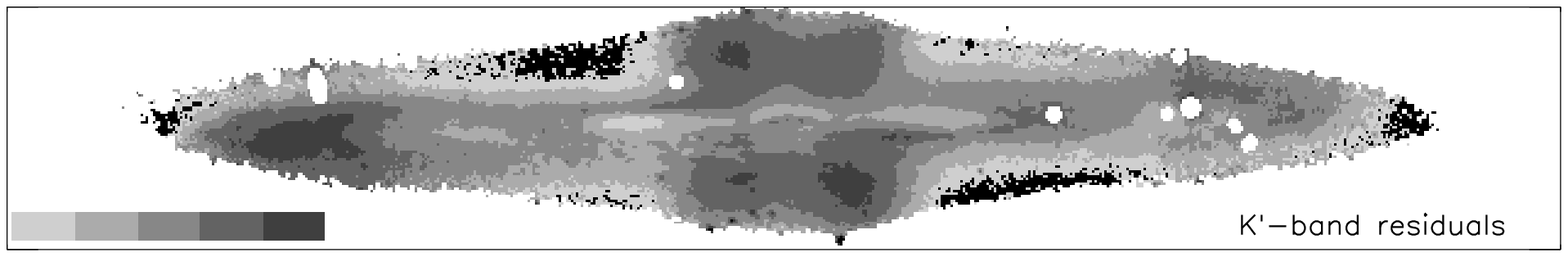}
\caption{NGC 5746. Image order is the same as in Fig.~\ref{n4013_f}.
The horizontal size of the box is 8\arcmin.}
\end{figure*}
}

\onlfig{9}{
\begin{figure*}
\centering
\includegraphics[width=17cm]{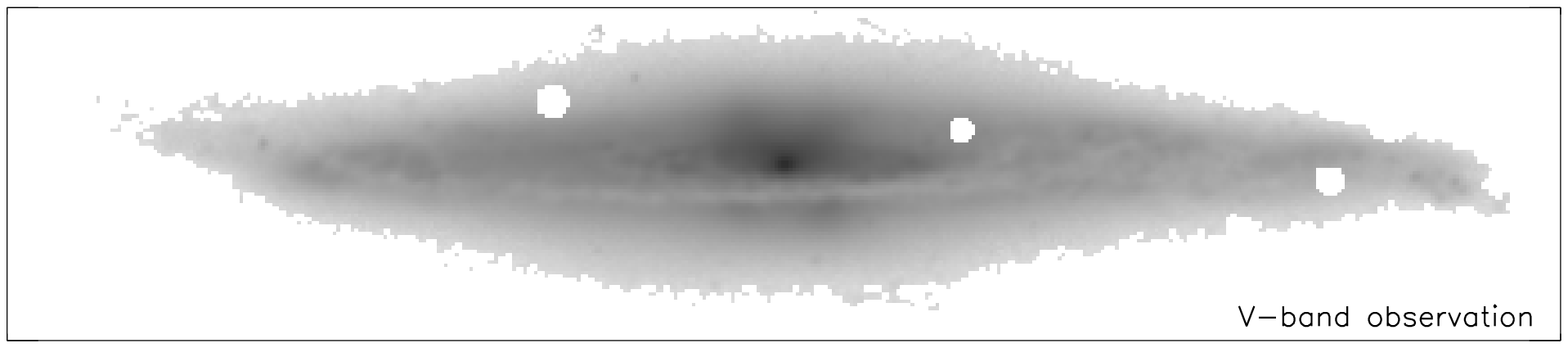}

\vspace{0.05cm}
\includegraphics[width=17cm]{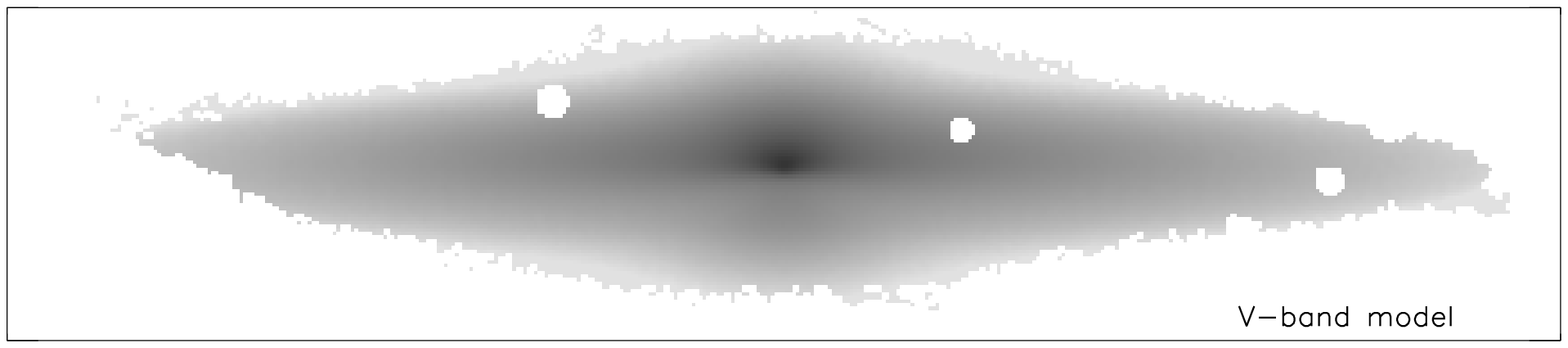}

\vspace{0.05cm}
\includegraphics[width=17cm]{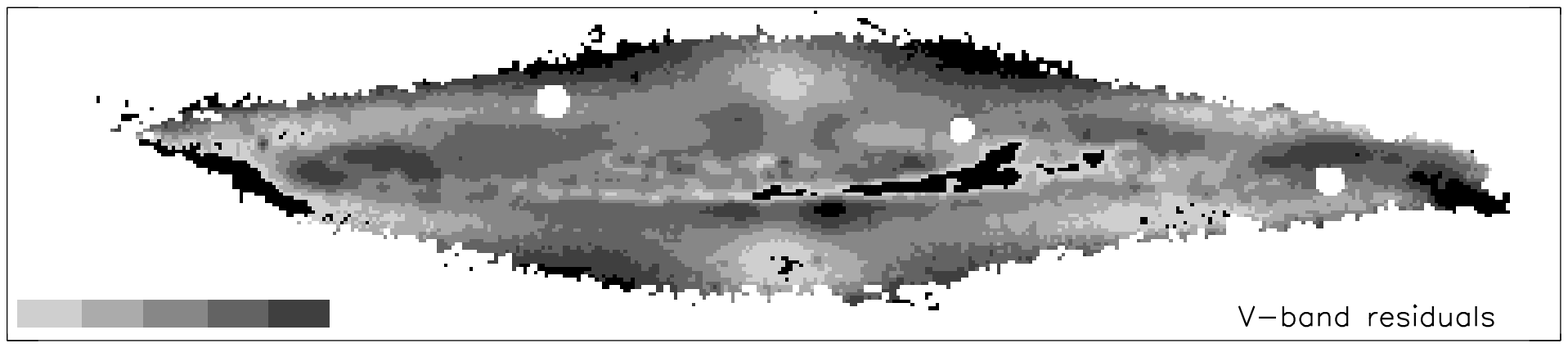}

\vspace{0.3cm}
\includegraphics[width=17cm]{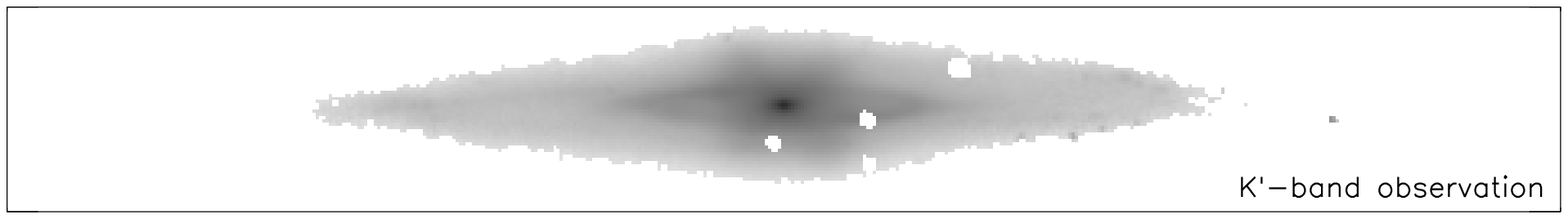}

\vspace{0.05cm}
\includegraphics[width=17cm]{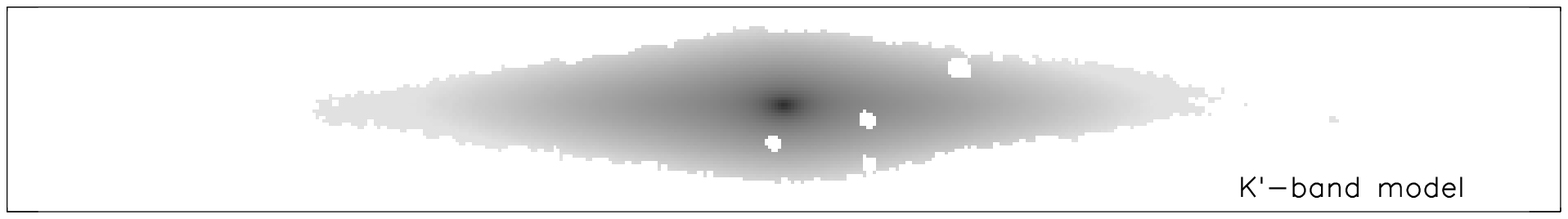}

\vspace{0.05cm}
\includegraphics[width=17cm]{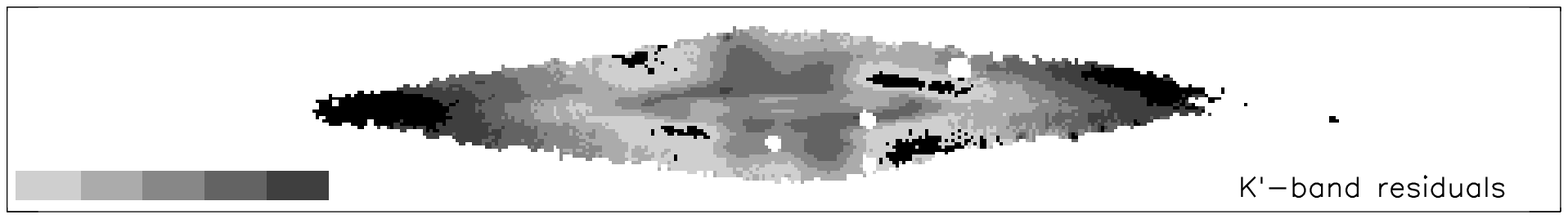}
\caption{NGC 5965. Image order is the same as in Fig.~\ref{n4013_f}.
The horizontal size of the box is 6\arcmin.}
\end{figure*}
}

\onlfig{10}{
\begin{figure*}
\centering
\includegraphics[width=17cm]{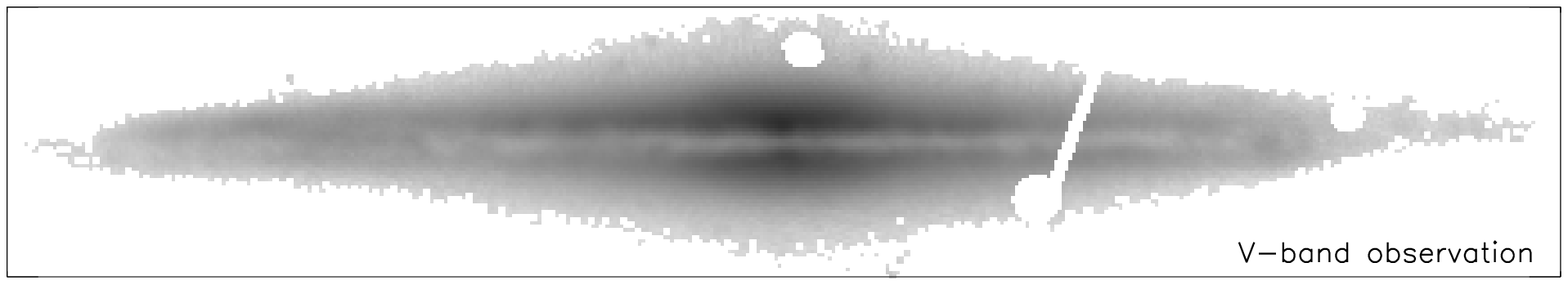}

\vspace{0.05cm}
\includegraphics[width=17cm]{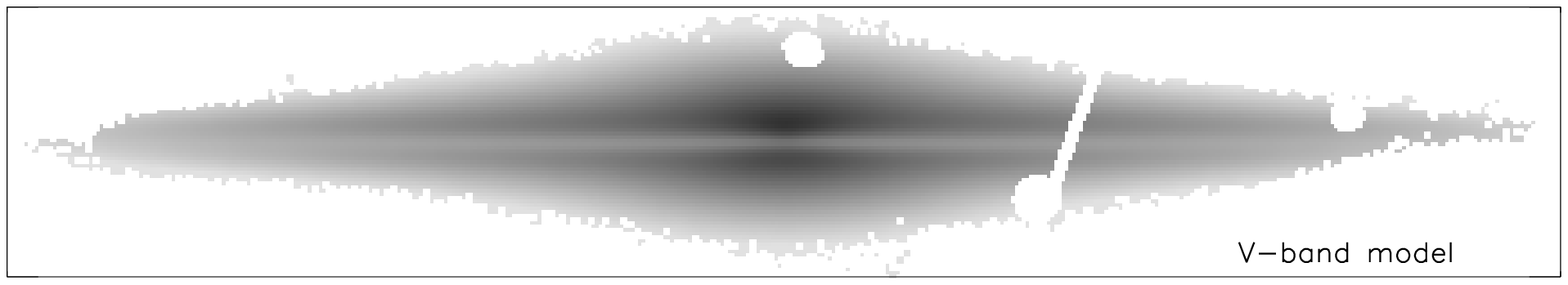}

\vspace{0.05cm}
\includegraphics[width=17cm]{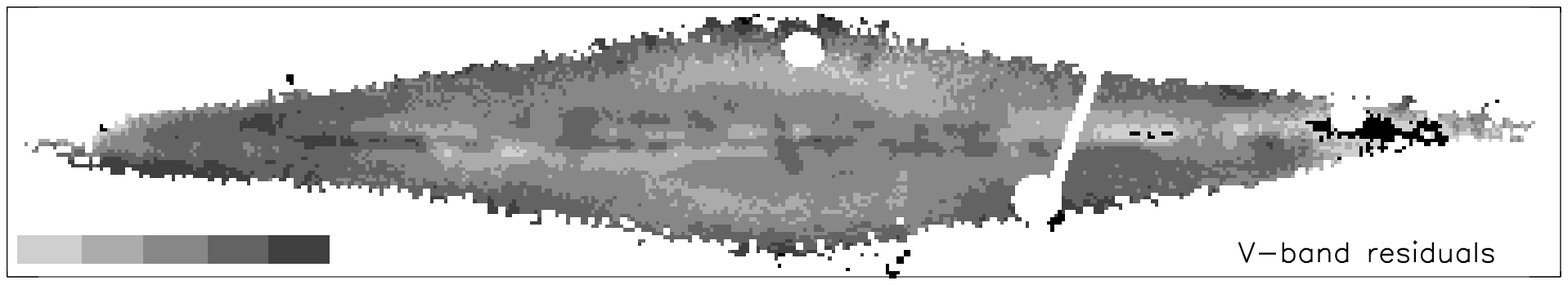}

\vspace{0.3cm}
\includegraphics[width=17cm]{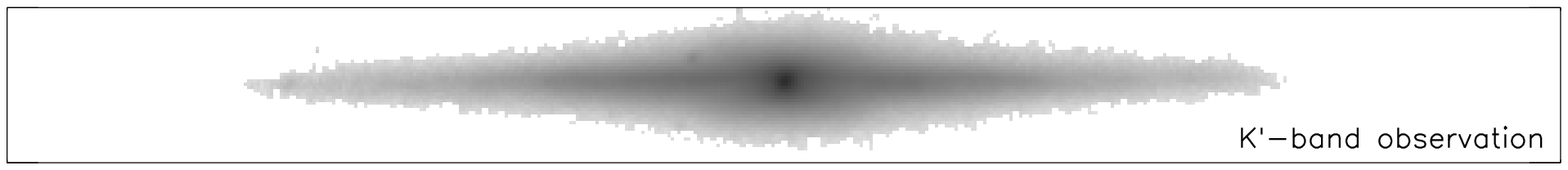}

\vspace{0.05cm}
\includegraphics[width=17cm]{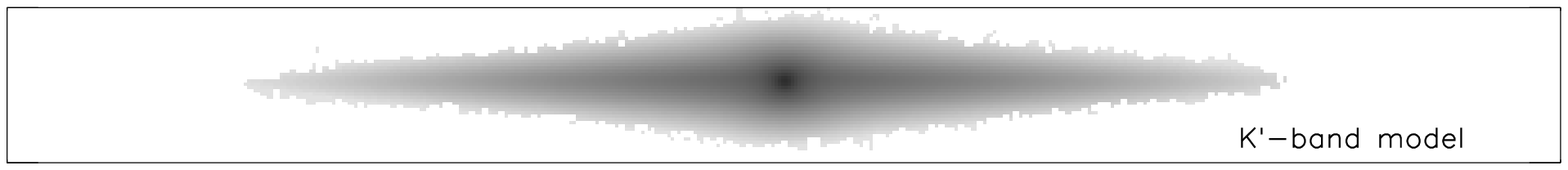}

\vspace{0.05cm}
\includegraphics[width=17cm]{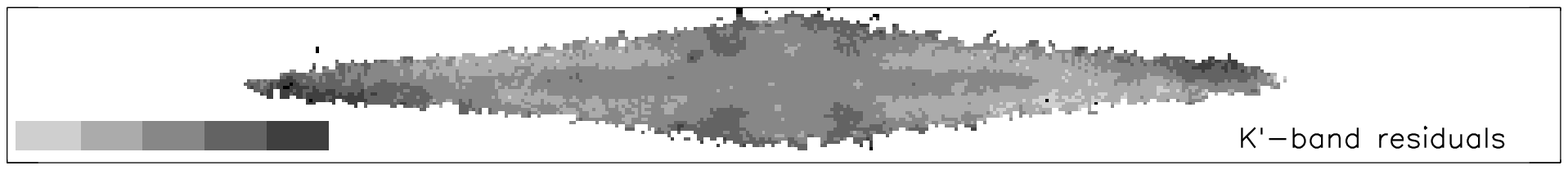}
\caption{UGC 4277. Image order is the same as in Fig.~\ref{n4013_f}.
The horizontal size of the box is 4\arcmin.}
\label{u4277_f}
\end{figure*}
}

\onlfig{11}{
\begin{figure*}
\centering
\includegraphics[width=8.5cm]{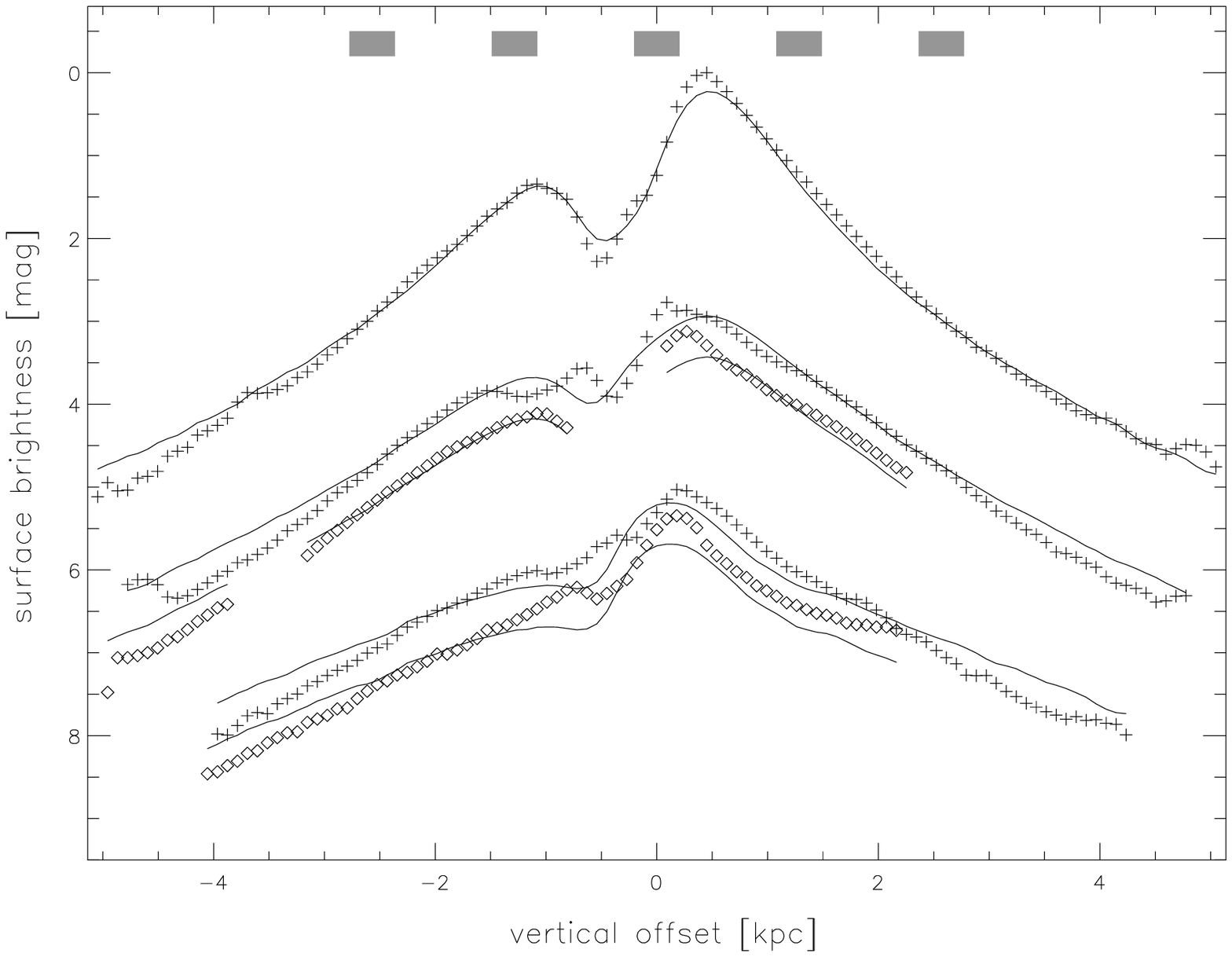}
\includegraphics[width=8.5cm]{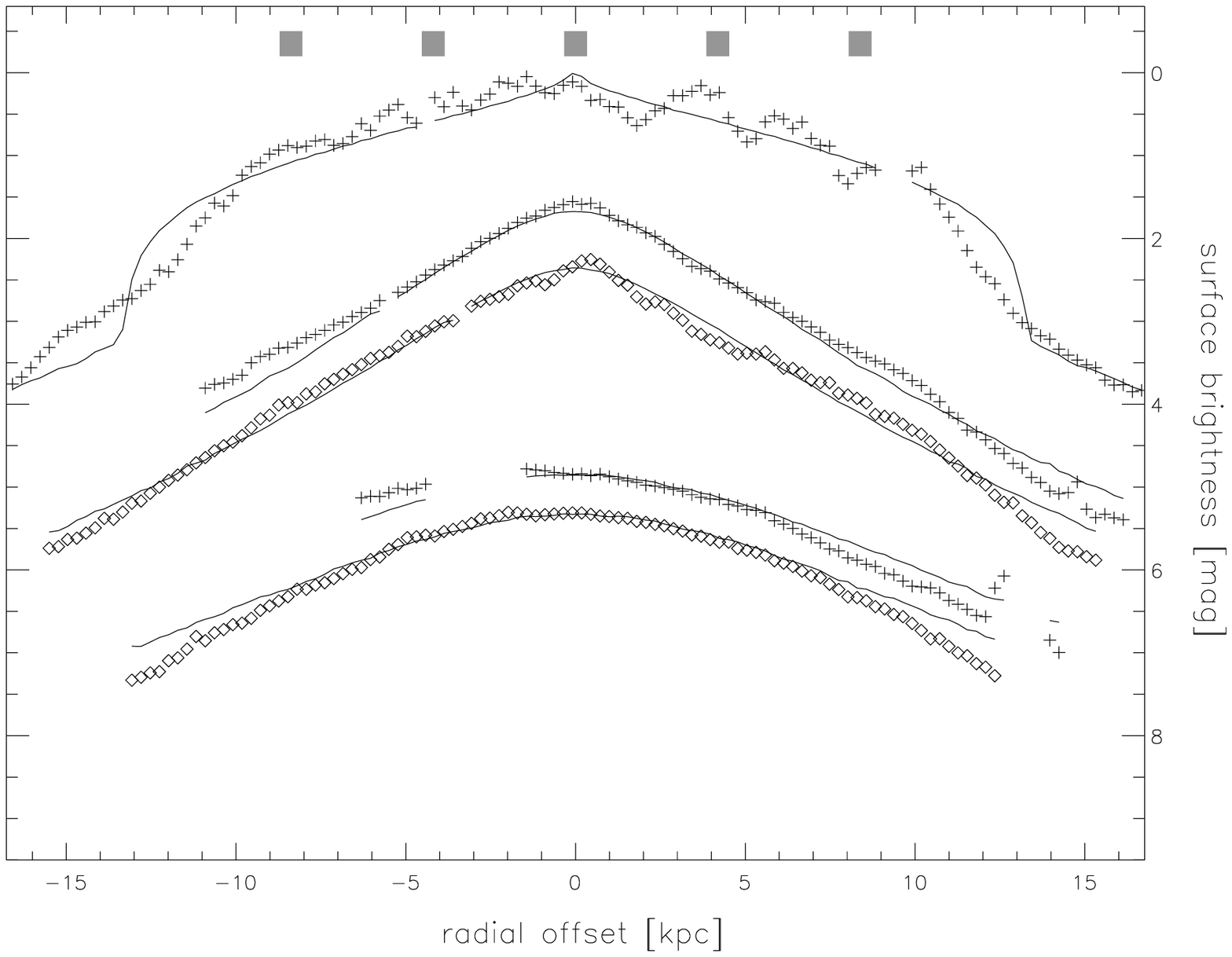}
\vspace{0.05cm}
\includegraphics[width=8.5cm]{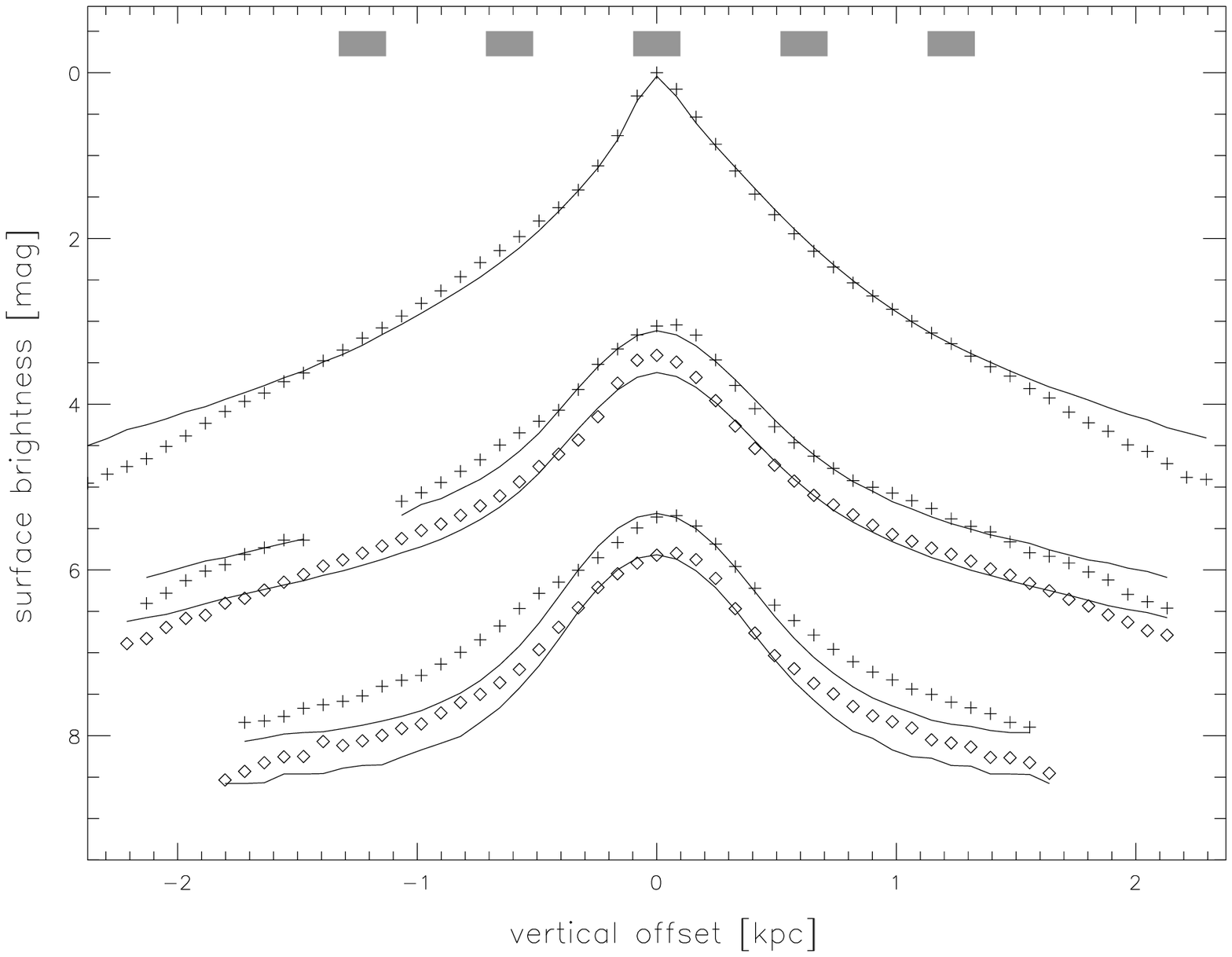}
\includegraphics[width=8.5cm]{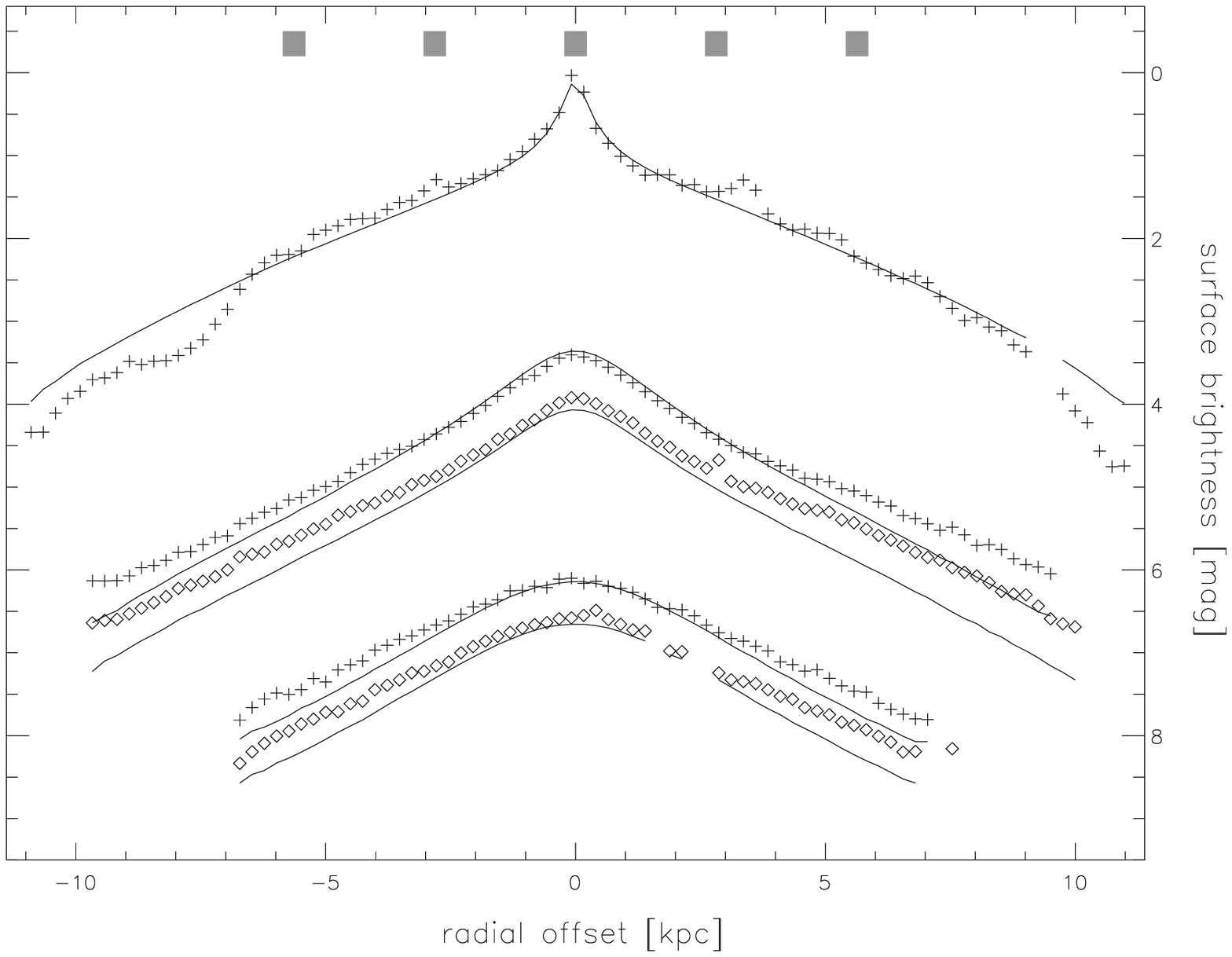}
\caption{Same as Fig.~\ref{n4013_p}, but for NGC 4217.}
\label{n4217_p}
\end{figure*}
}

\onlfig{12}{
\begin{figure*}
\centering
\includegraphics[width=8.5cm]{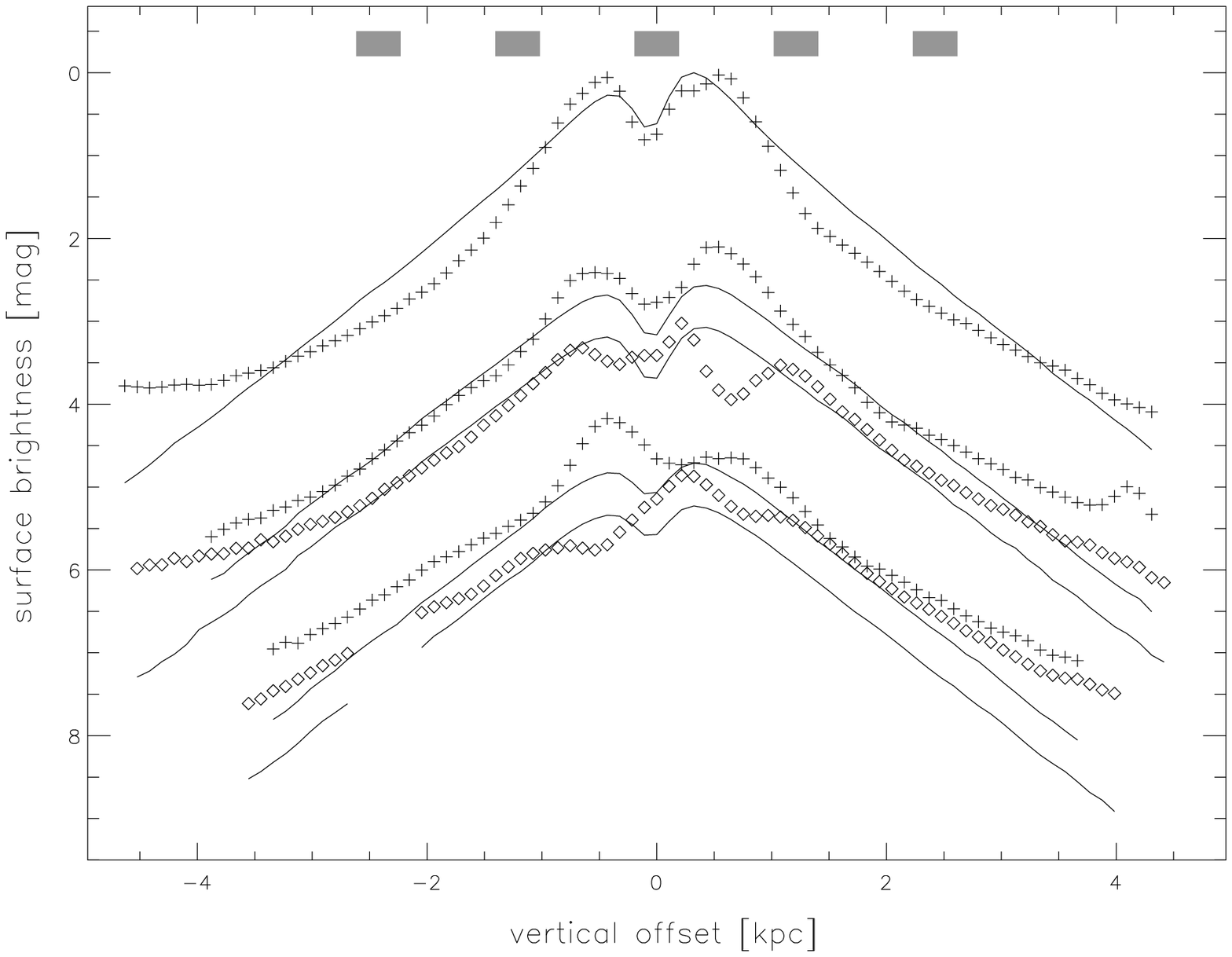}
\includegraphics[width=8.5cm]{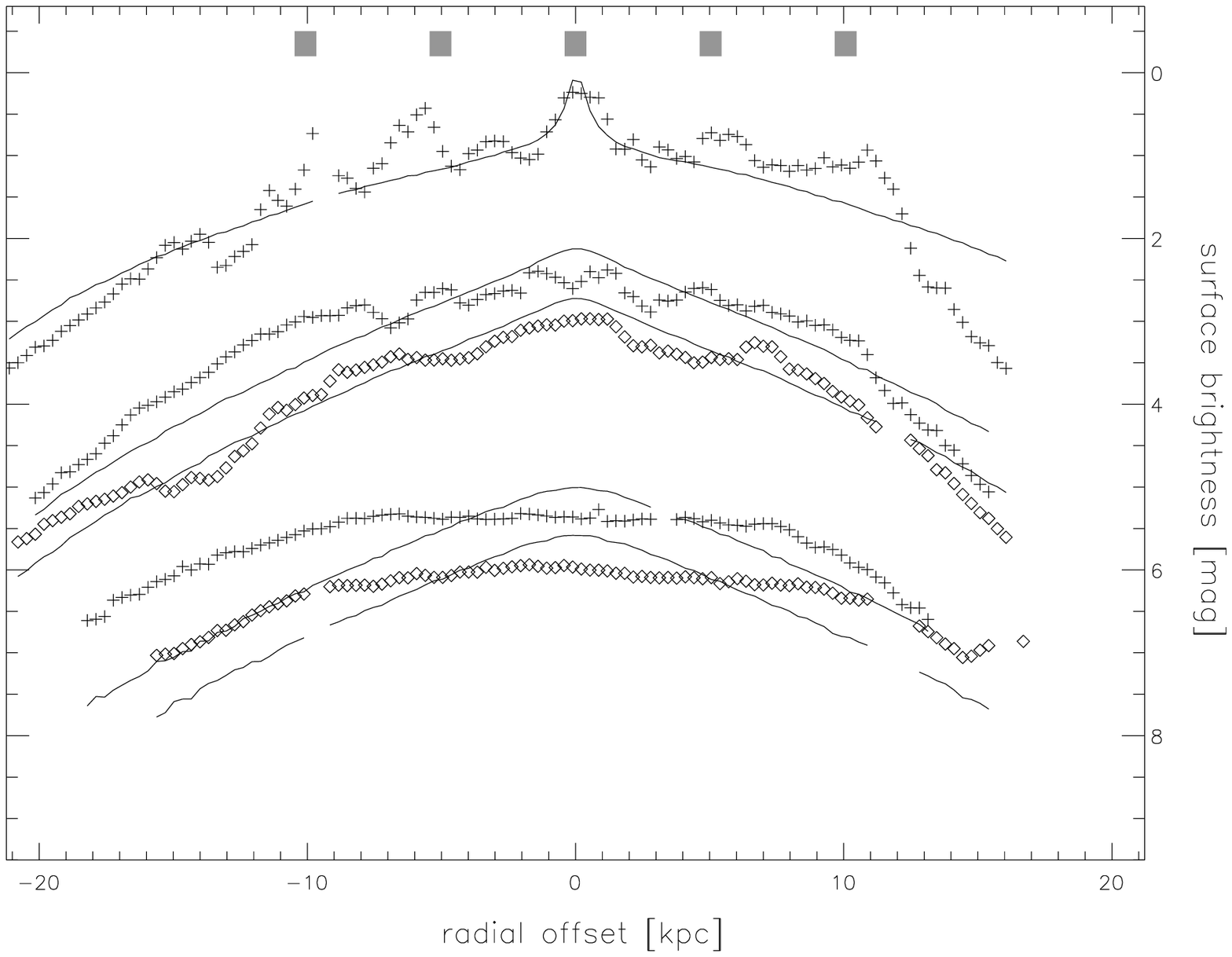}
\vspace{0.05cm}
\includegraphics[width=8.5cm]{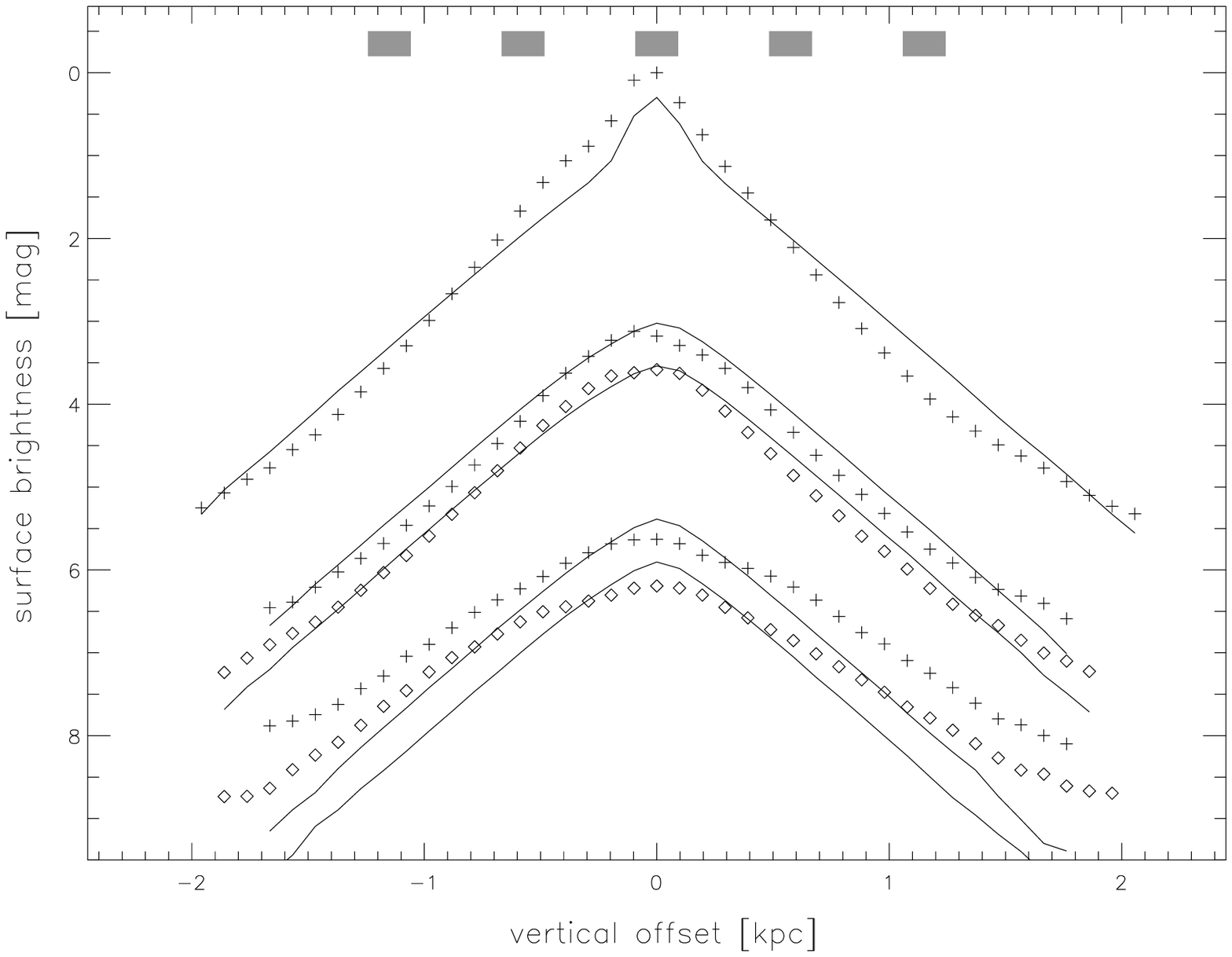}
\includegraphics[width=8.5cm]{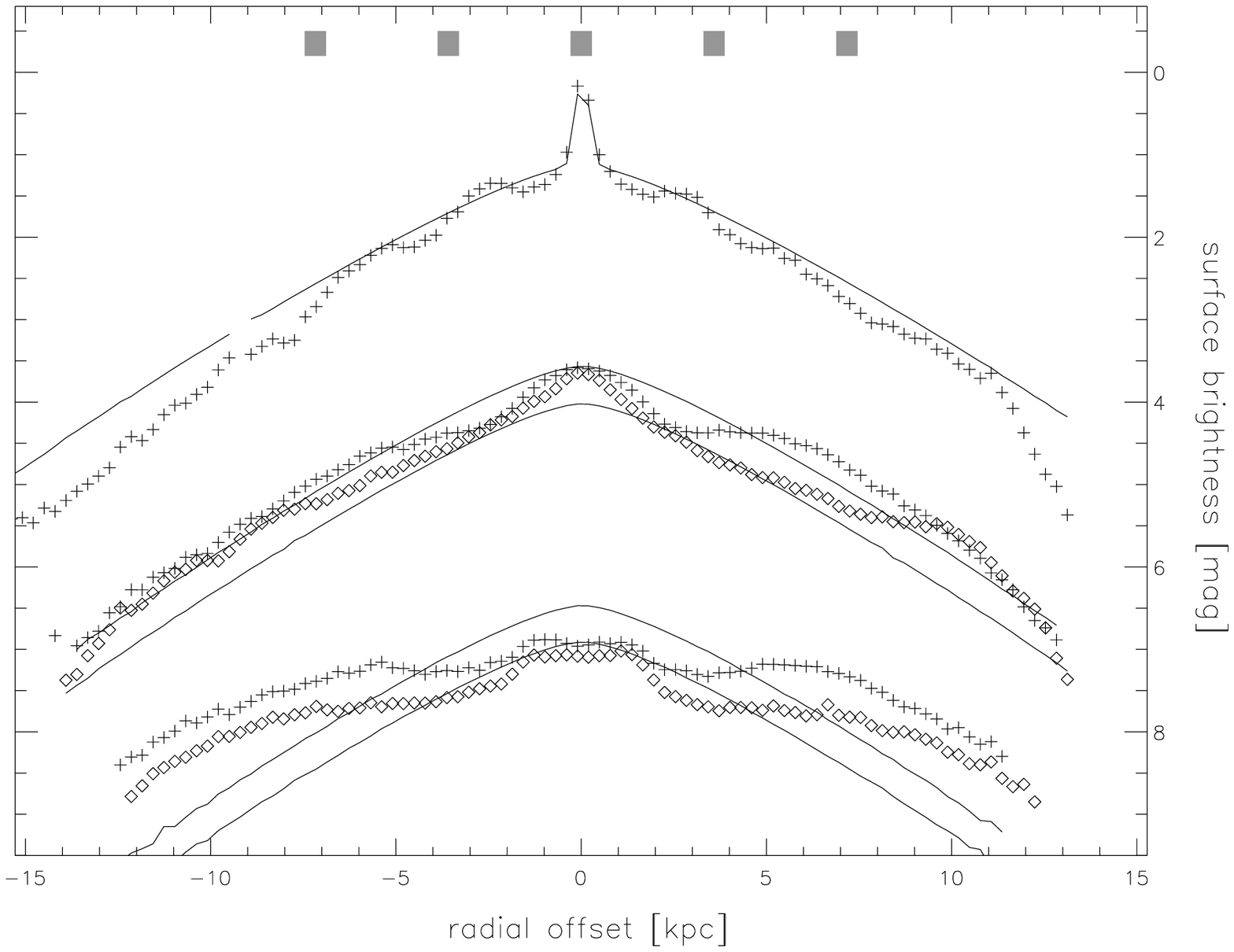}
\caption{Same as Fig.~\ref{n4013_p}, but for NGC 4302.}
\end{figure*}
}

\onlfig{13}{
\begin{figure*}
\centering
\includegraphics[width=8.5cm]{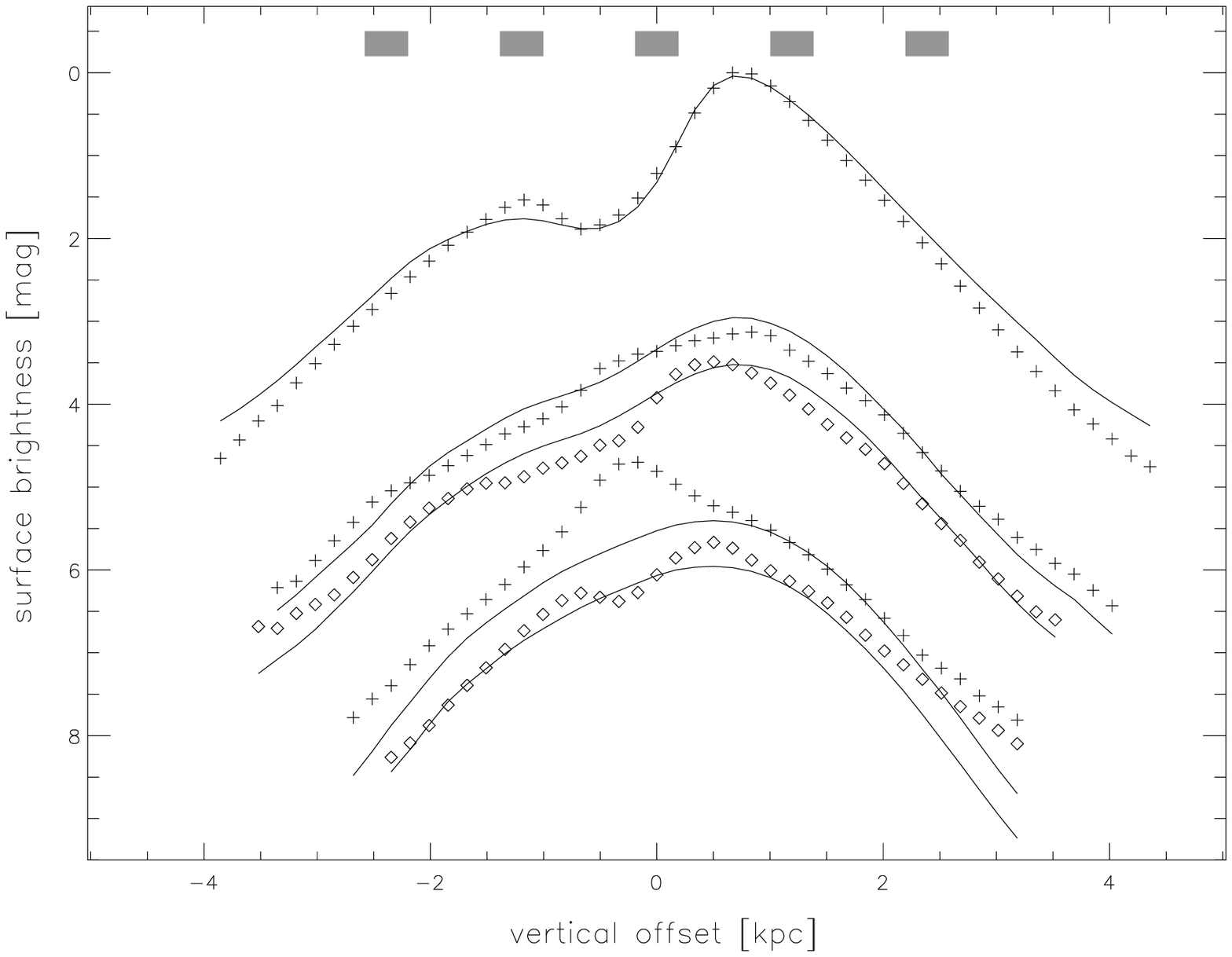}
\includegraphics[width=8.5cm]{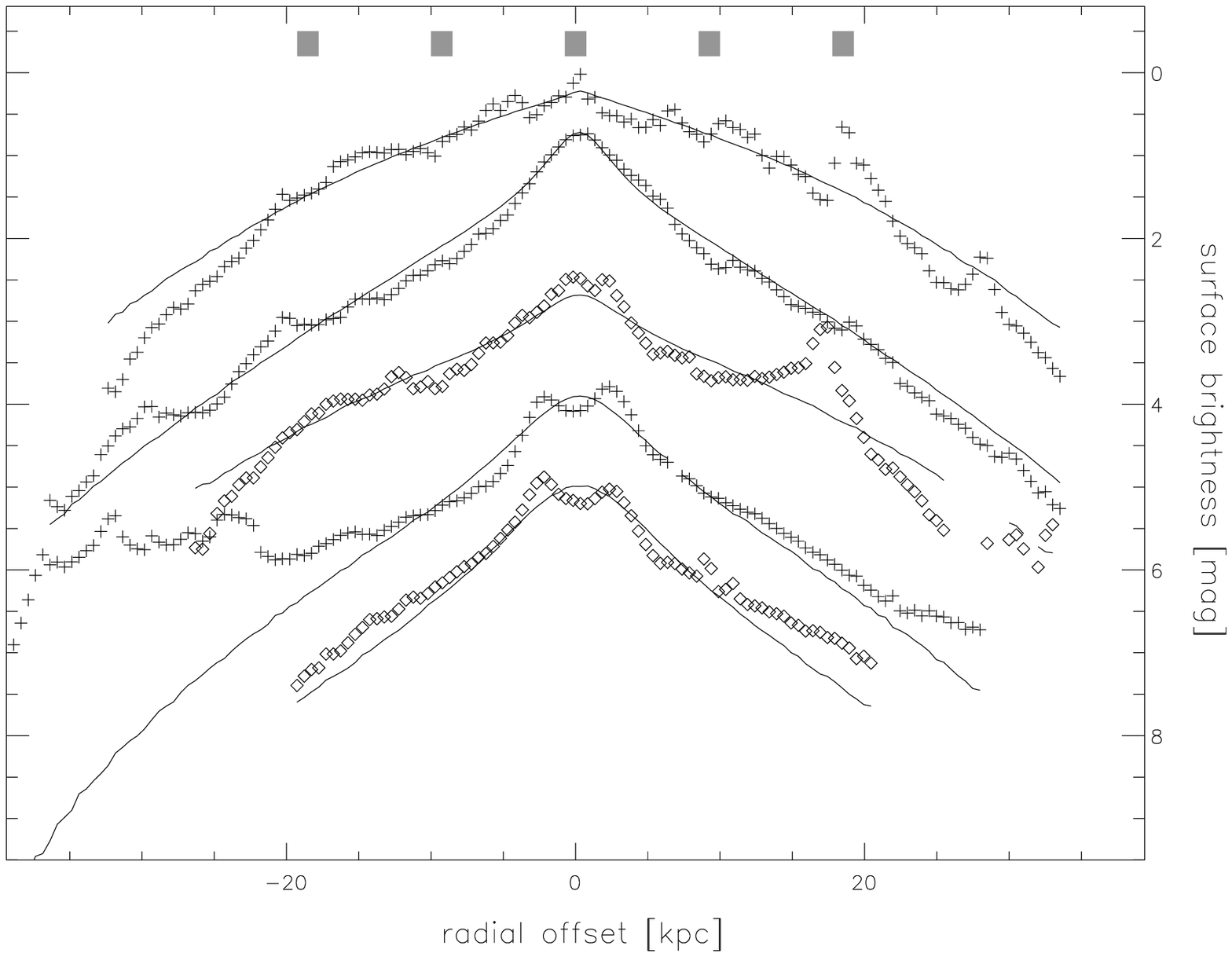}
\vspace{0.05cm}
\includegraphics[width=8.5cm]{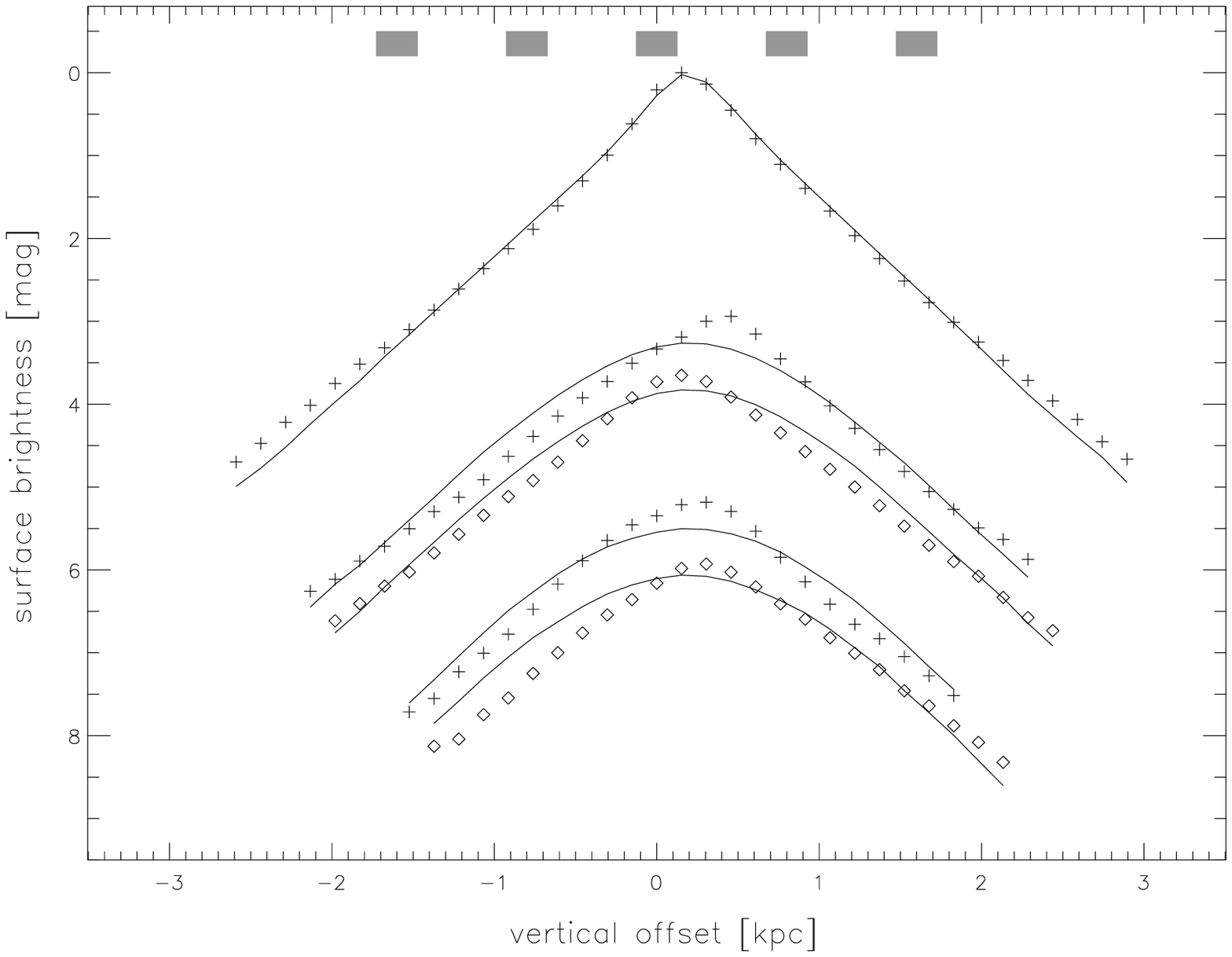}
\includegraphics[width=8.5cm]{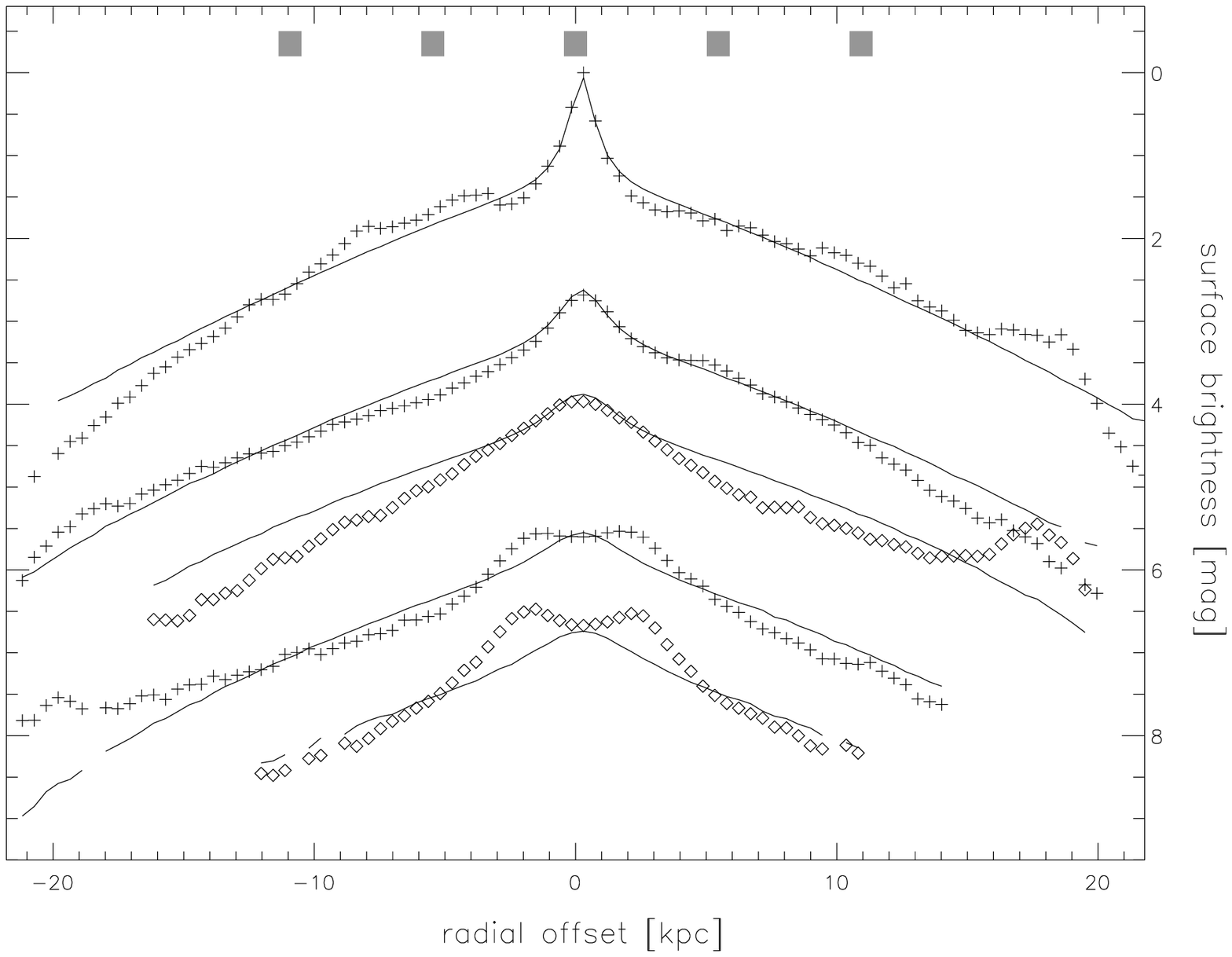}
\caption{Same as Fig.~\ref{n4013_p}, but for NGC 5529.}
\end{figure*}
}

\onlfig{14}{
\begin{figure*}
\centering
\includegraphics[width=8.5cm]{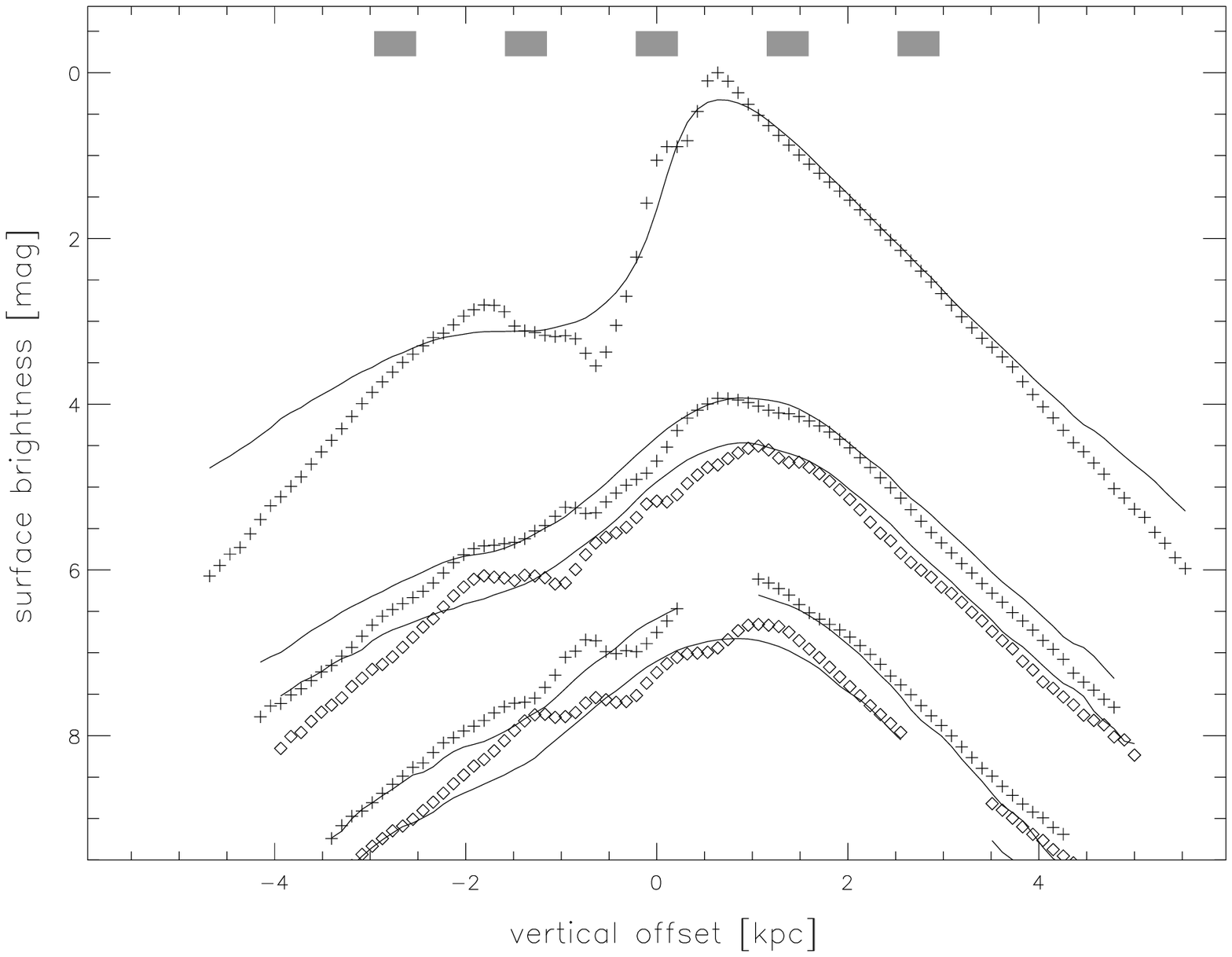}
\includegraphics[width=8.5cm]{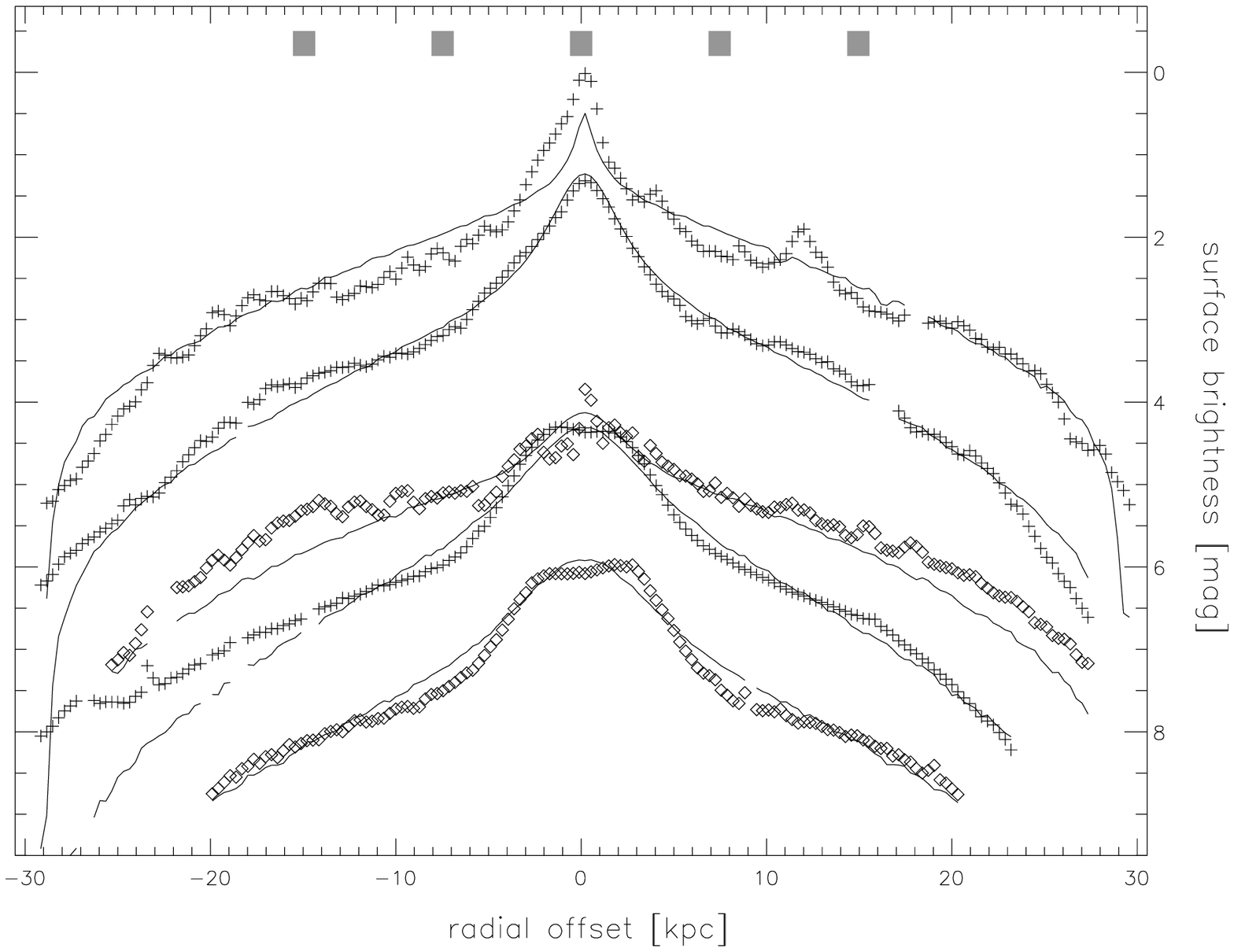}
\vspace{0.05cm}
\includegraphics[width=8.5cm]{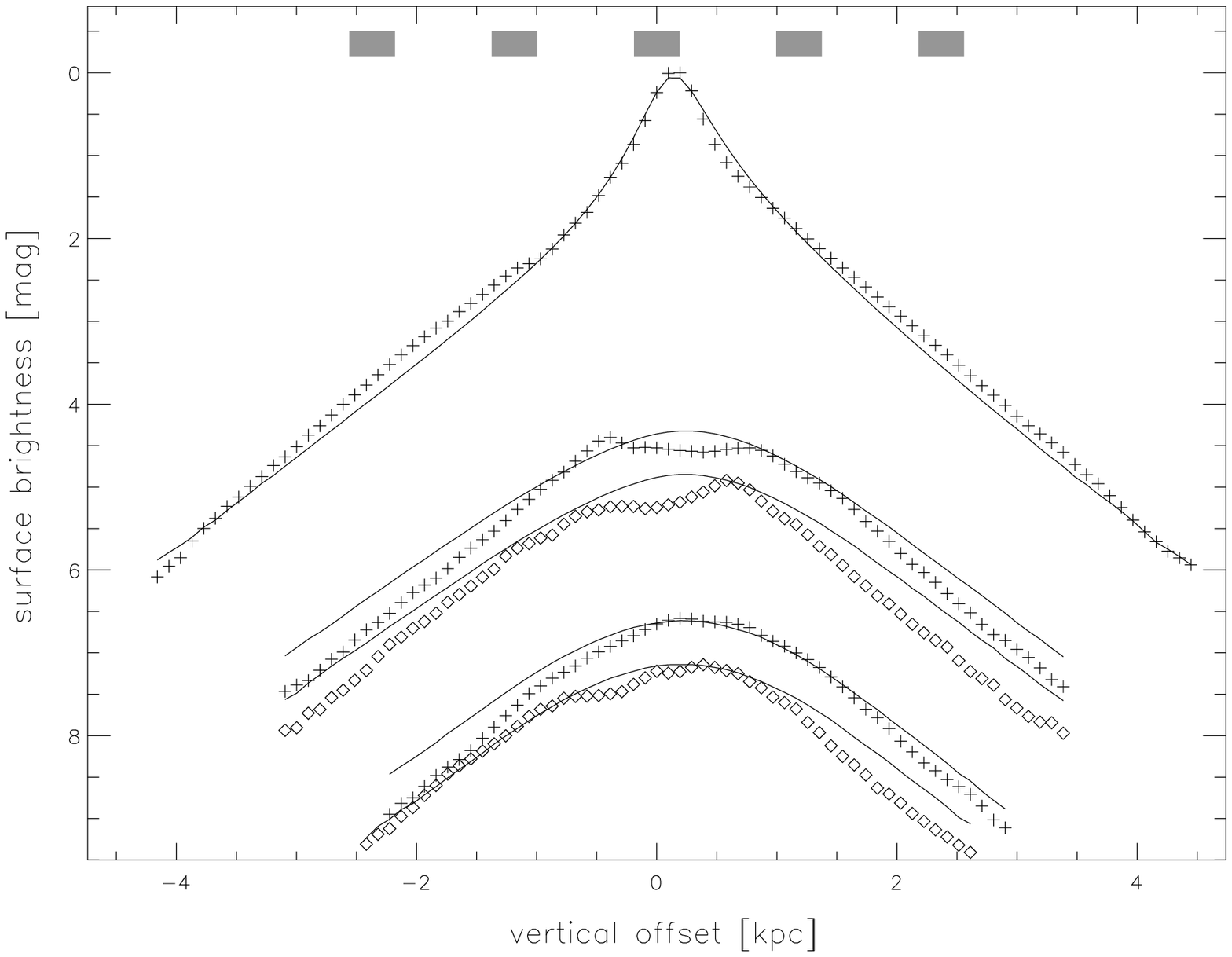}
\includegraphics[width=8.5cm]{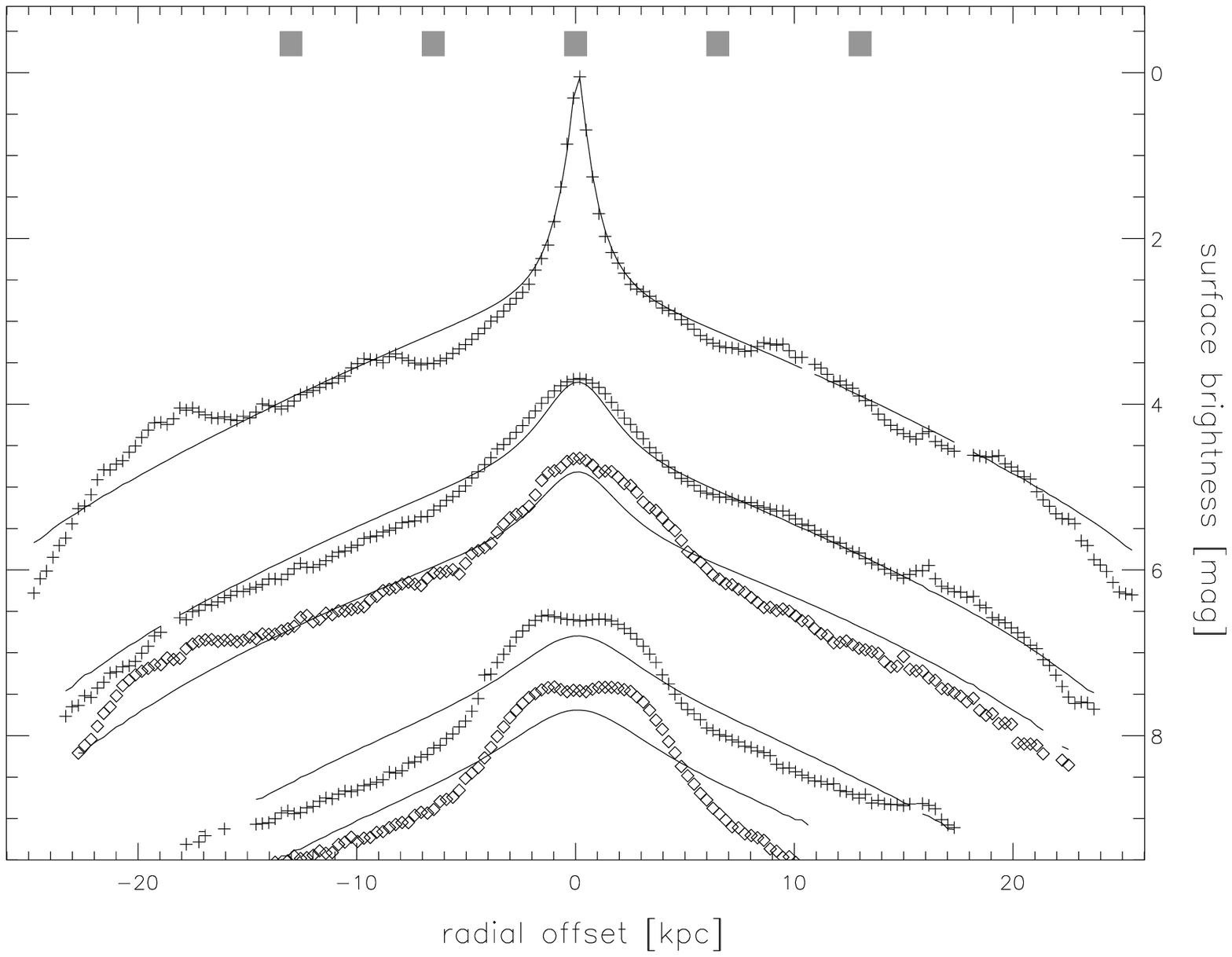}
\caption{Same as Fig.~\ref{n4013_p}, but for NGC 5746.}
\end{figure*}
}

\onlfig{15}{
\begin{figure*}
\centering
\includegraphics[width=8.5cm]{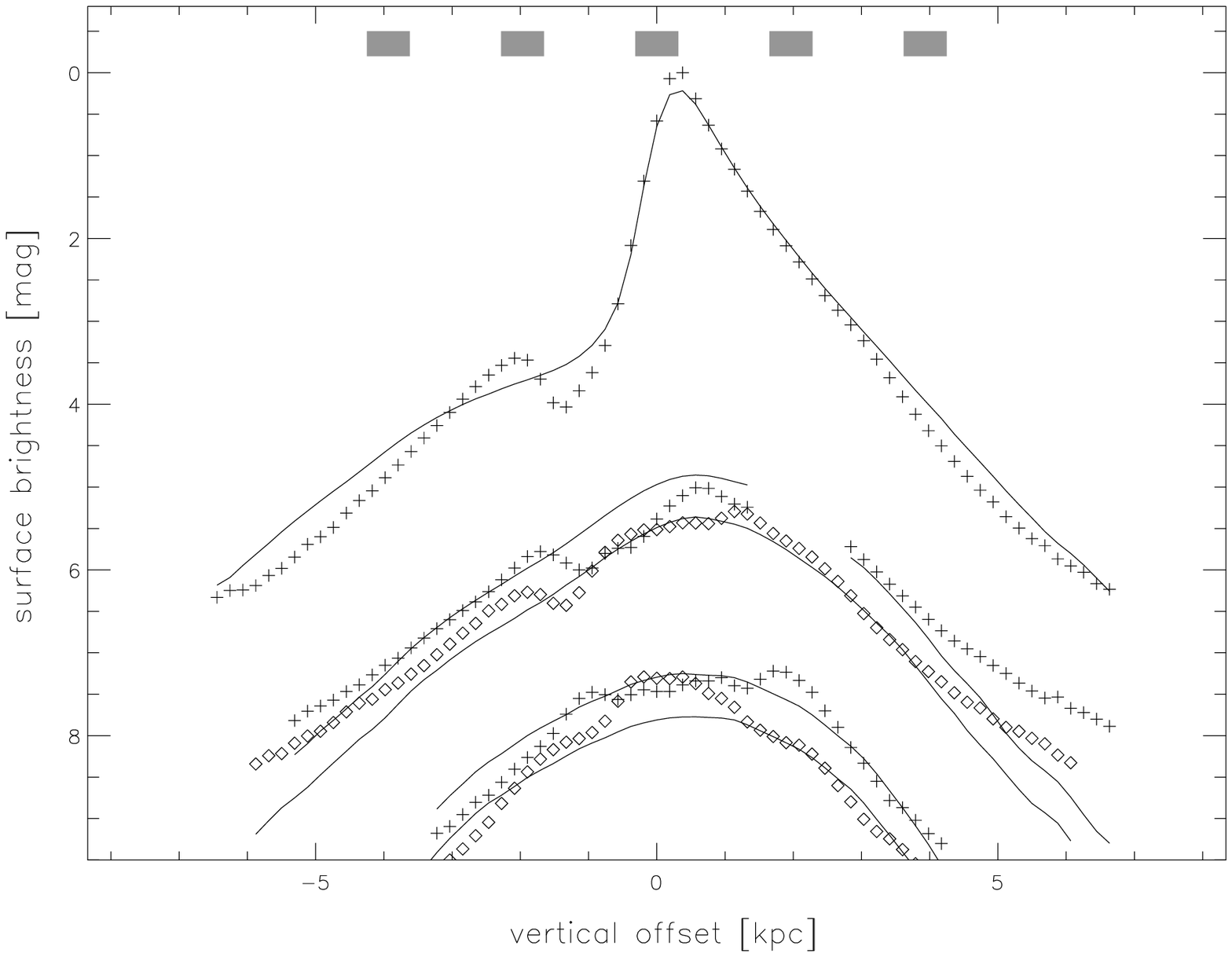}
\includegraphics[width=8.5cm]{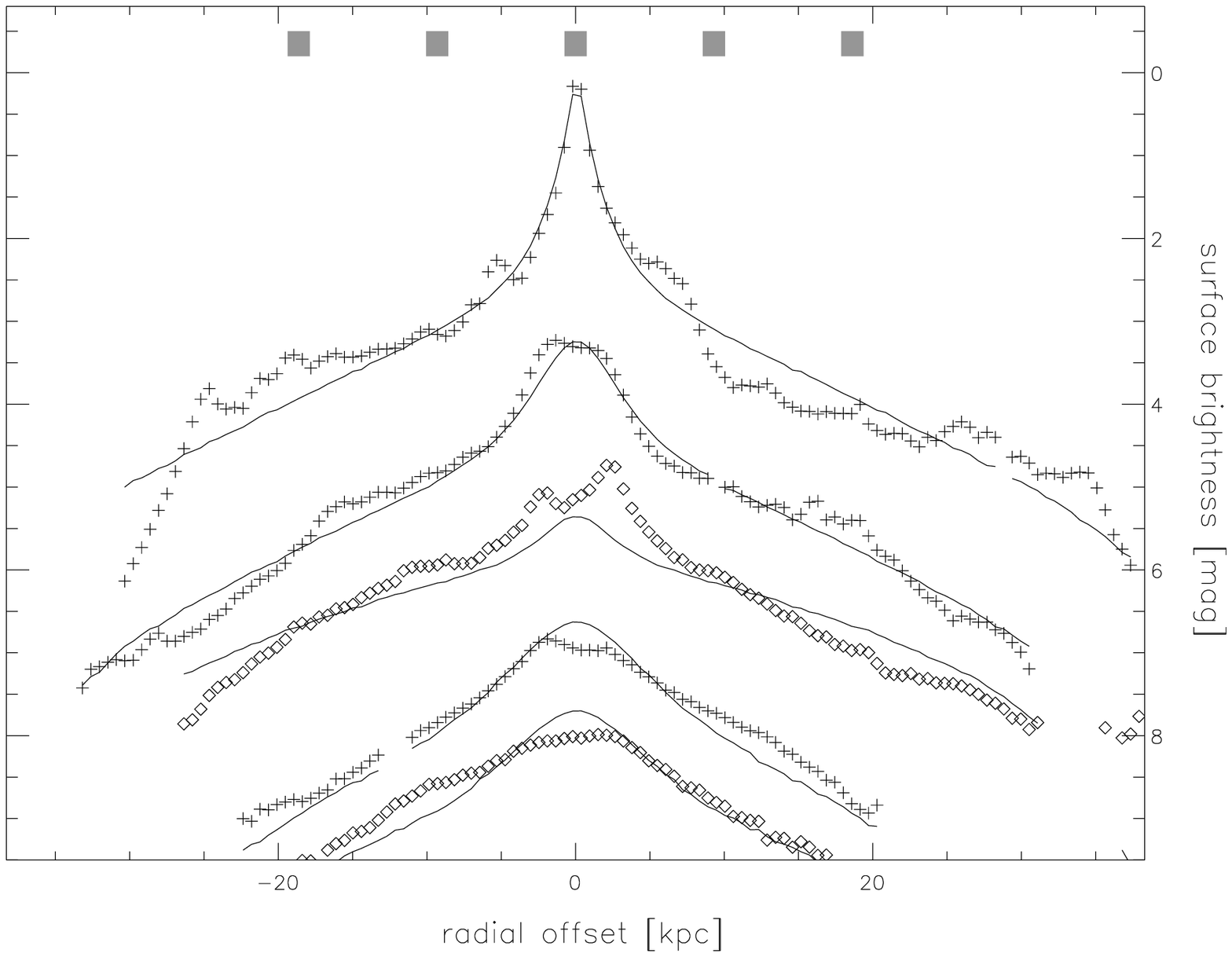}
\vspace{0.05cm}
\includegraphics[width=8.5cm]{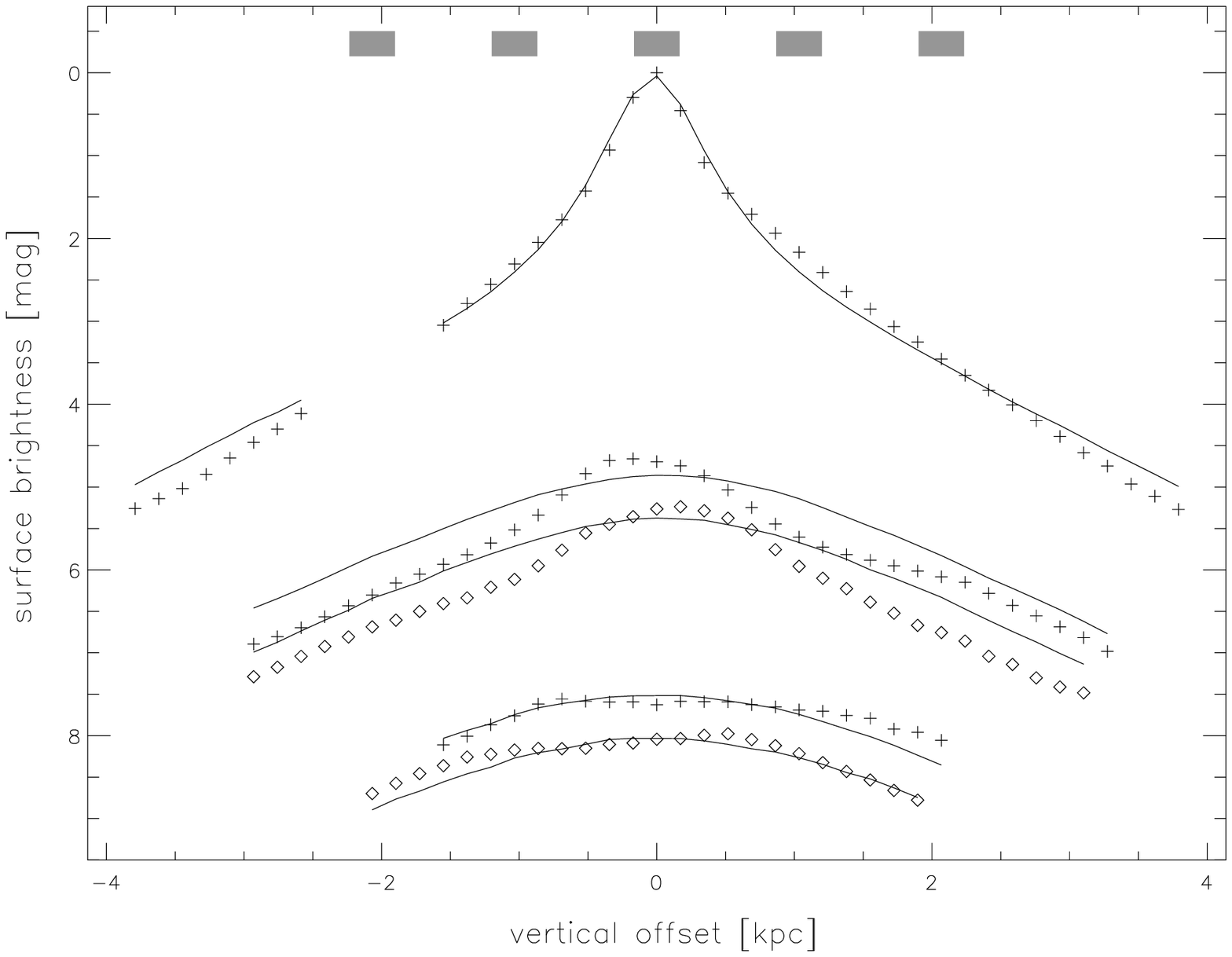}
\includegraphics[width=8.5cm]{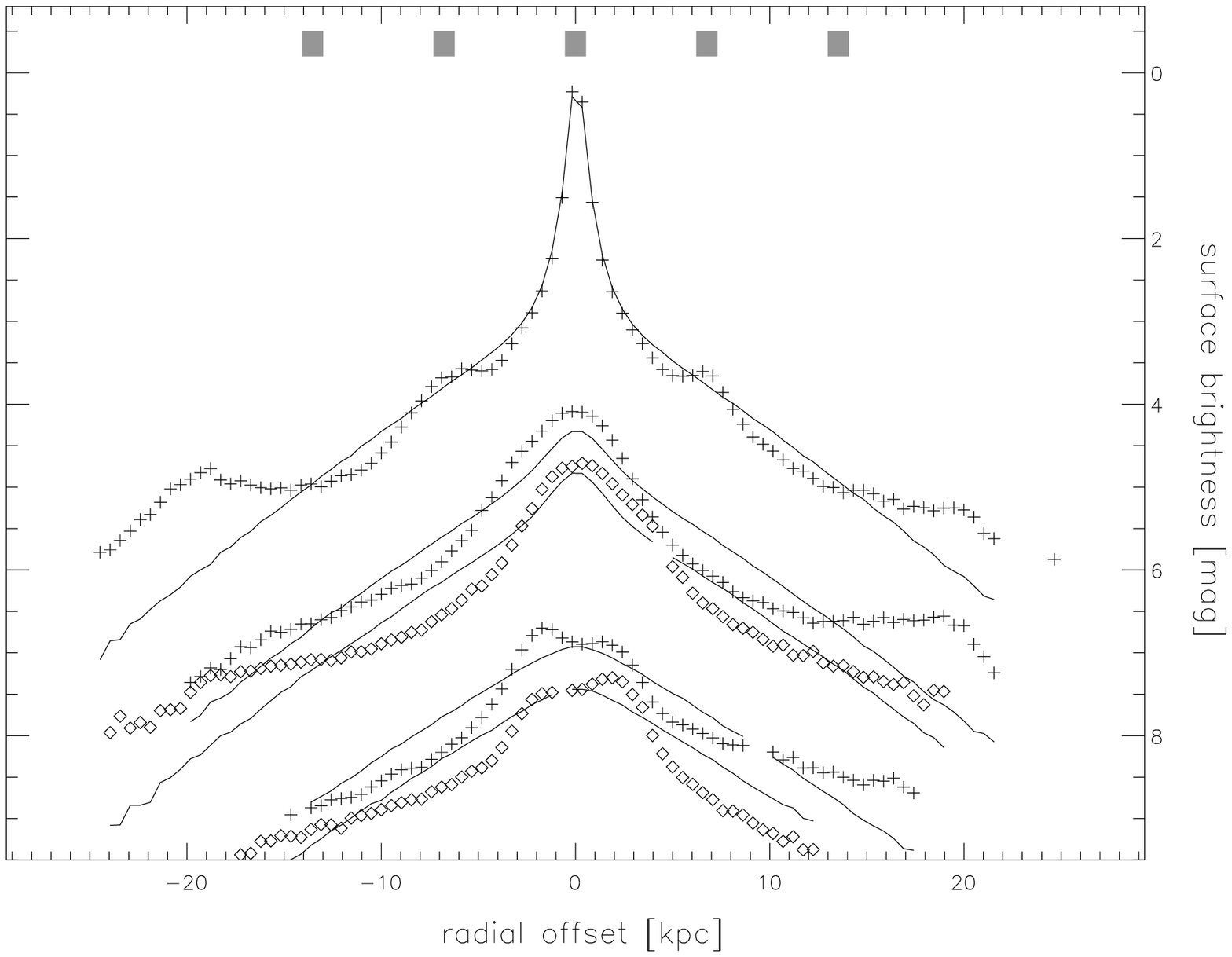}
\caption{Same as Fig.~\ref{n4013_p}, but for NGC 5965.}
\end{figure*}
}

\onlfig{16}{
\begin{figure*}
\centering
\includegraphics[width=8.5cm]{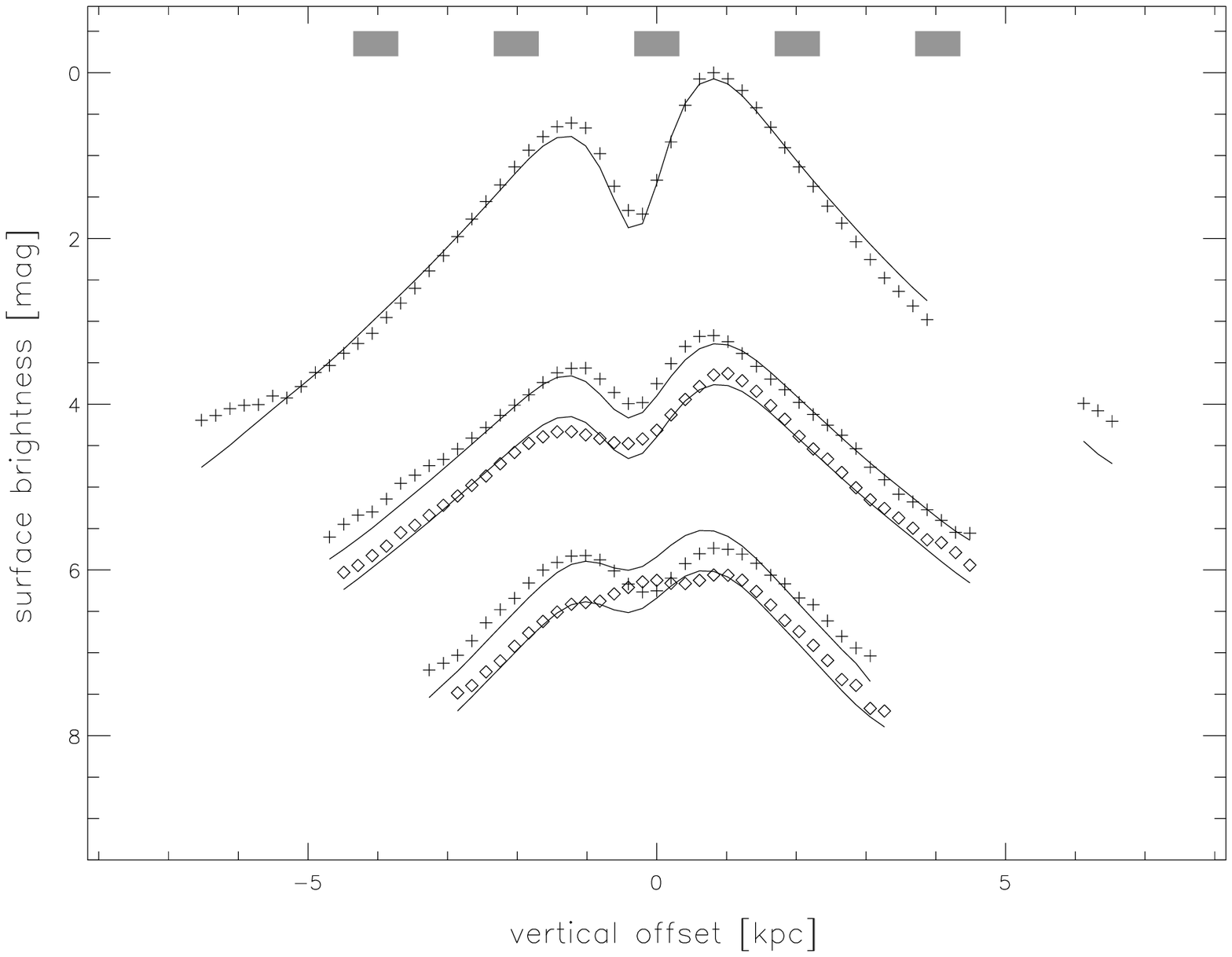}
\includegraphics[width=8.5cm]{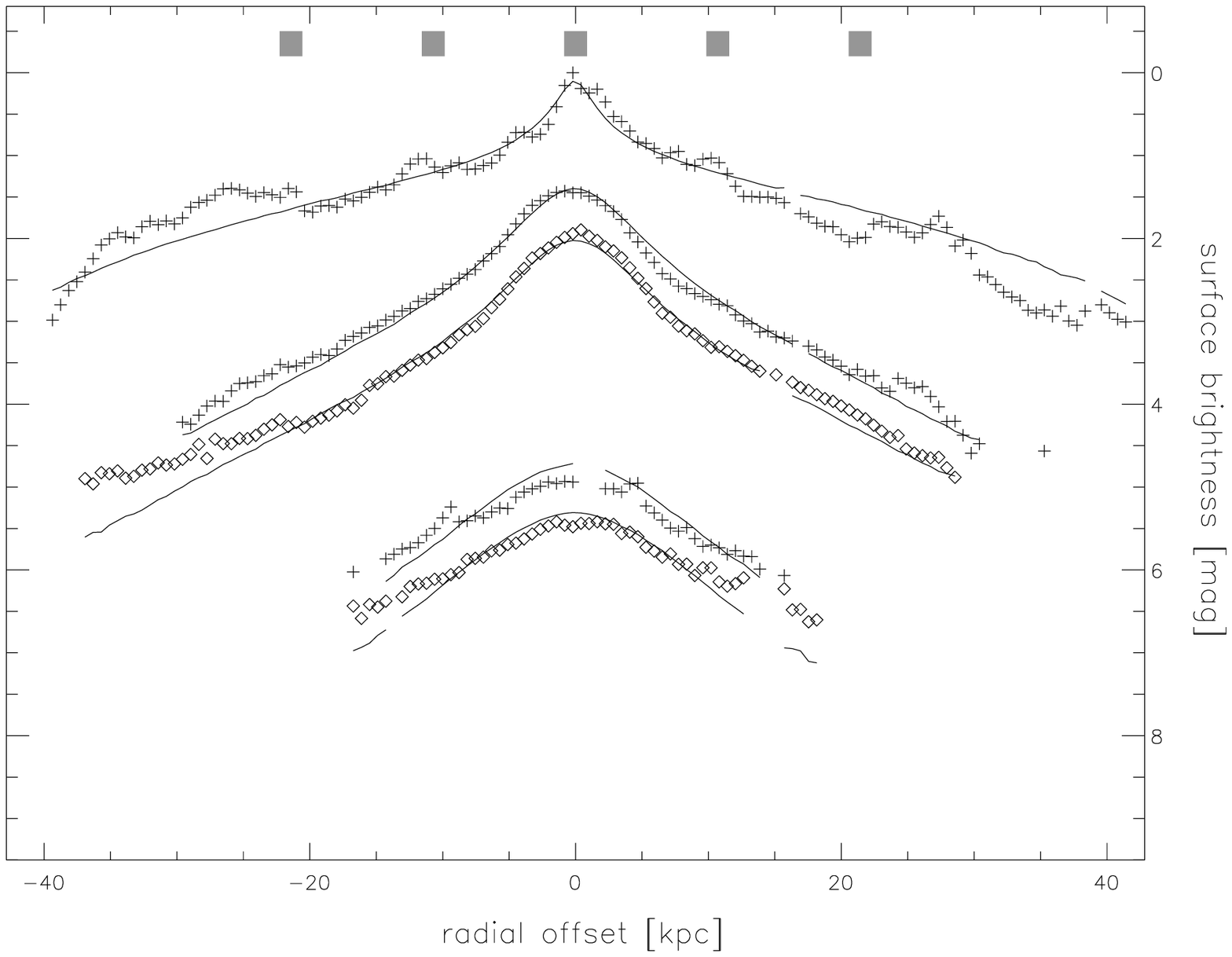}
\vspace{0.05cm}
\includegraphics[width=8.5cm]{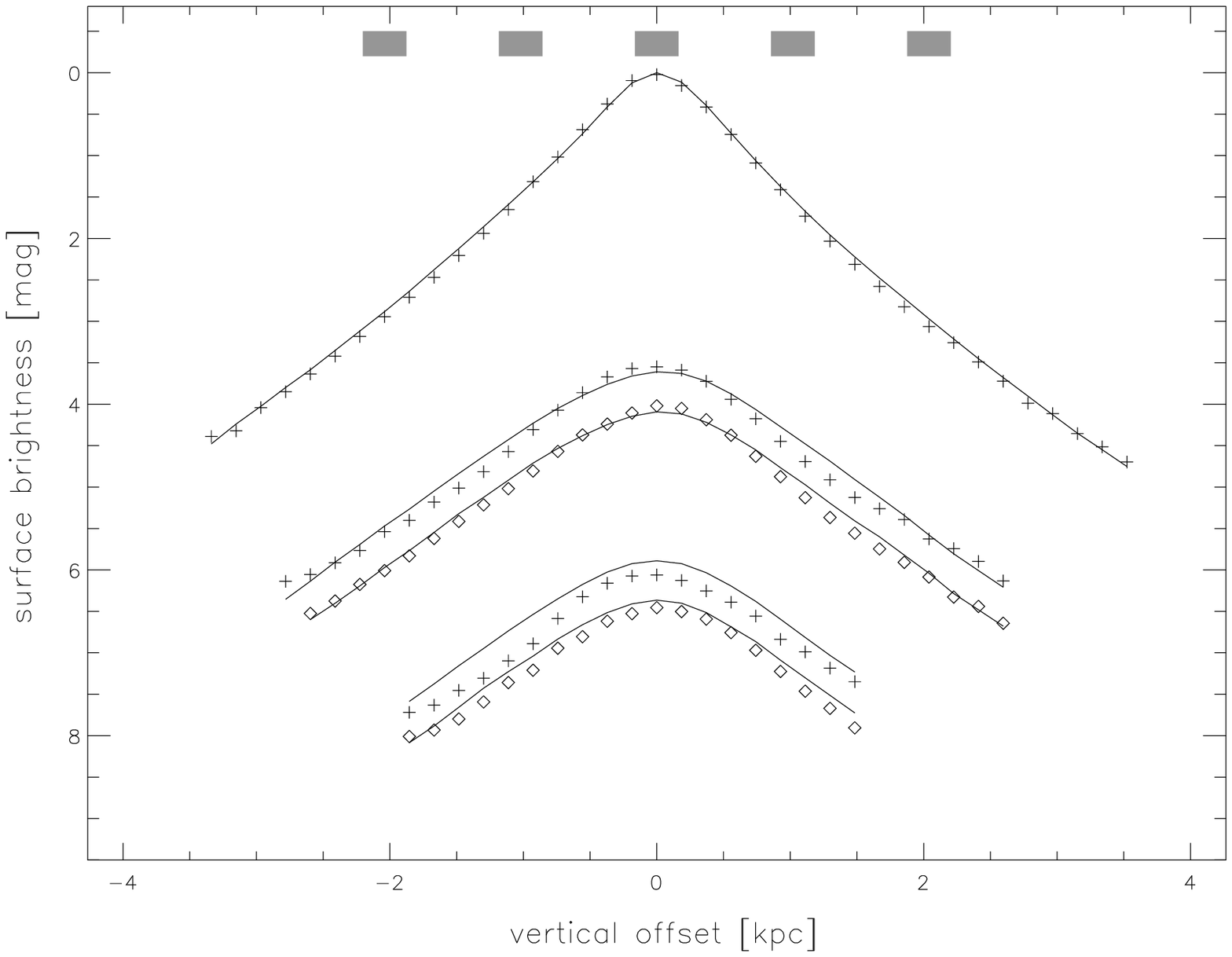}
\includegraphics[width=8.5cm]{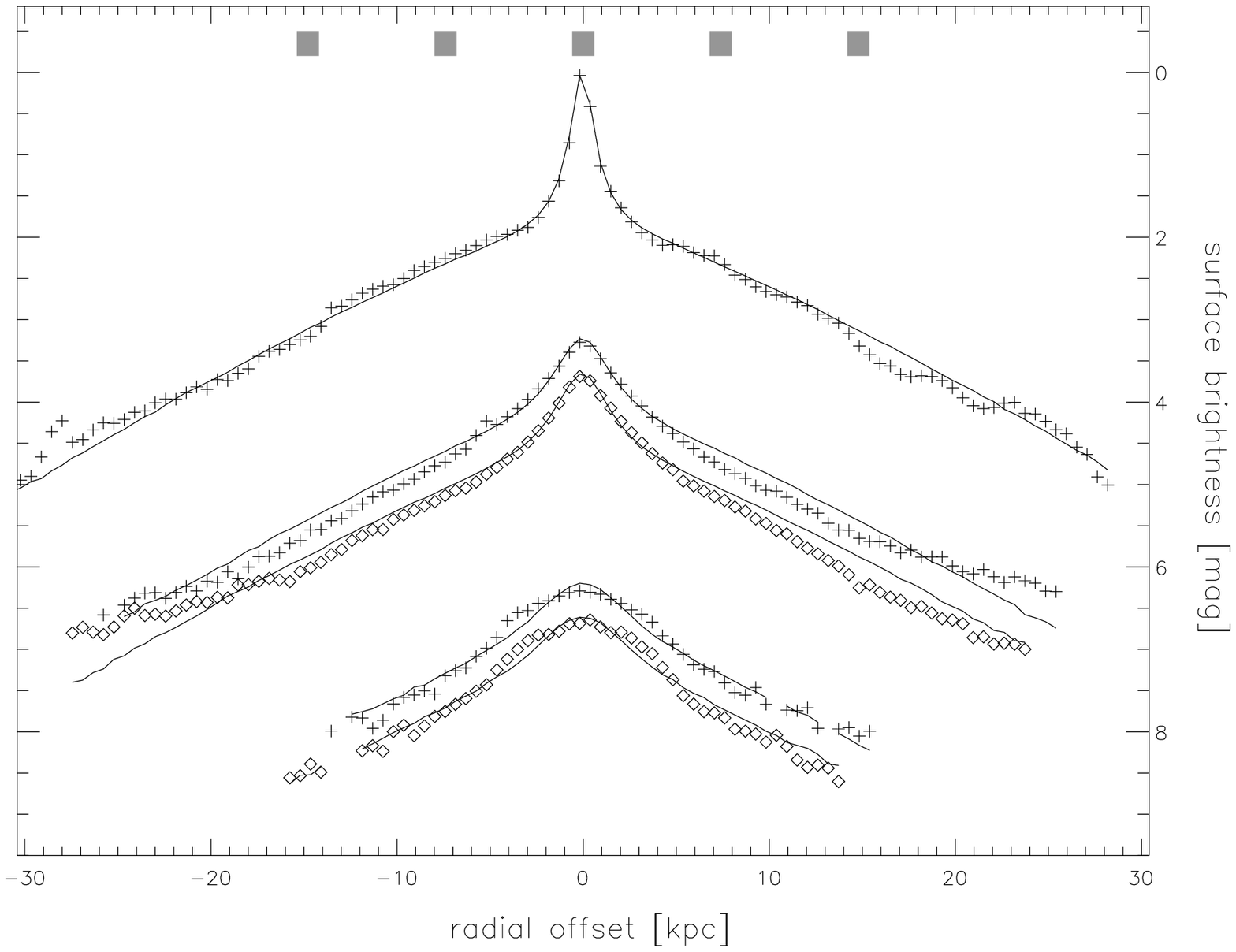}
\caption{Same as Fig.~\ref{n4013_p}, but for UGC 4277.}
\label{u4277_p}
\end{figure*}
}

\section{Summary \& discussion}
\label{summary}

In the present work a sample of seven nearby edge-on galaxies, observed
in the V and K'-band, has been analyzed using a radiative transfer model,
in order to constrain the parameters describing the stellar and dust 
distributions. The dust and stellar disks have been fitted with two smooth,
independent, exponential disks, and the bulge with a spheroidal distribution
following Sersic profiles with $n=2$ and 4. Hereafter I summarize and discuss 
the main results of the work:

\renewcommand{\labelenumi}{(\roman{enumi})}
\begin{enumerate}

\item When applying the fitting technique to mock images of the same extent 
and S/N as the observations, it became evident that the parameters cannot be 
constrained better than a few tens percent, because of the model degeneracies.
In particular, it is possible to obtain good fits even using faster radiative
transfer model which neglect scattering. For V-band models, this leads only 
to a small underestimate of the dust disk opacity, and to variation of the 
other fitted parameters within the accuracy of full fits including scattering.
Furthermore, the differences between scattering and non scattering models
are smaller than the typical residuals obtained in fits of real images.

\item In the V-band, the parameters obtained for the dust disk are generally 
consistent with those derived by \citet{XilourisSub1998} on a different sample
of objects: the dust disk has a larger radial scalelength than stars 
($h_\mathrm{d}/h_\mathrm{s} \sim 1.5$); it has a smaller vertical scalelength
($z_\mathrm{d}/z_\mathrm{s} \sim 1/3$); it is almost transparent when seen
face-on ($\tau_0 =0.5-1.5$). However, discrepancies exists when fits to single
objects are compared to other works in literature, possibly a result of different
models, fitting techniques and image coverage. It is worth noting that the largest
values for $h_\mathrm{d}$ have been obtained for two objects, NGC~5746 and 
NGC~5965, which clearly show the presence of a ring structure in the K'-band
image. If dust is distributed in a ring (indeed, extinction along the ring is
seen in the K' image of NGC~5746), the displacement of the maximum extinction out 
of the center of the galaxy could be the reason for the larger $h_\mathrm{d}$, 
when the fit is performed using the simple exponential disk of Eqn.~\ref{dustdisk}.

A similar effect could be caused by the more external clouds, if the clumpy 
structure of the dust disk dominates over its smooth component. 
\citet{MisiriotisA&A2002} have analyzed the effects of clumping on radiative
transfer fits and concluded that a clumpy distribution leads to a {\em smaller}
$h_\mathrm{d}$. The different result may be due to the fact, however, then their 
model was unable to produce high resolution images with discernible dust clouds.

\item The properties of the stellar disk are similar to those derived from
bulge/disk decomposition of less inclined galaxies. In particular, the stellar disk
has a larger $h_\mathrm{s}$ in the V-band than in the K'-band, suggesting that the 
color gradient along disks is due to the intrinsic properties of the stellar
populations rather than to the effect of extinction. The bulge component
proved to be more complex than the assumed spheroidal model, with an 'X'-like
residual betraying a box/peanut morphology in all but one galaxy. I caveat here
that the departure of the bulge shape from model assumptions can cause 
uncertainties in the derivation of the dust disk properties, since the extinction 
lane is deeper in regions where the bulge contributes most.

\item Images in the K'-band show little dust extinction, mostly consisting in
minor axis asymmetries and in clumpy structures. I have found no trace of the
second, thinner, more massive and smooth dust disk which has been introduced by 
\citet{PopescuA&A2000} and \citet{MisiriotisA&A2001} to explain the sub-mm dust
emission in NGC~891 and NGC~5907. Since the problem of the discrepancy in
{\em energy balance} between the radiation absorbed by dust from stars and
that emitted in the infrared persists, we are left with the second hypothesis
in \citet{PopescuA&A2000}: dust could be preferentially distributed in optically 
thick quiescent or star-forming clouds. Furthermore, dust in dense clouds could
have a larger emissivity than what usually assumed for Milky Way dust, thus
relieving part of the energy balance discrepancy \citep[see, e.g. ][for an 
alternative solution to the deficiency of sub-mm emission in models of 
edge-on galaxies]{DasyraA&A2005}. Realistic models for dust emission in a 
complex disk are needed, in order to ascertain the relative contribution of 
diffuse and clumpy dust to the infrared spectral energy distribution.
\end{enumerate}

Clearly, the uncertainties in the derivation of the structure of the dust disk
could be reduced by using a dust tracer more {\em direct} than its extinction 
on starlight. In particular, observations of dust emission in the sub-mm appear 
to be more easy to interpret in terms of dust density, because of the relatively
high resolution and of the small dependence of the Rayleigh-Jeans spectrum on the 
dust temperature gradient. In the next few years, the advent of high-sensitivity, 
second generation sub-mm bolometer-arrays like SCUBA-2 \citep{HollandProc2006} 
and LABOCA \citep{KreysaProc2003}, will allow to perform observations of spirals 
over a much wider extent along the disk than previously possible. Equally
important are the infrared observations of nearby galaxies carried on with
the Spitzer satellite: a large number of nearby spirals has been already
observed for the SINGS survey \citep{KennicuttPASP2003}, and analysis is ongoing
on a sample of local edge-on objects (Roelof de Jong and Benne Holwerda, private 
communication). Very promising for a high resolution study of the dust distribution 
appears to be the correlation between the PAHs emission, the ISM density 
\citep{ReganApJ2006} and the sub-mm emission \citep{HaasA&A2002}.

\begin{acknowledgements}
I am grateful to several people for their suggestions and help 
during the observations and the analysis conducted for this work:
Emmanuel Xilouris, Filippo Mannucci,  Edvige Corbelli, Andrea Ferrara, 
Francesca Ghinassi, Carlo Giovanardi, Leslie Hunt and Raffaella Schneider.
This publication makes use of data products from the Two Micron All Sky 
Survey, which is a joint project of the University of Massachusetts and 
the Infrared Processing and Analysis Center/California Institute of 
Technology, funded by the National Aeronautics and Space Administration 
and the National Science Foundation.
This research has made use of the NASA/IPAC Extragalactic Database 
(NED) which is operated by the Jet Propulsion Laboratory, California 
Institute of Technology, under contract with the National Aeronautics 
and Space Administration.
\end{acknowledgements}

\bibliographystyle{aa}
\bibliography{/home/voltumna/sbianchi/tex/DUST}

\end{document}